\documentclass[prd,nofootinbib,preprint,superscriptaddress]{revtex4-1}
\pdfoutput=1

\usepackage{graphicx}
\usepackage{epsfig}
\usepackage{amsmath}
\usepackage{amsfonts}
\usepackage{amssymb}
\usepackage{color}
\usepackage[bottom]{footmisc}
\usepackage{array,multirow,makecell}
\usepackage{slashed}
\usepackage[colorlinks,citecolor=blue]{hyperref}
\usepackage{float}
\usepackage{ulem}
\usepackage[utf8]{inputenc}
\usepackage[english]{babel}
\usepackage{url}
\usepackage{hyperref}
\usepackage{xcolor}
\usepackage{subfigure}
\usepackage{lipsum}
\usepackage[bottom]{footmisc}
\usepackage{pifont}

\usepackage[most]{tcolorbox}
\newtcbox{\mymath}[1][]{%
    nobeforeafter, math upper, tcbox raise base,
    enhanced, colframe=blue!30!black,
    colback=blue!30, boxrule=1pt,
    #1}

\usepackage{blindtext}

\def\lsim{\raise0.3ex\hbox{$\;<$\kern-0.75em\raise-1.1ex\hbox{$\sim\;$}}}
\def\gsim{\raise0.3ex\hbox{$\;>$\kern-0.75em\raise-1.1ex\hbox{$\sim\;$}}}

\def\bea{\begin{eqnarray}}
\def\eea{\end{eqnarray}}

\newcommand{\baz}{\begin{array}{cc}}

\newcommand{\bav}{\begin{array}{cccc}}

\begin{document}


\title{Non-thermal Origin of Asymmetric Dark Matter from Inflaton and Primordial Black Holes}


\author{Basabendu Barman}
\email{basabendu88barman@gmail.com}
\affiliation{Centro de Investigaciones, Universidad Antonio Nariño\\
Carrera 3 este \# 47A-15, Bogotá, Colombia}

\author{Debasish Borah}
\email{dborah@iitg.ac.in}
\affiliation{Department of Physics, Indian Institute of Technology Guwahati, Assam 781039, India}

\author{Suruj Jyoti Das}
\email{suruj@iitg.ac.in}
\affiliation{Department of Physics, Indian Institute of Technology Guwahati, Assam 781039, India}

\author{Rishav Roshan}
\email{rishav@prl.res.in}
\affiliation{Physical Research Laboratory, Ahmedabad-380009, Gujarat, India}

\begin{abstract}
We study the possibility of cogenesis of baryon and dark matter (DM) from the out-of-equilibrium CP violating decay of right handed neutrino (RHN) that are dominantly of non-thermal origin. While the RHN and its heavier partners can take part in light neutrino mass generation via Type-I seesaw mechanism, the decay of RHN into dark and visible sectors can create respective asymmetries simultaneously. The non-thermal sources of RHN considered are {\bf (a)} on-shell decay of inflaton, and {\bf (b)} evaporation of ultralight primordial black holes (PBH). After setting up the complete set of Boltzmann equations in both these scenarios, we constrain the resulting parameter space of the particle physics setup, along with inflaton and PBH sectors from the requirement of generating correct (asymmetric) DM abundance and baryon asymmetry, while being in agreement with other relevant cosmological bounds. Scenario {\bf (a)} links the common origin of DM and baryon asymmetry to post-inflationary reheating via RHNs produced in inflaton decay, whereas in scenario {\bf (b)} we find enhancement of baryon and DM abundance, compared to the purely thermal scenarios, in presence of PBH with appropriate mass and initial fraction. Although the minimal setup itself is very predictive with observational consequences, details of the UV completion of the dark sector can offer several complementary probes.

\end{abstract}
%
\begin{flushright}
  PI/UAN-2021-705FT \\
\end{flushright}
\maketitle

{
  \hypersetup{linkcolor=black}
  \tableofcontents
}

\section{Introduction}
\label{sec:intro}
Observational evidences~\cite{Zwicky:1933gu, Zwicky:1937zza, Rubin:1970zza, Clowe:2006eq} suggest that the present universe is composed of about $32\%$ matter out of which only $\sim 5\%$ is in the form of visible matter or baryons. The remaining $\sim 27\%$ is in the form of a mysterious, non-luminous and non-baryonic form of matter, popularly known as the dark matter (DM). In terms of density 
parameter $\Omega_{\rm DM}$ and reduced Hubble constant $h = \text{Hubble Parameter}/(100 \;\text{km} ~\text{s}^{-1} 
\text{Mpc}^{-1})$, the present DM abundance is conventionally reported as \cite{Aghanim:2018eyx}

\begin{equation}
\Omega_{\text{DM}} h^2 = 0.120\pm 0.001
\label{dm_relic}
\end{equation}

\noindent at 68\% CL. The visible or baryonic matter content is also highly asymmetric, giving rise to the longstanding puzzle of baryon asymmetry of the universe (BAU). While any non-zero primordial asymmetry chosen as an initial condition will be diluted by the exponentially expanding phase of inflation, it is also natural for the universe to start in a baryon symmetric manner. The observed BAU is quantitatively quoted as the ratio of excess of baryons over anti-baryons to photon \cite{Aghanim:2018eyx} 
\begin{equation}
\eta_B = \frac{n_{B}-n_{\overline{B}}}{n_{\gamma}} \simeq 6.2 \times 10^{-10}.
\label{etaBobs}
\end{equation}
The quoted value of baryon to photon ratio based on the cosmic microwave background (CMB) measurements, agrees with the big bang nucleosynthesis (BBN) estimates as well ~\cite{Zyla:2020zbs}. While the origin of this asymmetry is not known, the particle nature of DM is also a mystery. However, we do know that none of the standard model (SM) particles satisfy the criteria of a viable particle DM candidate. In addition, the SM also fails to satisfy the criteria (known as Sakharov's conditions~\cite{Sakharov:1967dj}) in adequate amounts to dynamically generate the observed BAU. This has led to several beyond the SM (BSM) proposals offering possible solutions to these puzzles. As it is well known, the weakly interacting massive particle (WIMP) paradigm has been the most widely studied particle DM scenario \cite{Kolb:1990vq, Jungman:1995df, Bertone:2004pz, Feng:2010gw, Arcadi:2017kky, Roszkowski:2017nbc}, while out-of-equilibrium decay of a heavy particle leading to the generation of baryon asymmetry has been a very well known mechanism for baryogenesis \cite{Weinberg:1979bt, Kolb:1979qa}. One interesting possibility to achieve baryogenesis via lepton sector physics is known as leptogenesis \cite{Fukugita:1986hr} where, instead of creating a baryon asymmetry directly, a lepton asymmetry is generated first which subsequently gets converted into baryon asymmetry by the $(B+L)$-violating electroweak sphaleron transitions~\cite{Kuzmin:1985mm}. 

The popular BSM frameworks mentioned above can certainly explain the origin of DM and BAU independently, but an intriguing observation is the similarity in their abundances $\Omega_{\rm DM} \approx 5\,\Omega_{\rm Baryon}$, within the same order of magnitude. Ignoring the possibility of a numerical or mere cosmic coincidence, one has to provide a dynamical origin behind such a serendipity. There have been several works in pursuit of finding a common origin for DM and baryon asymmetry, a brief review of which can be found in \cite{Boucenna:2013wba}. This broadly falls into two categories. In the first one, the usual mechanism for baryogenesis is extended to the dark sector assuming the dark sector to be asymmetric \cite{Nussinov:1985xr, Davoudiasl:2012uw, Petraki:2013wwa, Zurek:2013wia}. In typical asymmetric dark matter (ADM) scenario, the same out-of-equilibrium decay of a heavy particle into baryon and dark sector can give rise to asymmetries in the two sectors of similar order of magnitudes $n_B-n_{\overline{B}} \sim \lvert n_{\rm DM}-n_{\overline{ \rm DM}} \rvert$. The second approach is to produce such asymmetries through annihilations \cite{Yoshimura:1978ex, Barr:1979wb, Baldes:2014gca}, where one or more particles involved in the process eventually go out of thermal equilibrium to generate a net asymmetry. The so-called WIMPy baryogenesis \cite{Cui:2011ab, Bernal:2012gv, Bernal:2013bga} belongs to this category, where a DM particle freezes out to generate its own relic abundance while simultaneously producing an asymmetry in the baryon sector. The idea extended to leptogenesis is called WIMPy leptogenesis \cite{Kumar:2013uca, Racker:2014uga, Dasgupta:2016odo, Borah:2018uci, Borah:2019epq, Dasgupta:2019lha}.

Motivated by these, we consider a simple realisation of the asymmetric dark matter scenario where out-of-equilibrium decay of heavy right handed neutrinos (RHN) can play the role in generating both dark and visible sector asymmetries simultaneously \cite{Falkowski:2011xh,DuttaBanik:2020vfr}. The same RHNs can also give rise to light neutrino masses via Type-I seesaw mechanism \cite{Mohapatra:1979ia,Yanagida:1979as,GellMann:1980vs,Glashow:1979nm}. In such generic seesaw models, the scale of thermal leptogenesis remains high pushing the scale of RHN to a very high scale $M >10^9$ GeV, known as the {\it Davidson-Ibarra bound}~\cite{Davidson:2002qv}. However, the reheat temperature of the universe after inflation $(T_{\rm RH})$ can be much lower forbidding the thermal production of RHNs. In such a scenario, the RHNs can still be produced non-thermally from the inflaton field, leading to the scenario of non-thermal leptogenesis~\cite{Lazarides:1991wu,Murayama:1992ua,Kolb:1996jt,Giudice:1999fb,Asaka:1999yd,Asaka:1999jb,Hamaguchi:2001gw,Jeannerot:2001qu,Fujii:2002jw,Giudice:2003jh,Pascoli:2003rq,Asaka:2002zu,Panotopoulos:2006wj,HahnWoernle:2008pq,Buchmuller:2013dja,Croon:2019dfw,Borah:2020wyc,Samanta:2020gdw, Barman:2021tgt}. Also, there exists no experimental evidence to suggest that the universe was radiation dominated prior to the BBN era and existence of some non-standard cosmological phase can alter the predictions of high scale phenomena like leptogenesis. A recent review of such non-standard cosmology can be found in \cite{Allahverdi:2020bys}, while its effects on leptogenesis have been discussed in several works including \cite{Abdallah:2012nm, Dutta:2018zkg, Chen:2019etb, Mahanta:2019sfo, Konar:2020vuu, Chang:2021ose, JyotiDas:2021shi}. In the present work we consider two different sources of non-thermal RHNs namely, (a) inflaton decay, where we assume that the inflaton decays {\it only} into a pair of RHNs and (b) evaporation of primordial black holes (PBH), and in each case we study study the consequences for generation of dark and visible sector asymmetries. While the first case connects the cogenesis of DM and baryons to the post-inflationary reheating via non-thermal RHNs produced from inflaton, the presence of PBH with appropriate mass and initial fraction can lead to enhancement of baryon and asymmetric DM abundance compared to the purely thermal case.

This paper is organised as follows. In section \ref{sec:model} we discuss our minimal setup. Section~\ref{sec:asdm} and~\ref{sec:asdm2} are dedicated to the details of asymmetric dark matter production from inflaton and primordial black holes respectively. In section~\ref{sec:uvmodel}, we sketch a possible UV completion of the dark sector, and finally conclude in section~\ref{sec:concl}.

\section{The Minimal Setup}
\label{sec:model}

We start with a toy model to motivate our scenario and later provide its possible UV-complete manifestation. The minimum ingredients to generate the dark and visible sector asymmetries are adopted from \cite{Falkowski:2011xh}. The SM particle content is extended by two RHNs sufficient to fit light neutrino data via Type-I seesaw mechanism along with two dark sector particles namely a singlet scalar $\mathcal{S}$, a singlet Dirac fermion $\chi$ both odd under an in-built $\mathbb{Z}_2$ symmetry. The out-of-equilibrium CP-violating decay of these RHNs to the visible as well as to the dark sector produce asymmetry in both the sectors simultaneously. The interaction Lagrangian for such a scenario can be expressed as

\bea\begin{aligned}
& -\mathcal{L}\supset\frac{1}{2}\,M_N\,\overline{N^c}N+y_N\,\overline{N}\,\tilde{H}^{\dagger}\,\ell\,+m_\chi\,\overline{\chi}\,\chi+y_\chi\,\overline{N}\,\mathcal{S}\,\chi \, +{\rm h.c.}\,,     
    \end{aligned}\label{eq:lgrng}
\eea

\noindent where we extend the SM particle spectrum by adding two generations of RHN $N_i$, singlet under the SM gauge symmetry. The SM leptons are denoted by $\ell$. A lepton number is assigned to the RHN such that its Majorana mass term is lepton number violating. Although three copies of RHNs are considered in typical Type-I seesaw model, two are sufficient to fit light neutrino data. The details of light neutrino mass generation and parametrisation of the lepton-RHN coupling in terms of observed  parameters of neutrino mass and mixing, are given in Appendix~\ref{appen1}.

The singlet fermion $\chi$ plays the role of viable DM candidate and carries a lepton number same as that of $N$. We consider this fermion to be vector like such that a bare mass term $m_\chi$ can be assigned to it without violating any $U(1)$ symmetry like global lepton number. To ensure the stability of the DM we impose an ad-hoc $\mathbb{Z}_2$ symmetry under which both the singlet scalar $\mathcal{S}$ and the single fermion $\chi$ are odd, while all other particles are even. This also implies, $m_\chi<m_\mathcal{S}<M_i$ such that the singlet fermion is the only DM in the present particle spectrum\footnote{Possibility of either of them to be DM, depending on the mass hierarchy, has been addressed in~\cite{Falkowski:2011xh}.}. Note that, the scalar $\mathcal{S}$ is devoid of any vacuum expectation value (VEV) such that there is no mixing between the DM and the RHN that may lead to DM decay\footnote{Non-zero VEV can result in DM decay into SM states that can have observational consequences~\cite{Kusenko:2009up,Falkowski:2011xh}.}. The detailed phenomenology of this minimal setup has been discussed in  \cite{Falkowski:2011xh} by considering thermal RHNs. For thermal RHNs, the subsequent phenomenology is insensitive to early universe histories. In this work, we consider non-thermal RHNs to be the dominant source of asymmetries. As we discuss below, inflaton and primordial black holes can play non-trivial roles in producing such non-thermal RHNs in the early universe leading to subsequent asymmetries in visible and dark sectors.


\section{Asymmetric Dark Matter from Inflaton Decay}\label{sec:asdm}
In the standard vanilla leptogenesis \cite{Buchmuller:2004nz} as well as in minimal ADM framework, the decaying particles (RHNs in our setup) are produced thermally from the SM bath. However, the lower bound on RHN mass in such scenarios (Davidson-Ibarra bound, mentioned before), leads to a lower bound on the reheat temperature $T_{\rm RH} \geq 10^{10}$ GeV~\cite{Davidson:2002qv, Buchmuller:2005eh} so that the RHNs can be produced from the thermal bath. While there is no observational evidence to suggest such a high reheat temperature, one also faces the gravitino  overproduction problem in supersymmetric scenarios for such high $T_{\rm RH}$ \cite{Kawasaki:2004qu}. One suitable alternative is to consider non-thermal production of RHNs. In this section, we consider the inflaton decay into RHNs as a possible source. Our set-up is based on the assumption that the inflaton field $\varphi$ decays {\it only} into a pair of RHNs~\cite{Lazarides:1991wu,Hahn-Woernle:2008tsk,Barman:2021tgt} via the coupling

\bea\begin{aligned}
& -\mathcal{L}\supset y_\varphi\,\varphi\,\,\overline{N^c}\,N \,,     
    \end{aligned}\label{eq:lgrng1a}
\eea

We consider inflaton coupling to other particles like DM $\chi$: $\varphi\,\overline{\chi}\,\chi$, singlet scalar $\mathcal{S}$: $\varphi S S$, SM Higgs $H$: $\varphi H^{\dagger} H$ to be absent for simplicity. This ensures the inflaton to transfer its entire energy density into the RHNs which subsequently decays into other light degrees of freedom (DOF), leading to the required reheating of the universe. Without going into the details of the dynamics of inflation, in the present scenario we look into the post slow-roll era when the inflaton energy starts converting into the energy of the heavy neutrinos. This helps us to perform the analysis without worrying about the details of inflationary model or inflation potential. Similar approach can be found in \cite{Garcia:2020eof} which considered the separation of the period of inflation from reheating along with references for specific models of inflation which allow this possibility.

\begin{figure}[htb!]
$$
\includegraphics[scale=0.35]{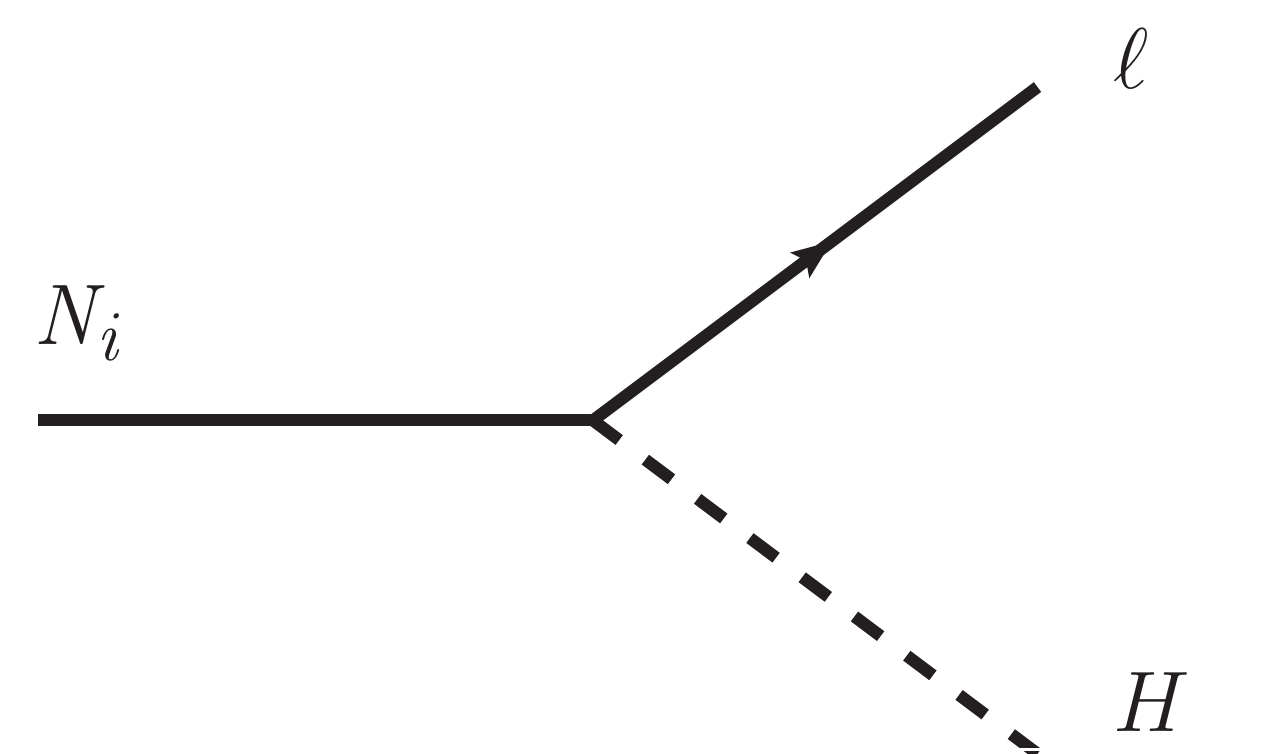}~~
\includegraphics[scale=0.35]{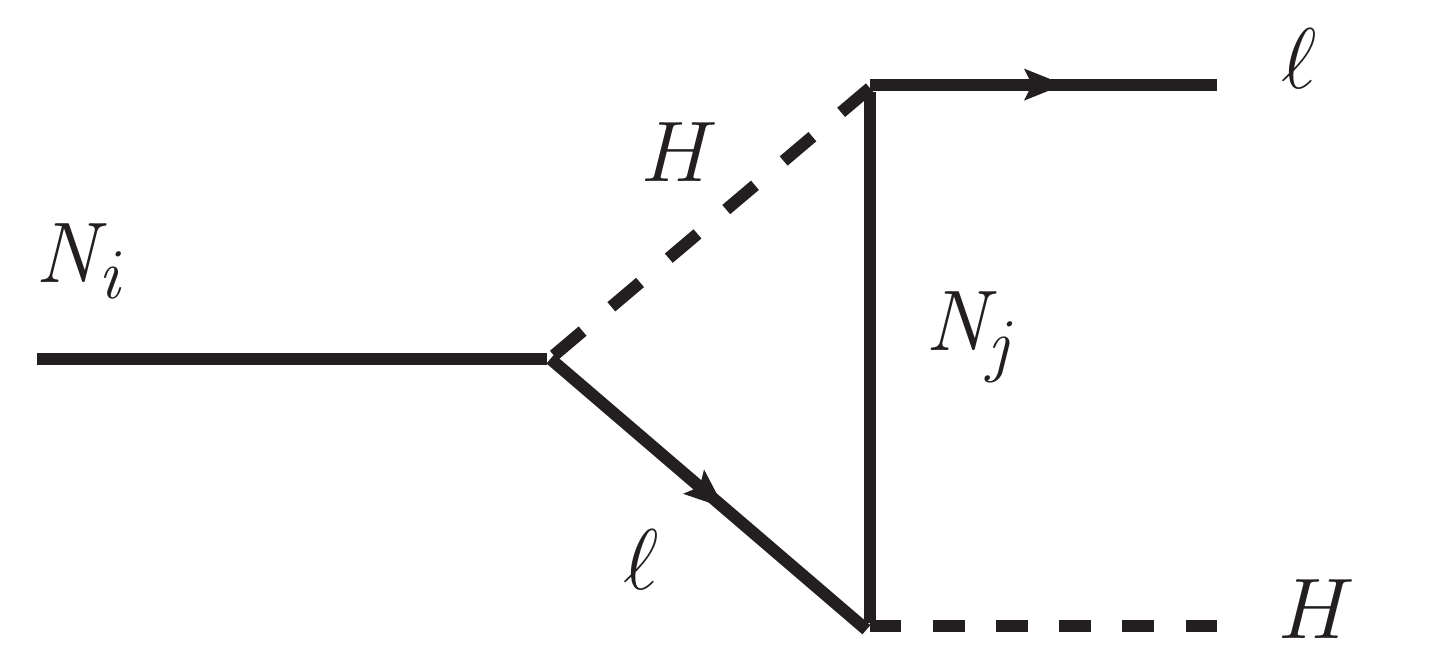}
$$
$$
\includegraphics[scale=0.35]{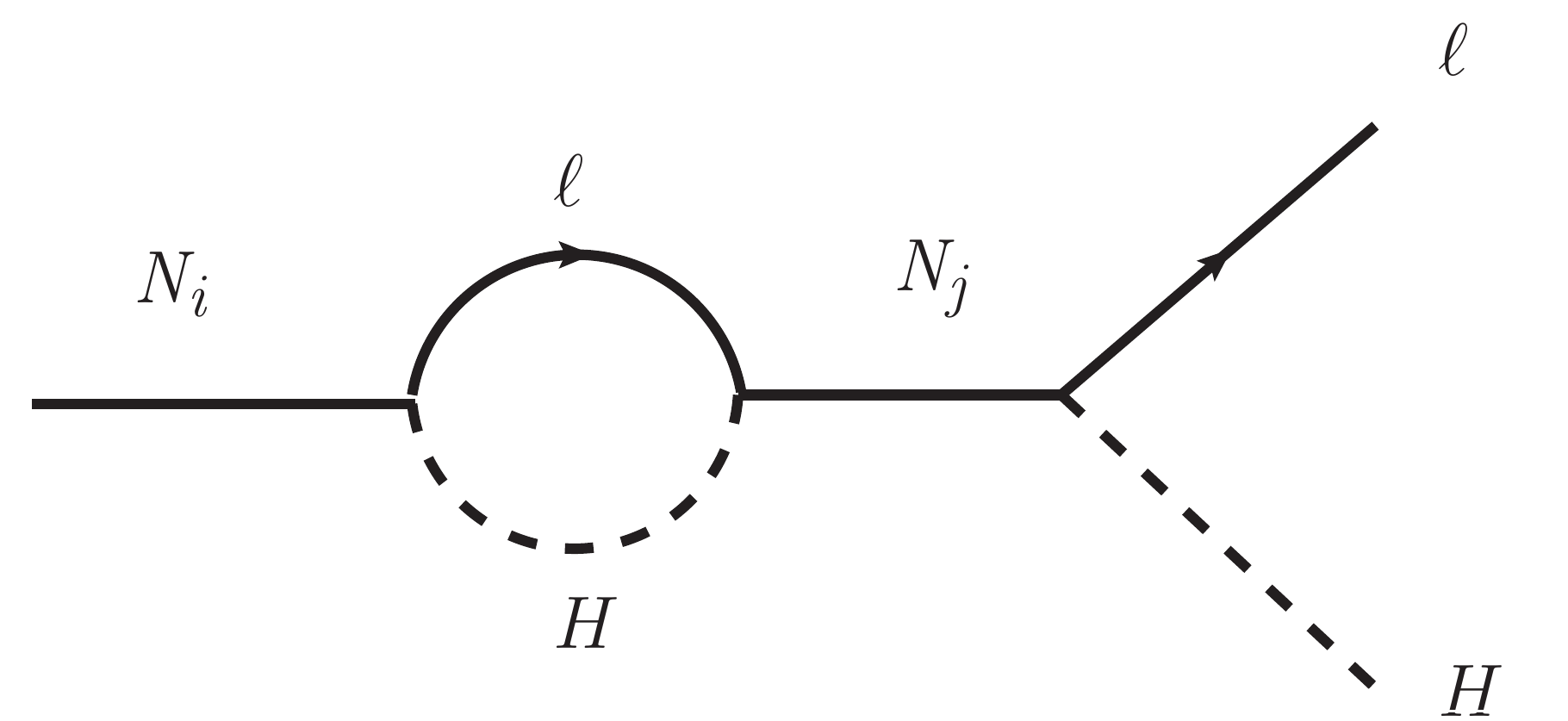}~~
\includegraphics[scale=0.35]{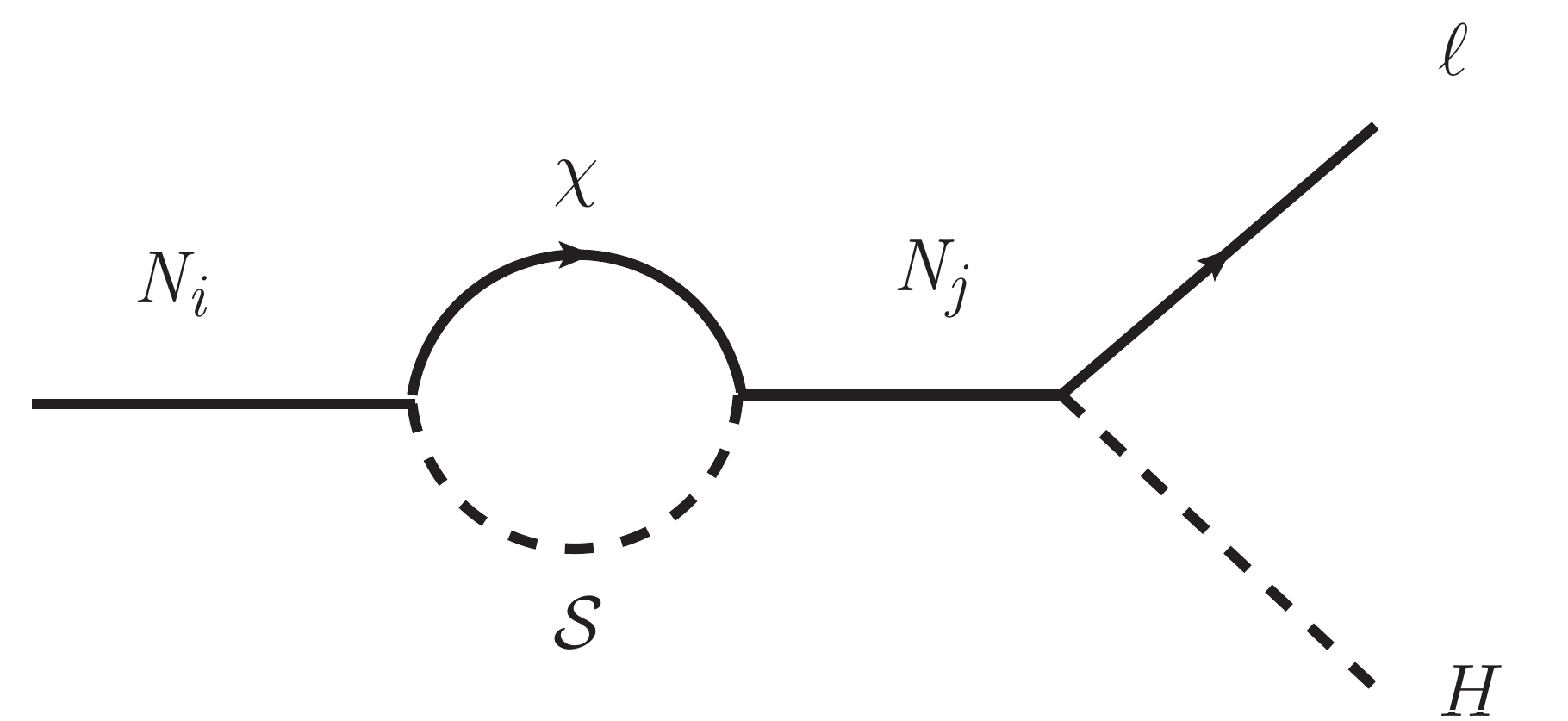}
$$
\caption{Tree level, vertex and the self-energy diagrams required for the generation of the asymmetry in the lepton sector.}\label{fig:feyn-vis}
\end{figure}

\begin{figure}[htb!]
$$
\includegraphics[scale=0.35]{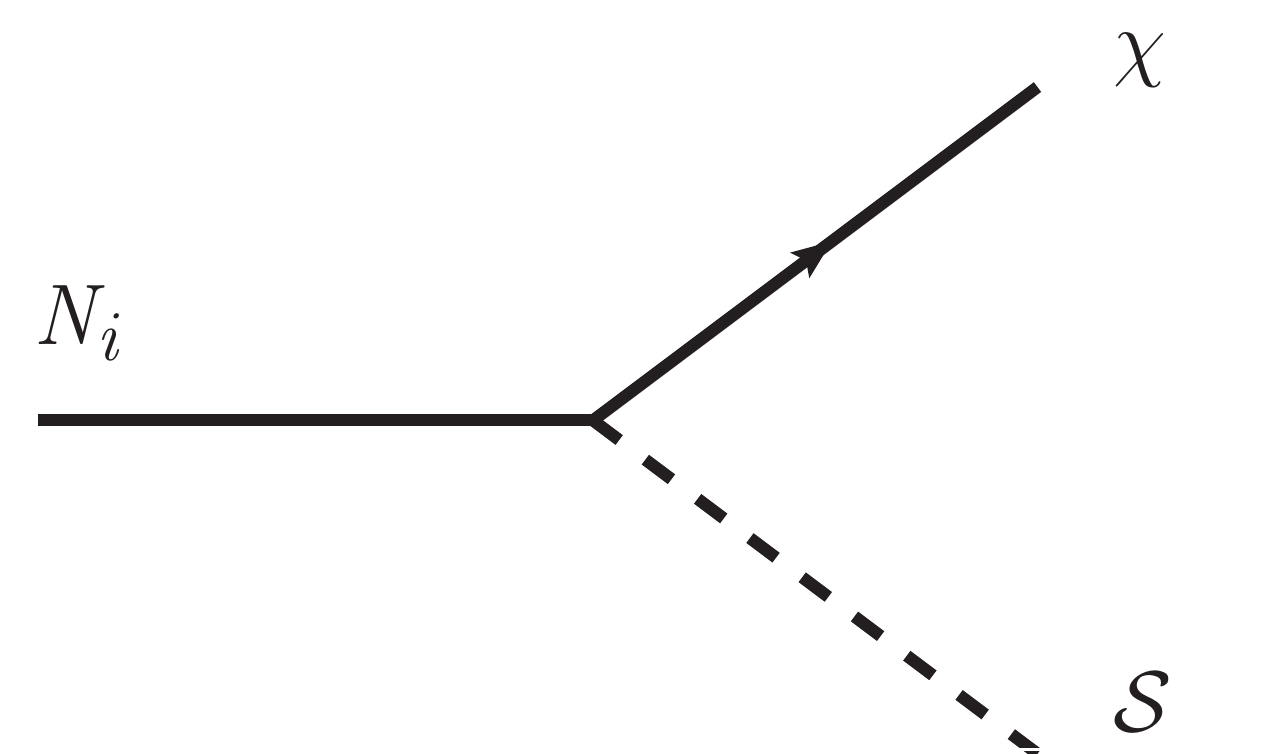}~~~
\includegraphics[scale=0.35]{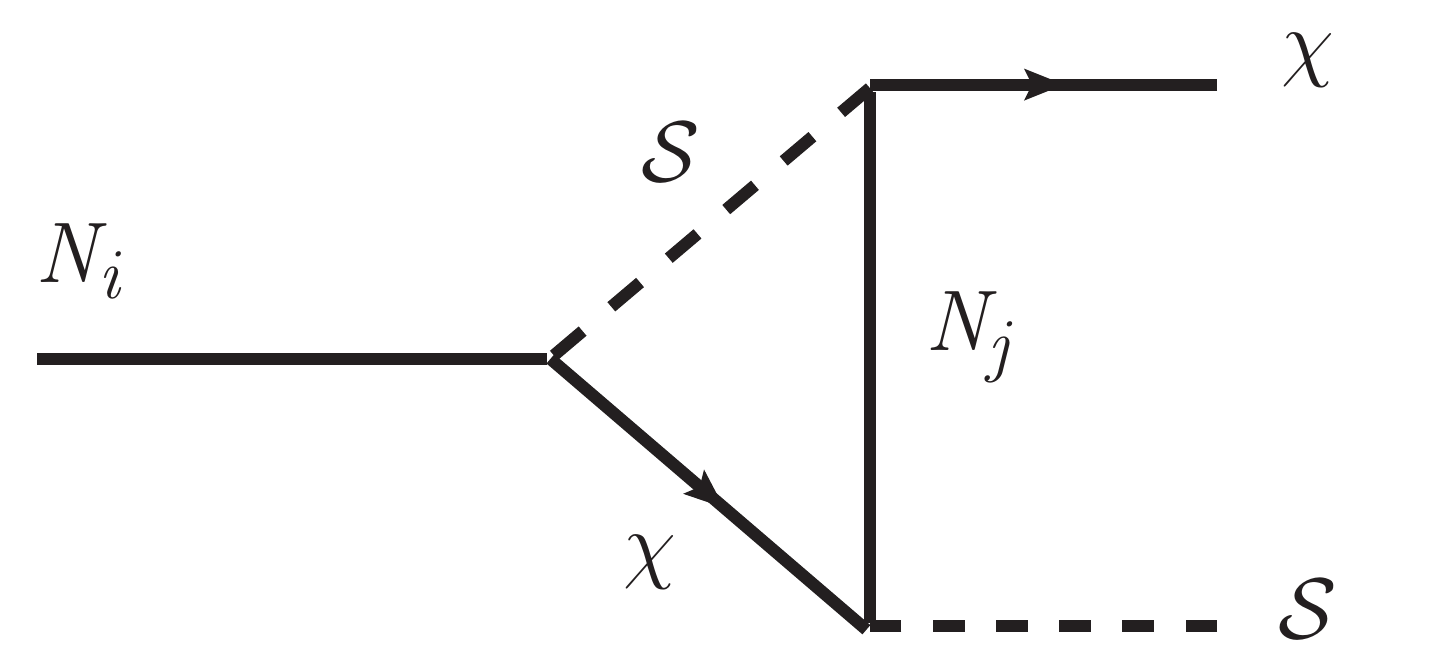}
$$
$$
\includegraphics[scale=0.35]{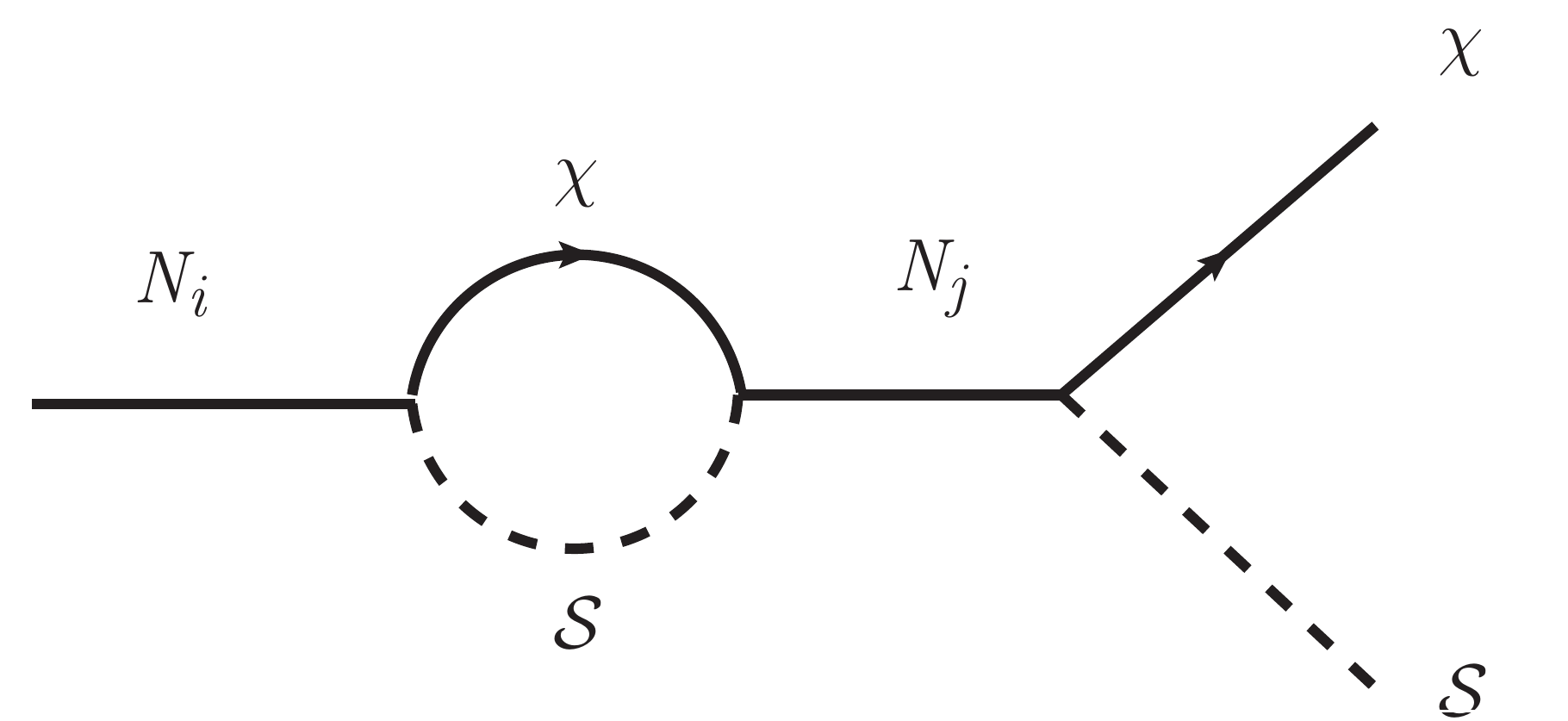}~~~
\includegraphics[scale=0.35]{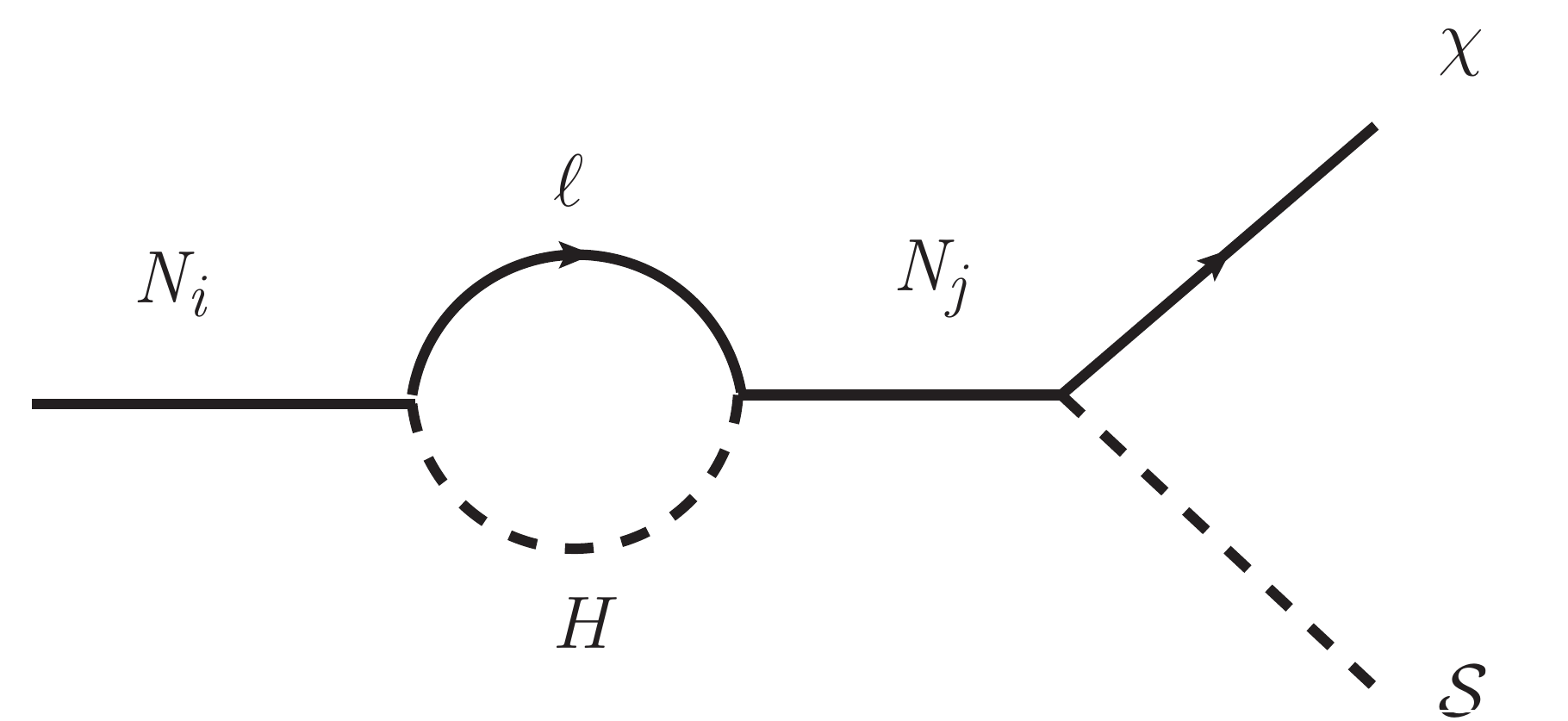}
$$
\caption{Same as Fig.~\ref{fig:feyn-vis} but for the generation of the asymmetry in the dark sector.}\label{fig:feyn-dark}
\end{figure}

As advocated in the beginning, we are interested in the scenario where the asymmetry in the visible and in the DM sector are simultaneously generated from the non-thermal decay of RHNs where the latter originate solely from the decay of inflaton. Assuming the symmetric component of the DM being washed out, it is the asymmetry $n_\chi-n_{\bar{\chi}}$ in number densities of DM particles that determines the DM abundance in late universe. Since the DM $\chi$ carries a lepton number, hence lepton number asymmetries are generated in both the sector. We also assume $M_1\ll M_2$ i.e, we can integrate out $N_2$ and consider contribution only from $N_1$ decay.  The decay of $N_1$ to the SM final states also generate the thermal bath in this process. In Fig.~\ref{fig:feyn-vis} and Fig.~\ref{fig:feyn-dark} we depict the relevant Feynman diagrams that generate these asymmetries. Below we express the CP asymmetries produced in the two sectors from the decay of the lightest RHN $N_1$~\cite{Falkowski:2011xh}:

\bea
\epsilon_{\Delta L}  &= & \frac{\sum_{\alpha}[\Gamma(N_1 \rightarrow l_{\alpha}+H)-\Gamma(N_1 \rightarrow \overline{l_{\alpha}}+H^{*})]}{\Gamma_1}
\\
&\simeq& \frac{M_1}{8\,\pi} \frac{{\rm Im}[(3\,y_N^*\,y_N^T + y_\chi^*\,y_\chi^T)\,
M^{-1} y_N\, y_N^\dagger]_{11}}{[2\,y_N\,y_N^\dagger + y_\chi\,y_\chi^{\dagger}]_{11}}\,,
\label{eq:asyL}
\eea
and
\bea
\epsilon_{\Delta\chi} &=&
\frac{\Gamma(N_1 \rightarrow \chi+S)-\Gamma(N_1 \rightarrow \bar{\chi}+S^{*})}{\Gamma_1} 
\\
& \simeq &  \frac{M_1}{8\,\pi} \frac{{\rm Im}[(y_N^*\,y_N^T + y_\chi^*\,y_\chi^T) 
M^{-1} y_\chi\,y_\chi^\dagger]_{11}}{[2\,y_N\,y_N^\dagger + y_\chi\,y_\chi^{\dagger}]_{11}}\, ,
\label{eq:asyD}
\eea

where 

\bea
\Gamma_1=\frac{M_{1}}{16\,\pi}\,(2\,y_N\,y_N^{\dagger}+y_\chi\,y_\chi^{\dagger})_{11}\,,
\label{eq:N1decay}
\eea

\noindent is the total decay width of $N_1$ and $M=\text{diag}(M_1\,,M_2)$ is the diagonal RHN mass matrix considering two RHNs. We choose hierarchical RHN mass spectrum with $M_2=50\,M_1$ such that the inflaton only decays into $N_1$. Since we have a single generation of 
$\chi$, the $y_{\chi}$ matrix can be taken, in general, to be of the form

\bea
{y_{\chi }} = \left(
\begin{array}{c}
 y_{\chi 1} \\
y_{\chi 2} \\
\end{array}
\right).
\label{yD}
\eea

\noindent For the analysis purpose, we assume $y_{\chi i}$ to be real and identical, denoted by $y_{\chi}$. It is noteworthy that even with real $y_\chi$ we are being able to generate adequate CP-violation thanks to the complex Yukawa couplings in the visible sector. Moreover, relative difference in the two asymmetries depend upon the branching ratio of $N_1$ decay and the washout, transfer effects in case of thermal leptogenesis. However, since we are considering non-thermal leptogenesis scenario where the RHN mass is larger than the reheating temperature, these effects are sub-dominant and we ignore them in our analysis.

\subsection{Evolution of Yields}\label{sec:cbeq}

\begin{figure}[htb!]
$$
\includegraphics[scale=0.14]{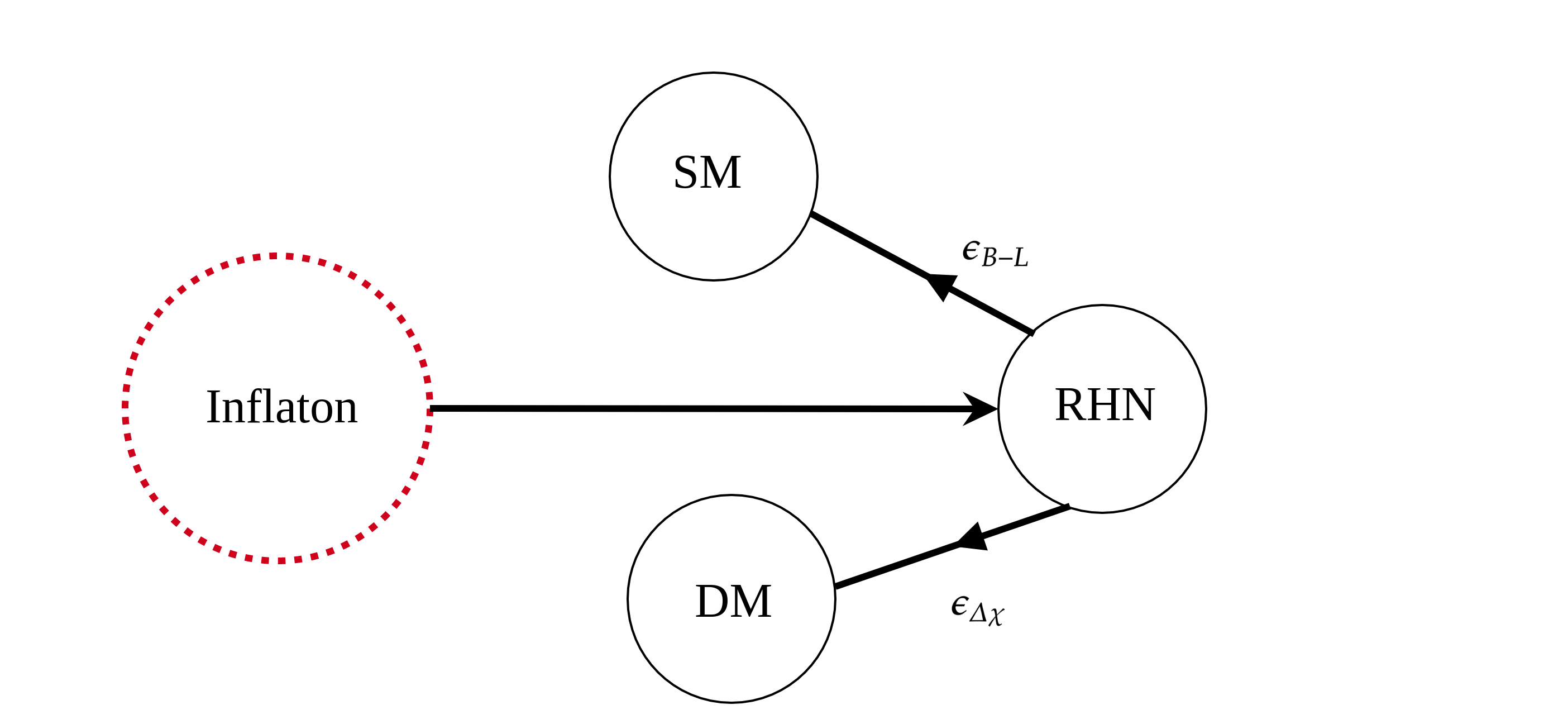}
$$
\caption{Schematic diagram of asymmetry production in the dark and in the visible sector from RHN produced from inflaton decay.}\label{fig:scheme}
\end{figure}

Since inflaton decays into RHNs only which subsequently decays into radiation while producing dark and visible sector asymmetries at the same time, it is expected that the evolution of yields for different components are interlinked, and can be expressed in terms of a set of coupled Boltzmann equations (BEQ). This is given in Eq.~\eqref{eq:cpld-beq1}. The setup can also be understood from the cartoon in Fig.~\ref{fig:scheme}, where we show how different components are linked to each other. The set of coupled BEQs governing the evolution of energy densities together with the dark and visible sector asymmetries is given by

\bea\begin{aligned}
& \dot{\rho_\varphi} = -3\,\mathcal{\mathcal{H}}\,\rho_\varphi-\Gamma_\varphi\,\rho_\varphi\\&
\dot{\rho_N} = -3\mathcal\,{\mathcal{H}}\,\rho_N+\Gamma_\varphi\,\rho_\varphi-\Gamma_N\,\rho_N\\&
\dot{n}_\text{B-L} = -3\,\mathcal{\mathcal{H}}\,n_\text{B-L}-\epsilon_{\Delta L}\,\Gamma_N\,n_N-\Gamma_\text{ID}\,n_\text{B-L}\\&
\dot{\rho}_R = -4\mathcal{H}\,\rho_R+\Gamma_N\,\rho_N\\&
\dot{n}_{\Delta\chi} = -3\mathcal{\mathcal{H}}\,n_{\Delta\chi}+\epsilon_{\Delta\chi}\,\Gamma_N\,n_N-\Gamma_\text{ID}\,n_{\Delta\chi}
    \end{aligned}\label{eq:cpld-beq1}
\eea

\noindent where $\Gamma_\text{ID}$ is the inverse decay rate.
 Here we would like to comment on the role of inverse decay and scattering in washing out the asymmetries generated. Choosing some benchmark values of the masses and couplings (which we shall also use in Sec.~\ref{sec:results}), we find $\Gamma_1\simeq 2\times 10^8\,\text{GeV}\ll M_1=10^{12}$ GeV, which implies we are always in the weak washout regime~\cite{Falkowski:2011xh}. We further note that, for the same set of parameters, $\Gamma_1<\mathcal{H}(T=M_1)$ holds, implying $N_1$ is out of thermal equilibrium with the SM bath at $T\geq M_1$~\cite{Bernal:2017zvx}, where $\mathcal{H}(T=M_1)\simeq 10^{11}$ GeV. This is anyway understandable as the Hubble rate is always faster (compared to pure radiation domination) with additional contribution. Even in the absence of the inverse decay, the $\Delta L=2$ scattering processes can be effective, since in that case the expression for washout does not have a Boltzmann suppression, rather scales as some power of $1/M_1$ (see Eq. (27) of \cite{McDonald:2016ehm}). However, we find that the final asymmetry does not alter even in the presence of these processes, simply because of very heavy RHN mass that suppresses the effect of these washouts. Hence, we do not consider them in the BEQ.

As we shall see in a moment, in the present scenario $M_1\sim T_\text{max}$, which is the maximum temperature of the thermal bath during inflation, and $T_\text{max}\gg T_\text{RH}$. Thus, for $T<T_\text{max}$, the RHNs are always out of thermal equilibrium. The third and the fifth equations determine the yield for the visible and dark sector asymmetries respectively, while the first, second and fourth equations determine the evolution of the energy densities for the inflaton, RHN and radiation respectively. We consider the RHNs to be non-relativistic, therefore $n_N=\rho_N/m_N$ holds, where in our case $m_N\equiv M_1$.  Note that this assumption does not strictly hold as the energy of the RHNs at the production is $m_\varphi/2> M_1$, and hence they are relativistic. Therefore, the RHNs can either decay while they are still relativistic or after becoming non-relativistic. Now, in the present scenario with high scale leptogenesis, we have $\Gamma_N\gg\Gamma_\varphi$, which shows, the RHNs decay instantaneously after their production, and thus the physical reheating temperature of the universe remains the same as the one obtained by considering instantaneous decay of the inflaton. Subsequently, the final $B-L$ asymmetry in this case does not change as the decay rate of the RHN is modified by a factor of $M_1/E_N$, still pertaining to the fact that $\Gamma_N\times M_1/E_N\gg\Gamma_\varphi$ for our choice of parameters, where $E_N=m_\varphi/2$~\cite{Senami:2008yw}\footnote{In~\cite{Giudice:1999fb} the authors discussed non-thermal leptogensis considering the RHNs decay after becoming non-relativistic.}. For low scale leptogenesis where RHNs are relatively long-lived, their loss of energy before decay can be important and appropriate dilution factor needs to be incorporated depending upon their relativistic or non-relativistic nature. A complete numerical analysis, carefully considering the decay and red-shift of energy density of relativistic RHNs is not the goal of the present study and we keep it as for future studies. 

The inflaton decays {\it only} into a pair of lightest RHNs, the decay width is thus given by

\bea
\Gamma_\varphi\simeq \frac{y_\varphi^2}{4\,\pi}\,m_\varphi\,.
\eea

\noindent Since we are considering a hierarchical mass spectrum for the heavy neutrinos $M_{2}\gg M_1$ (with $2\,M_1<m_\varphi$), hence potential effects of $N_{2}$ can be neglected\footnote{Baryon asymmetry can be generated by the second-lightest RHN in certain areas of parameter space as discussed in the context of thermal leptogenesis~\cite{DiBari:2005st}.}. 


It is convenient to make suitable variable transformation while solving Eq.~\eqref{eq:cpld-beq1}. We make the following set of transformations by scaling the energy and number densities with the scale factor $a$~\cite{HahnWoernle:2008pq,Giudice:2000ex}

\bea\begin{aligned}
& \widetilde{\rho}_\varphi = \rho_\varphi a^3,\\&
\widetilde{\rho}_N = \rho_N a^3,\\&
\widetilde{N}_{\rm B-L} = n_{\rm B-L}a^3,\\&
\widetilde{\rho}_R = \rho_R a^4,\\&
X = n_{\Delta\chi} a^3\,.
    \end{aligned}\label{eq:var}
\eea

We also define $$\xi=\frac{a}{a_I}$$ as the ratio of scale factors and assume $a_I=1$. Note that, no physical result depends on this choice and hence $a_I$ can be chosen to be anything\footnote{In~\cite{Giudice:2000ex}, the authors have defined $a_I=T_\text{RH}^{-1}$.}. The factor $a_I$ is chosen as the initial value of the scale factor, while $\xi$ is the proxy to the time (temperature) variable. With these definitions, the Hubble parameter reads

\bea
\mathcal{\mathcal{H}} = \sqrt{\frac{8\,\pi}{3\,M_\text{pl}^2}\,\frac{\widetilde{\rho}_\varphi\,a_I\,\xi+\widetilde{\rho}_N\,a_I\,\xi+\widetilde{\rho}_R}{a_I^4\,\xi^4}}.
\eea

In terms of these rescaled variables, Eq.~\eqref{eq:cpld-beq1} can be written as

\bea\begin{aligned}
& \widetilde{\rho}_\varphi^{'} = -\frac{\Gamma_\varphi}{\mathcal{H}}\,\frac{\widetilde{\rho}_\varphi}{\xi},\\&
\widetilde{\rho}_N^{'} = \frac{\Gamma_\varphi}{\mathcal{H}}\frac{\widetilde{\rho}_\varphi}{\xi}-\frac{\Gamma_N}{\mathcal{H}\,\xi}\,\widetilde{\rho}_N,\\&
\widetilde{N}_{B-L}^{'} = \frac{\Gamma_N}{\mathcal{H}\,\xi}\,\epsilon_{\Delta L}\,\frac{\widetilde{\rho}_N}{M_1},\\&
\widetilde{\rho}_R^{'} = \frac{\Gamma_N a_I}{\mathcal{H}}\,\widetilde{\rho}_N,\\&
X^{'} = \frac{\Gamma_N}{\mathcal{H} \xi}\,\epsilon_{\Delta\chi}\,\frac{\widetilde{\rho}_N}{M_1}.
    \end{aligned}\label{eq:cpld-beq2}
\eea


\noindent In the above set of equations, prime in superscript corresponds to derivative with respect to $\xi=a/a_I$. Excepting for the inflaton, which has an initial energy density, all other quantities are being produced from different sources originating directly or indirectly from inflaton only. Hence we can set the initial conditions for different quantities appearing in Eq.~\eqref{eq:cpld-beq2} as
\bea\begin{aligned}
& \widetilde{N}_{(B-L)_I}=0;~\widetilde{\rho}_\varphi = \widetilde{\rho}_{\varphi_I};\\& \widetilde{\rho}_{R_I}=0;~\widetilde{\rho}_{N_I}=0;~X_I=0.       \end{aligned}
\eea
In obtaining the initial energy density for the inflaton we use 
\bea
\widetilde{\rho}_{\varphi_I} = \frac{3}{8\,\pi}\,M_\text{pl}^2\,H_I^2\equiv\rho_{\varphi_I}\,,
\label{eq:inf-init}
\eea

\noindent where the last line follows from the fact that $a_I=1$. Now, CMB observation puts a bound on the scale of inflation~\cite{Planck:2018jri}: $H_I < 2.5 \times 10^{-5}\,M_\text{pl}$, implying $\rho_{\varphi_I}\lesssim 3\times 10^{-6}\,M_\text{pl}^4$. The observed baryon asymmetry depends on the final $B-L$ abundance, which can be derived in terms of the redefined $\widetilde{N}_{B-L}$ via
\bea\begin{aligned}
& Y_{\rm B-L} =\frac{n_{\rm B-L}}{s}=\Biggl[\frac{45}{2\,\pi^2\,g_{\star s}}\Biggr] \Biggl(\frac{30}{\pi^2\,g_{\star\rho}}\Biggr)^{-3/4}\,\tilde{N}_{\rm B-L}\,\widetilde{\rho}_R^{-3/4}\,,
    \end{aligned}
\eea

\noindent where we have used

\bea
T = \Biggl(\frac{30\rho_R}{\pi^2 g_\star}\Biggr)^{1/4}
\eea

\noindent which is allowed since the heavy neutrinos are non-relativistic\footnote{If the produced heavy neutrinos were relativistic then $T = \Bigl(\frac{30\left(\rho_R+\rho_N\right)}{\pi^2 g_\star}\Bigr)^{1/4}$, which would be true if $m_\varphi\gg M_N$.}. Sphaleron interactions are in equilibrium in the temperature range between $\sim 100$ GeV and $10^{12}$ GeV, and they convert a fraction of a non-zero $B-L$ asymmetry into a baryon asymmetry via

\bea
Y_B\simeq a_\text{sph}\,Y_{B-L}=\frac{8\,N_F+4\,N_H}{22\,N_F+13\,N_H}\,Y_{B-L}\,,
\eea

\noindent where $N_F$ is the number of fermion generations and $N_H$ is the number of Higgs doublets, which in our case $N_F=3\,,N_H=1$ and $a_\text{sph}$ turns out to be 28/79. In leptogenesis, where purely a lepton asymmetry is generated, $B-L=-L$. This is converted into the baryon asymmetry via sphaleron transition~\cite{Buchmuller:2004nz,Buchmuller:2005eh}. Finally, the observed baryon asymmetry of the universe is given by~\cite{Planck:2018jri}

\bea
\eta_B=\frac{n_B-n_{\overline{B}}}{n_\gamma}\simeq 6.2\times 10^{-10}\,,Y_B\simeq 8.7\times 10^{-11}\,.\label{eq:etab}
\eea

As mentioned before, the DM abundance is set by the asymmetry in the DM sector that indicates the residual number density for the asymmetric DM, and is given by

\bea
\Omega_\chi\,h^2 \simeq 2.75\times 10^8\Bigl(\frac{m_\chi}{\text{GeV}}\Bigr)\,\mathcal{Y}_{\Delta\chi}\left(T_0\right)= 2.75\times 10^8\,m_\chi\,X\,\widetilde{\rho}_R^{-3/4}\,\Bigl(\frac{45}{2\,\pi^2\,g_{\star s}}\Bigr)\,\Bigl(\frac{\pi^2\,g_{\star\rho}}{30}\Bigr)^{3/4}\label{eq:dm-relic} 
\eea

\noindent where $\mathcal{Y}_{\Delta\chi}\left(T_0\right)\equiv\,n_{\Delta\chi}/s\left(T_0\right)$ is the yield of the asymmetry in the DM sector at the present temperature $T_0$ and $s$ is the entropy per comoving volume in the visible sector.

\subsection{Results and Discussions}\label{sec:results}

In this section we discuss the results of the first part of our analysis i.e., asymmetric DM from inflaton decay. The solution to the set of five coupled BEQ in Eq.~\eqref{eq:cpld-beq2} is numerically performed to obtain the yield of the different components. Note that, the minimal model provides us with the following free parameters

\bea
\{M_1\,,m_\varphi\,,m_\chi\,,y_\chi\,,y_\varphi\}\,.
\eea

\noindent However, in the following analysis we will always fix the mass of the inflaton and the RHN ensuring $M_1<m_\varphi/2$ such that the inflaton can always decay into a pair of RHNs on-shell. Hence the resulting parameter space is decided by three free parameters, namely the Yukawas $y_{\chi,\varphi}$ and the DM mass $m_\chi$. Before delving into a detailed parameter space scan, we first analyze the impact of different free parameters on the yield by fixing a few of them to benchmark values. 

\begin{figure}[htb!]
$$
\includegraphics[scale=0.4]{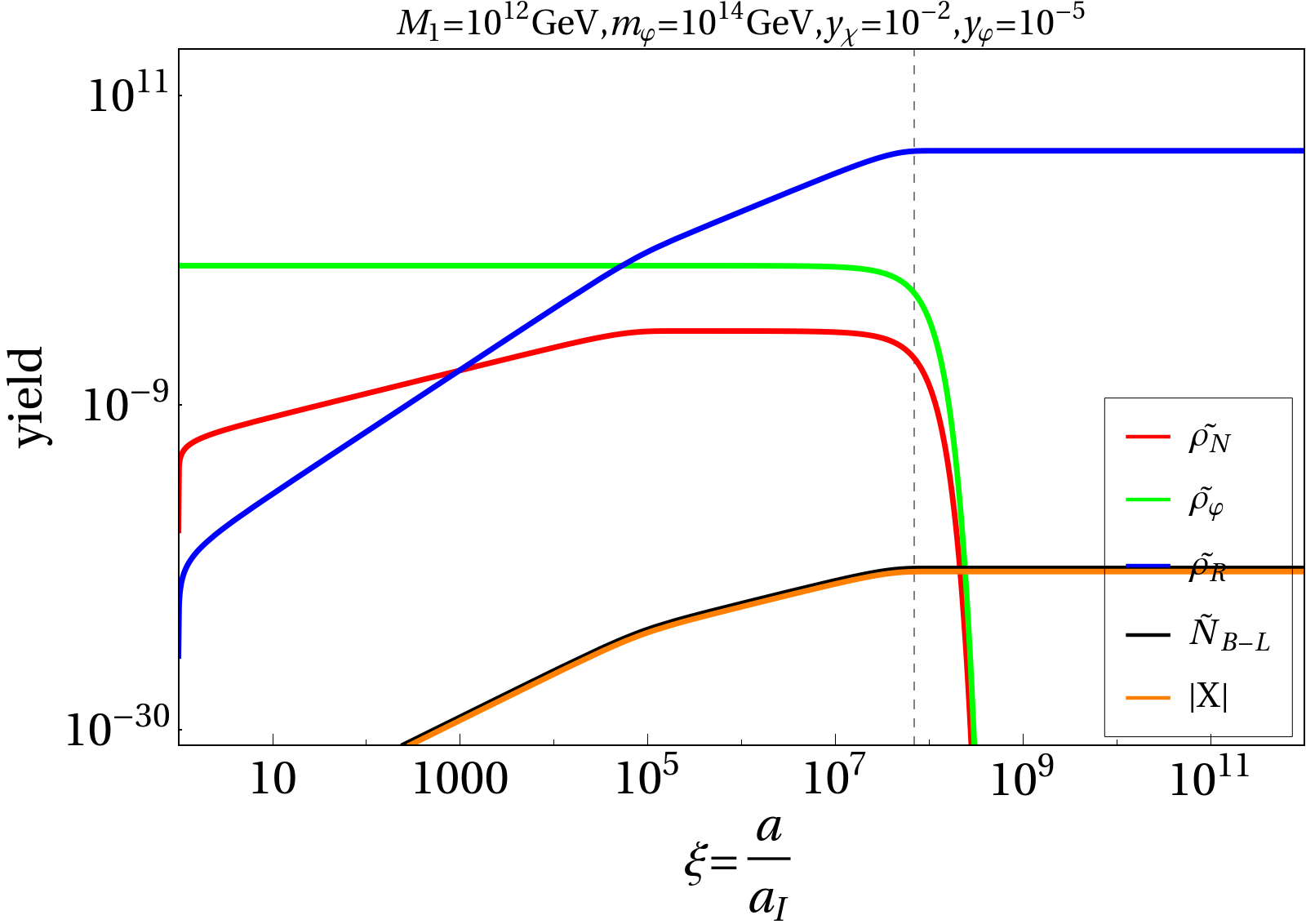}~~~~\includegraphics[scale=0.4]{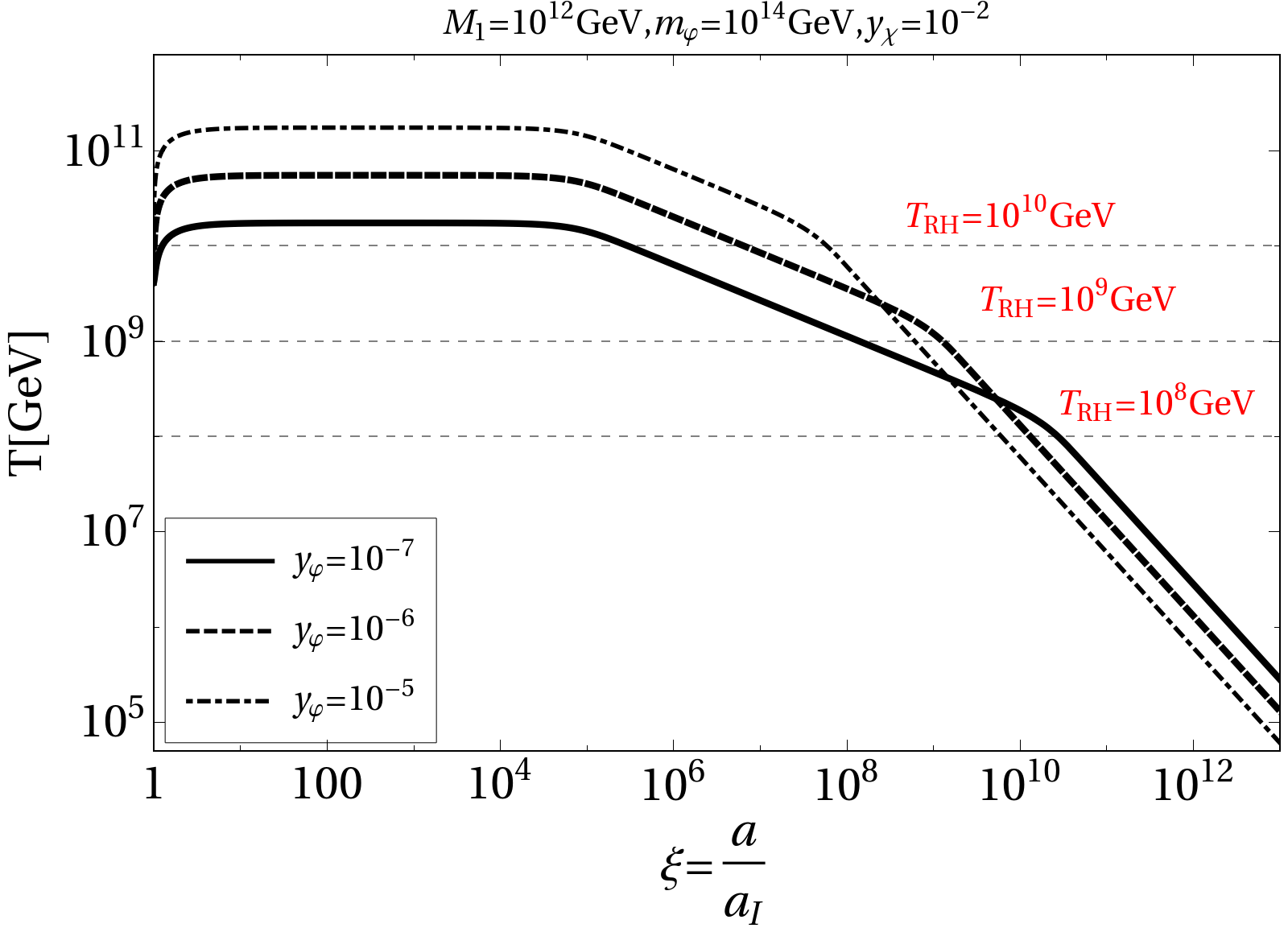}
$$
$$
\includegraphics[scale=0.4]{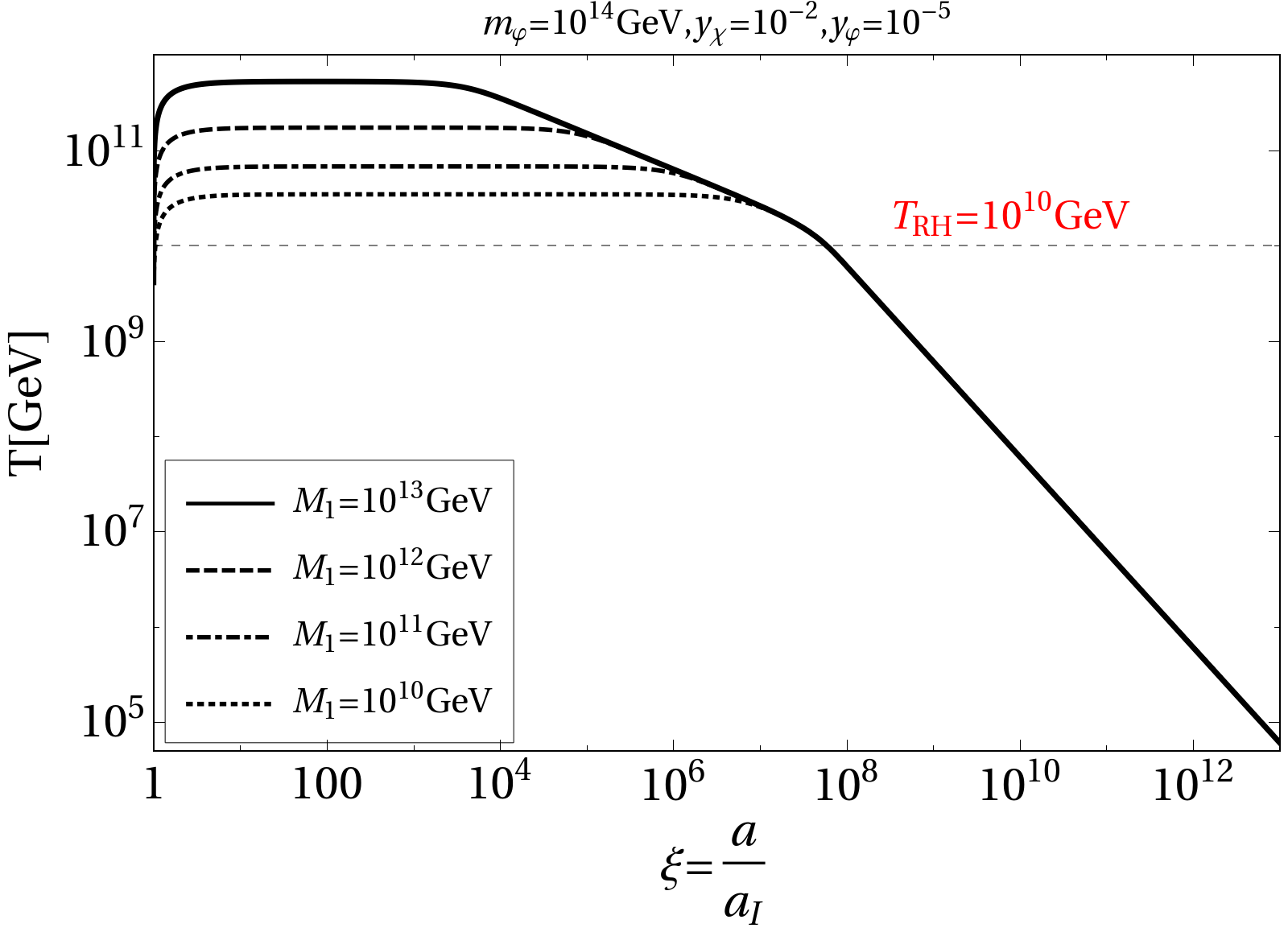}~~~~
\includegraphics[scale=0.4]{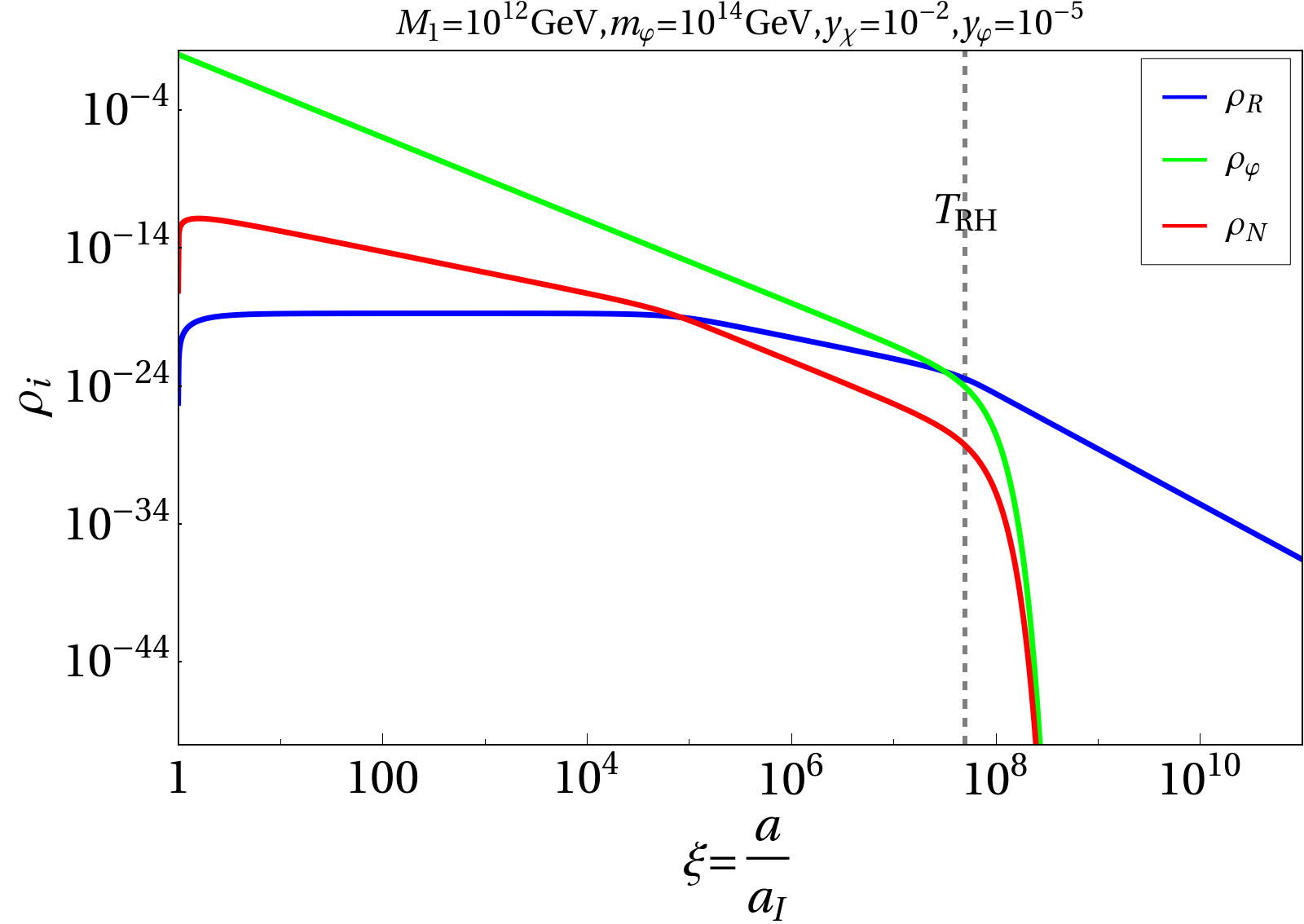}
$$
\caption{Top left panel: Yield of different components (normalized to $\widetilde{\rho}_{\varphi_I}$) as a function of $\xi=a/a_I$ for fixed values of the RHN mass $M_1=10^{12}$ GeV, inflaton mass $m_\varphi=10^{14}$ GeV and Yukawas $y_\chi=10^{-2},y_\varphi=10^{-5}$. Top right panel: Evolution of SM bath temperature $T$ with $\xi=a/a_I$ for different choices of $y_\varphi$ shown by different patterns. Bottom left panel: Same as top right panel but with different choices of the RHN mass shown by different curves. Bottom right panel: Energy densities as function of $\xi$ for fixed choices of other parameters as mentioned in the plot. }\label{fig:yld-T}
\end{figure}

In the top left panel of Fig.~\ref{fig:yld-T} we show yields as function of $\xi=a/a_I$ for $M_1=10^{12}$ GeV, $m_\varphi=10^{14}$ GeV, $y_\chi=10^{-2}$ and $y_\varphi=10^{-5}$. These choices are arbitrary and a different combination can also lead to the observed relic abundance for the DM or the baryon asymmetry of the universe. Here we see, as expected, the yields for radiation (in blue) and the asymmetries (in black and orange) start from very small values initially and then increase with time, whereas the same for the inflaton and the RHN diminish. In fact the RHN abundance also increases from small values at very early epochs, after the inflaton field starts decaying into them. The yields for radiation and asymmetries get saturated at $\xi\sim 6.5\times10^7$. This corresponds to the reheating temperature $T_\text{RH}\simeq 10^{10}$ GeV as one can read off from the top right panel plot where we show the evolution of SM bath temperature with the scale factor. Needless to mention, at $T=T_\text{RH}$ inflaton decay is completed and the universe enters into radiation dominated era. As a result, the radiation yield also saturate at the same time, as shown by the blue curve in top left panel plot. The completion of inflaton decay also implies that the RHNs do not get produced further beyond $\xi\sim 6.5\times10^7$, since the inflaton has a 100\% decay into the RHNs. Hence we see a sharp fall in the RHN yield (shown via red curve) due to its decay into leptons and DM. Consequently, the asymmetries both in the visible and the DM sector freeze in at the same time (shown in black and orange colour). This is interesting since inflaton decay plays the key role in deciding the dynamics of different components. One must also notice a bend near $\xi\sim 5.2\times 10^4$ in the yields for the RHN (and the asymmetries). Beyond this point the production of the RHNs via $\varphi\to N_1\,N_1$ process becomes comparable to its decay. As a consequence, we see a plateau region in RHN evolution before its yield starts falling sharply at $T=T_\text{RH}$, where the inflaton decay is over. Since RHN is the source for both the radiation and the asymmetries, hence we see a change in slopes of blue, black and orange coloured curves near $\xi\sim 5.2\times 10^4$. This practically corresponds to the maximum temperature attainable during reheating i.e.,  $T=T_\text{max}$.  Finally, note that the asymmetries in either sector evolve exactly in the same manner, since they have the same source. 
In the top right panel the evolution of the SM temperature is plotted against $\xi$, where we clearly see the effect of non-instantaneous decay of the inflaton, due to which the bath temperature rises up to $T=T_\text{max}$ at a very early time. One should note, for a given $y_\varphi$, initially,  the temperature is independent of $\xi$ (for $y_\varphi=10^{-5}$ this happens till $\xi\sim 10^5$). During this time the RHN number density is large enough such that $\Gamma_N\,\rho_N$ (for a given $M_1$, $\Gamma_N$ is fixed) becomes comparable to $4\,\mathcal{H}\,\rho_R$ that accounts for the dilution of radiation energy density due to expansion. Hence we do not see any visible change in the bath temperature as the radiation energy density remains approximately constant with time during this period cf. second equation in Eq.~\eqref{eq:cpld-beq1}. The temperature then evolves as $T\propto a^{-3/8}$, when the RHN decay rate becomes comparable to its production rate, as a consequence the RHN yield stops increasing further after a certain $\xi$ depending on $y_\varphi$ (for $y_\varphi=10^{-5}$ we see this pattern between $\xi\sim 10^5-10^7$), but the decay still goes on. Afterward, the temperature scales as $T\propto a^{-1}$ when the inflaton decay to RHN is complete, as a result of which subsequent RHN production ceases and there is no further production of radiation bath. This is where the radiation dominated era begins. The change of slope at $T=T_\text{RH}$ denotes this transition. Note that, a smaller $y_\varphi$ results in a smaller reheating temperature. This is expected since a larger $y_\varphi$ results in a larger decay width $\Gamma_{\varphi\to N_1\,N_1}\propto y_\varphi^2$ for the inflaton, giving rise to a smaller lifetime. This means the inflaton decay is complete earlier and radiation domination begins at a higher temperature i.e., a higher $T_\text{RH}$. On the other hand, for a fixed $y_\varphi$, the evolution of SM bath temperature does not get affected for different $y_\chi$ since it does not affect the inflaton decay width. The plateau becomes wider with smaller RHN mass since in that case the decay of the RHN is delayed and the RHN-dominated epoch gets longer as shown in the bottom left panel. It is important to note here that for fixed $y_\varphi$ and $y_\chi$, the ratio of $T_\text{RH}$ to $T_\text{max}$ is fixed, and not a free parameter. We find, in our scenario $T_\text{max}/T_\text{RH}$ can be maximum of $\sim\mathcal{O}(100)$ for $y_\varphi=10^{-7}$, while for larger $y_\varphi$, this ratio can be of $\sim\mathcal{O}(10)$ as can be seen from top right panel of Fig.~\ref{fig:yld-T}.  In either cases, we see $M_1\sim T_\text{max}$, implying the RHNs fall easily out of equilibrium at $T<T_\text{max}$, as explained earlier. Note that, with the chosen values of $y_\varphi$ we always have $\Gamma_\varphi\ll\Gamma_N$, therefore the RHNs decay instantaneously after having been produced in inflaton decays, implying the RHN dominated epoch occurs over a very short period of time. Lastly, in the bottom right panel we show the evolution of energy densities (scaled with respect to the initial inflaton energy density) of radiation, inflaton and RHN with $\xi$. The importance of this plot lies in the fact that the energy density due to radiation dominates over that due to inflaton at and after $\xi\sim 5.2 \times 10^7$, which, in other words, is the point beyond which the temperature scales as $a^{-1}$ i.e., radiation dominated era begins. This boundary is exactly where we can define the reheating temperature $T_\text{RH}$, as can be read off from the top right panel (dot-dashed curve). 

\begin{figure}[htb!]
$$
\includegraphics[scale=0.35]{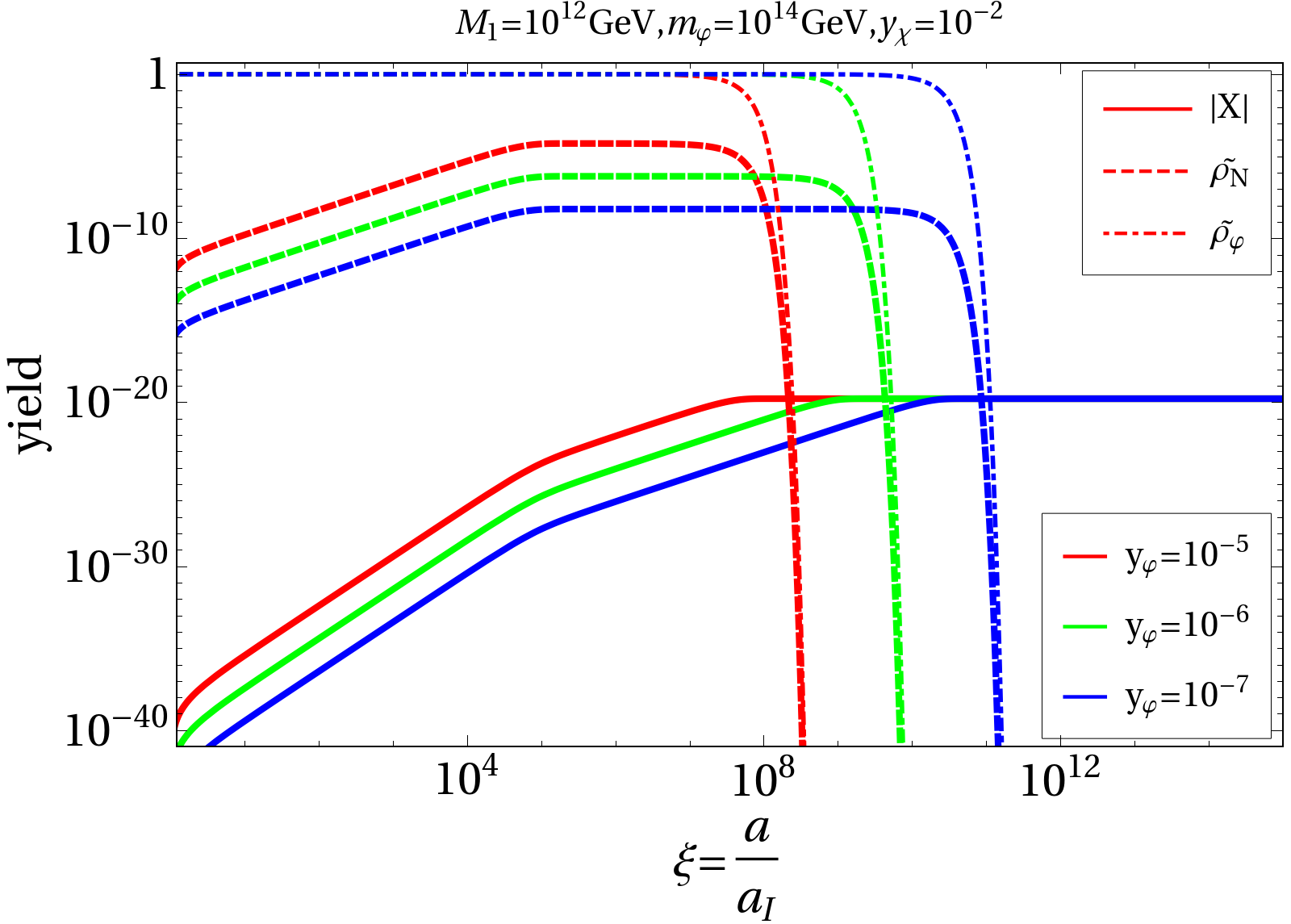}~~~~
\includegraphics[scale=0.35]{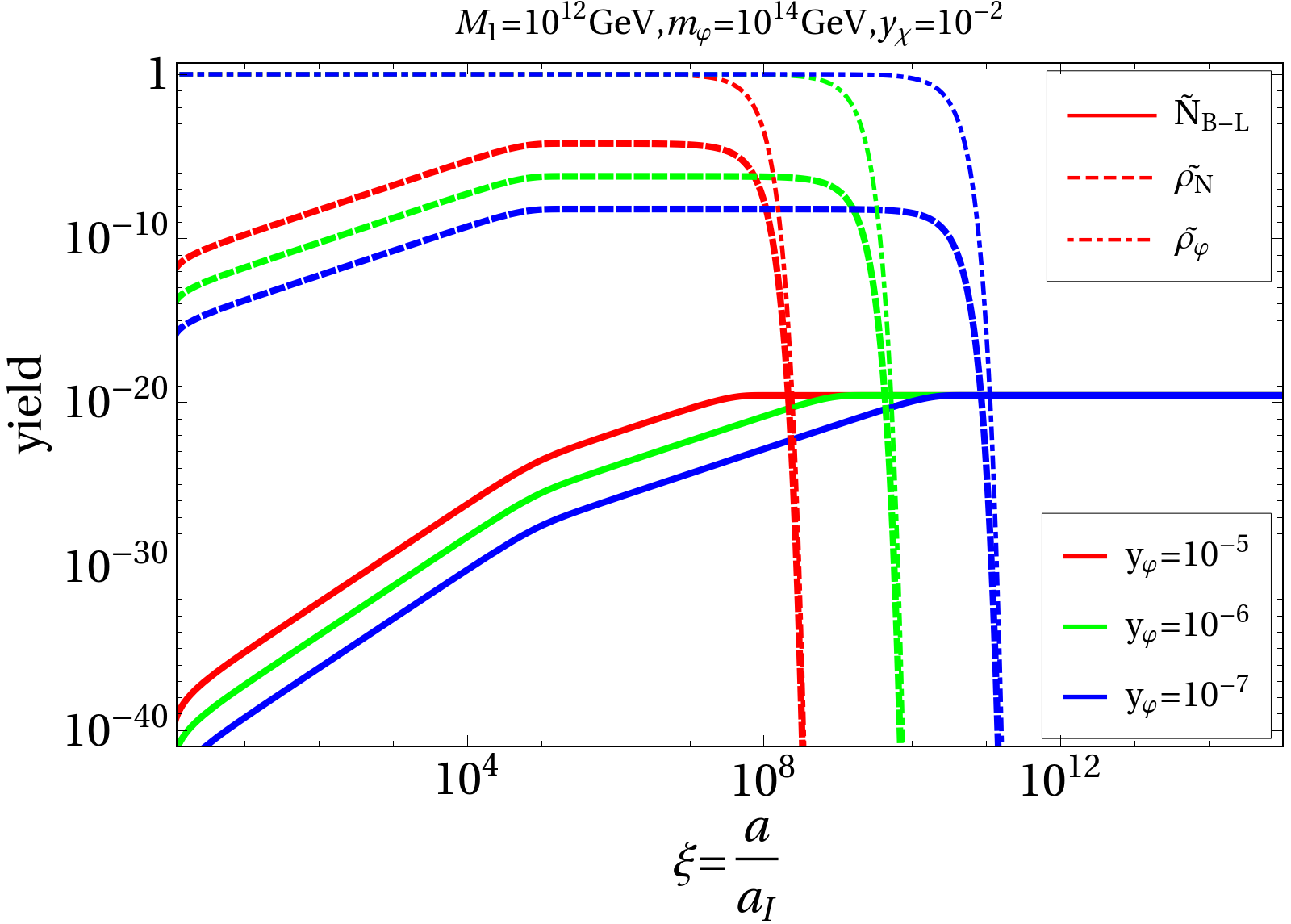}
$$
$$
\includegraphics[scale=0.35]{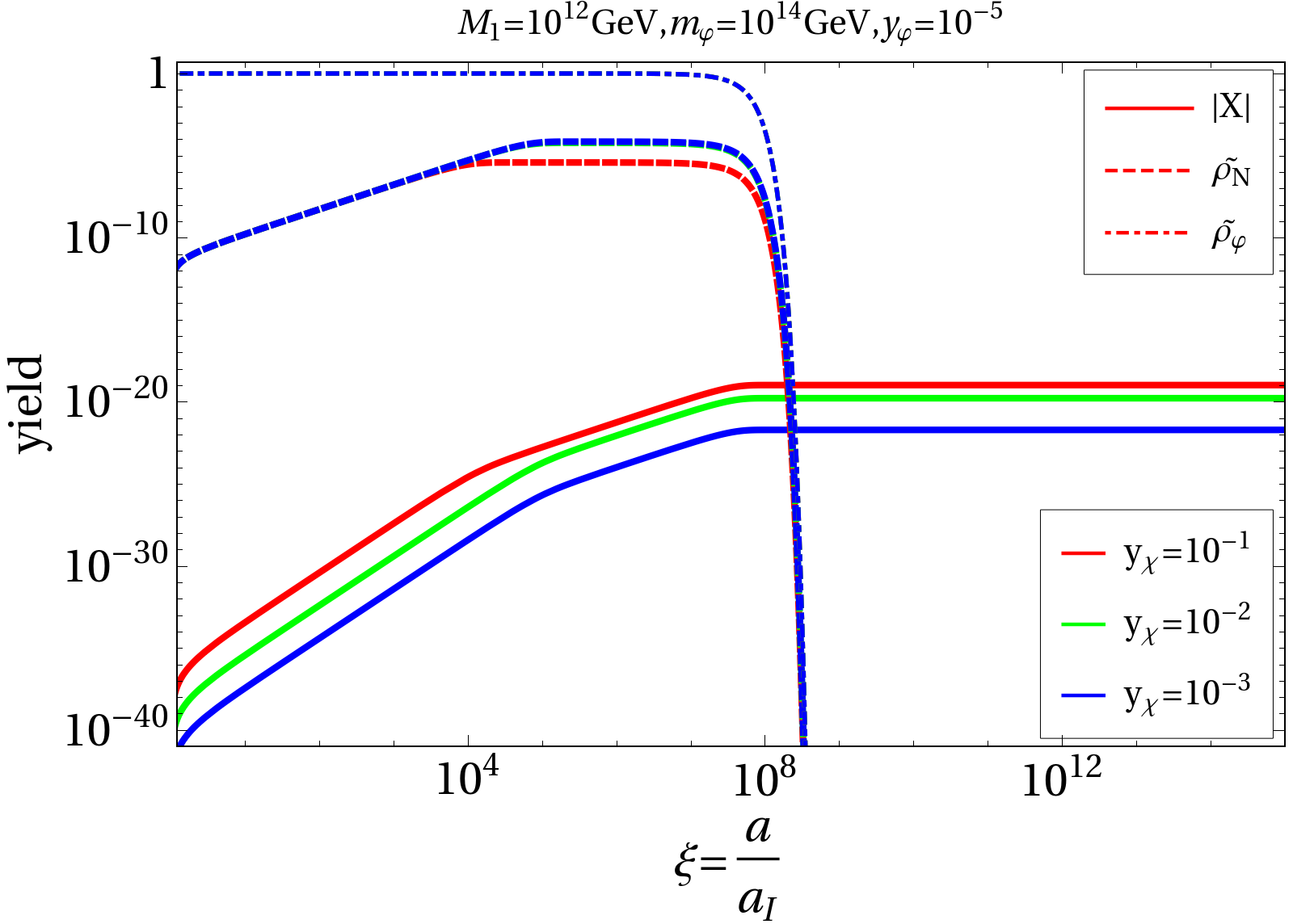}~~~~
\includegraphics[scale=0.35]{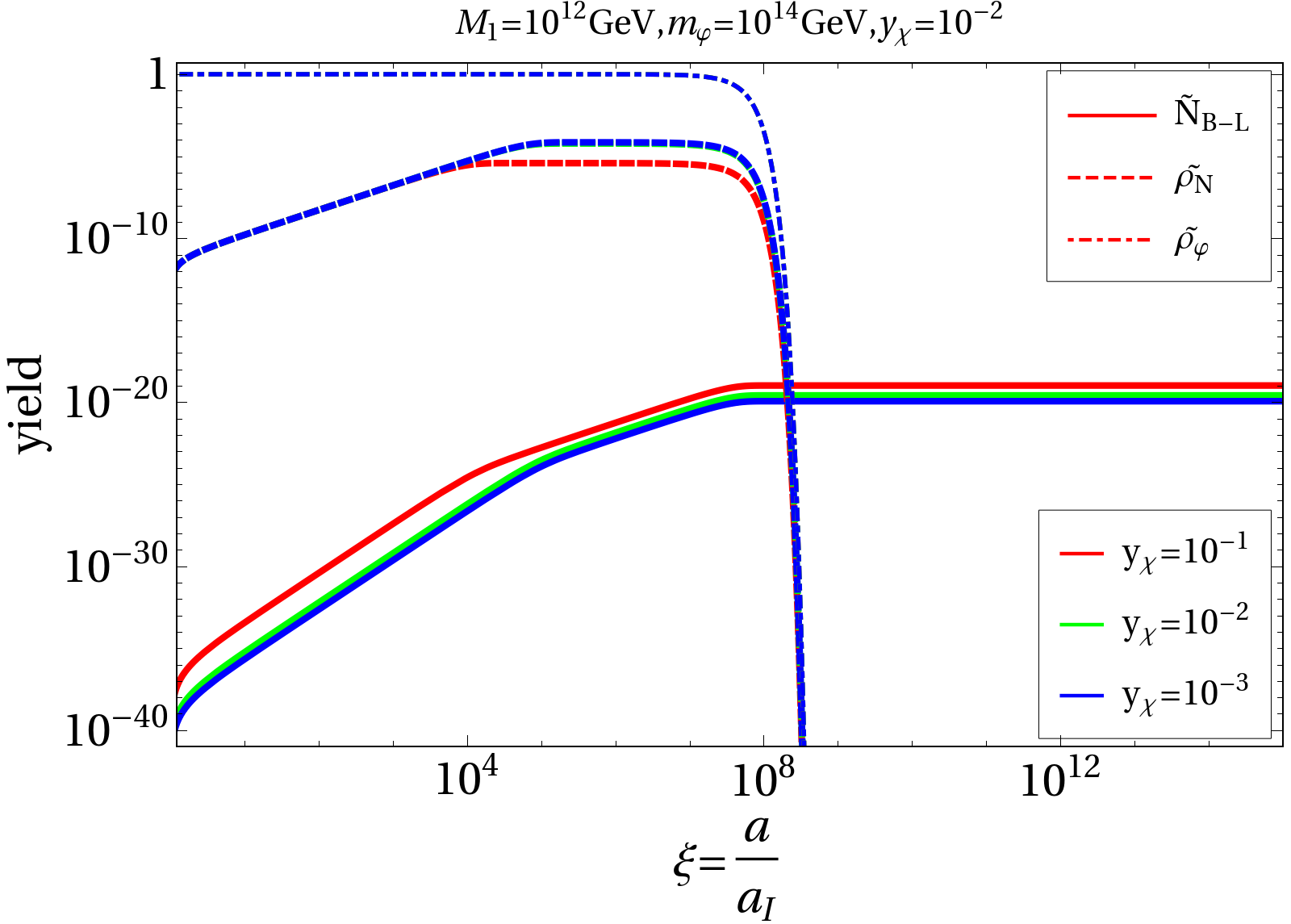}
$$
$$
\includegraphics[scale=0.35]{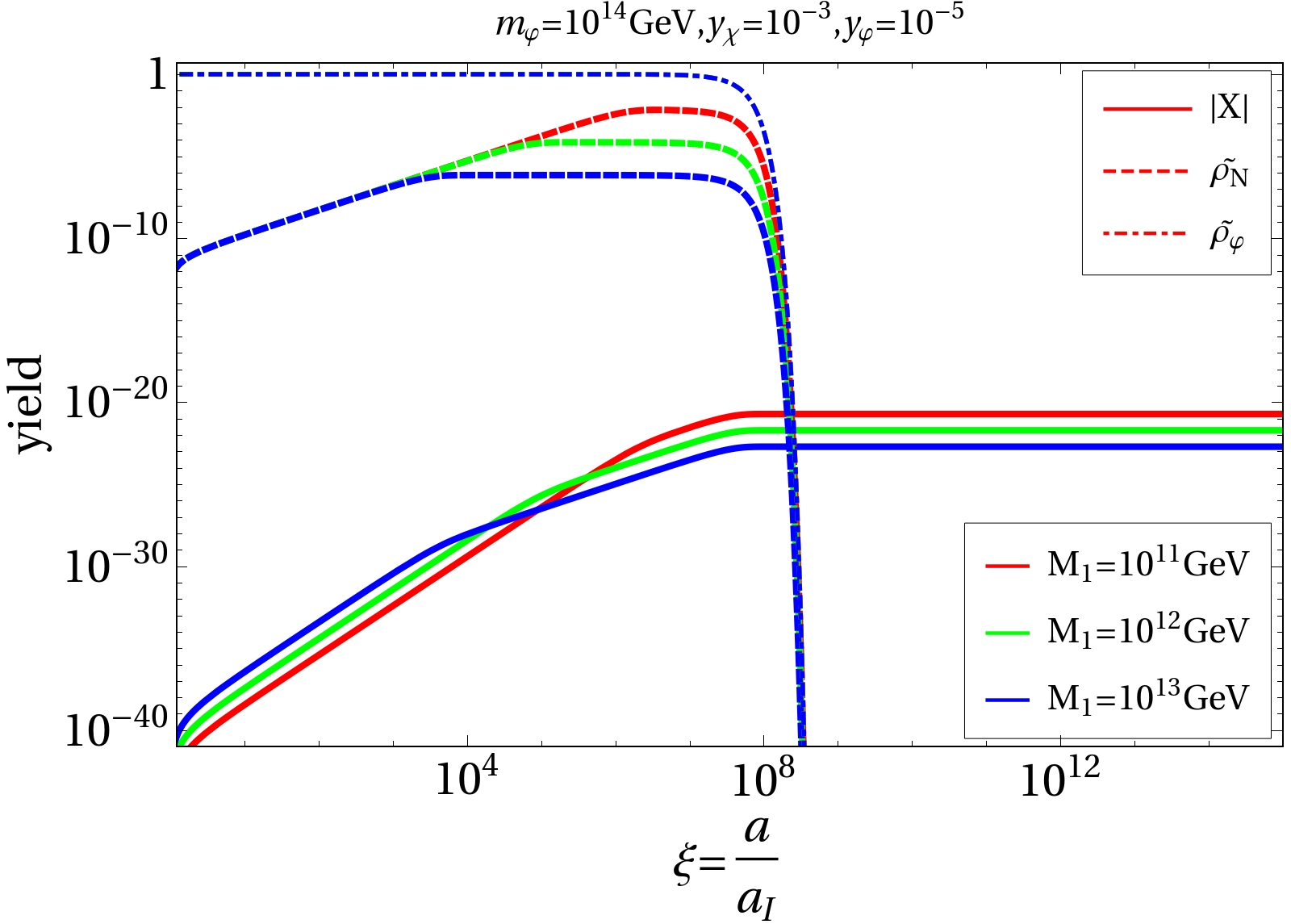}~~~~
\includegraphics[scale=0.35]{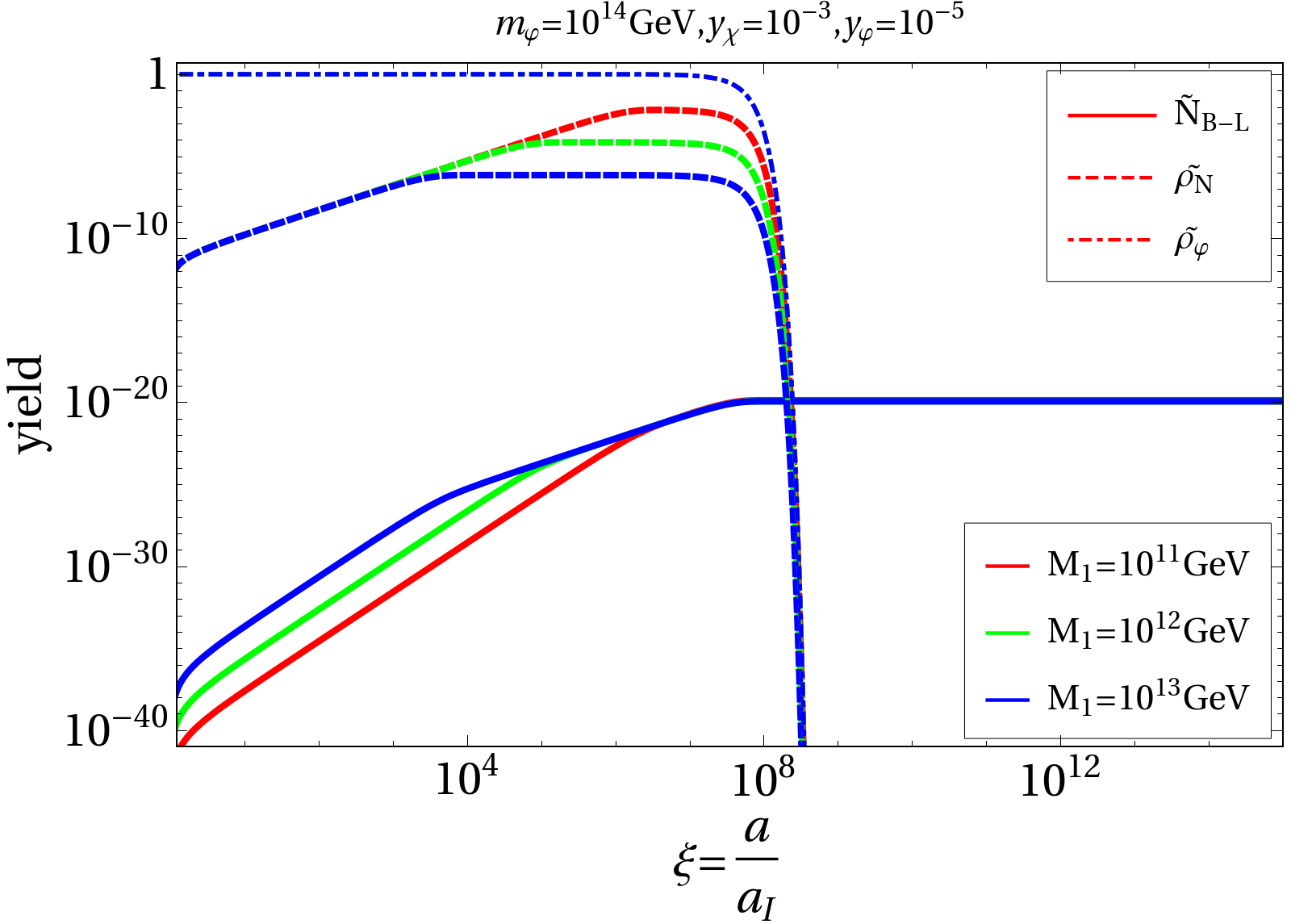}
$$
\caption{Top panel: Variation of inflaton energy, radiation energy, together with the DM asymmetry (left panel) and $B-L$ asymmetry (right panel) shown respectively by dot-dashed, dashed and solid curves (normalized to $\widetilde{\rho}_{\varphi_I}$). Here different colours correspond to different choices of $y_\varphi$, while $y_\chi$ is kept fixed along with the mass of the RHN. Middle panel: Same as top panel but for different choices of $y_\chi$ shown in different colours, while $y_\varphi$ is kept to a fixed value. Bottom panel: Same as top and middle but for different choices of the RHN mass $M_1$ shown in different colours, where $y_{\chi\,,\varphi}$ are fixed. In all these plots the inflaton mass if fixed to $m_\varphi=10^{14}$ GeV.}\label{fig:num_yld}
\end{figure}

The effect of the free parameters on the yield of DM and $B-L$ asymmetry are illustrated in Fig.~\ref{fig:num_yld}. We again fix the RHN mass $M_1=10^{12}$ GeV and the inflaton mass to $m_\varphi=10^{14}$ GeV. Then we plot the asymmetries as a function of $\xi=a/a_I$ considering different values of the Yukawas $\{y_\chi\,,y_\varphi\}$. In the top left panel we show how the DM asymmetry varies with $\xi$ for a fixed $y_\chi$ with different choices of $y_\varphi$ shown in different colours. Here we see, irrespective of the choice of $y_\varphi$, the asymptotic DM asymmetry remains the same. This is because, the asymmetry in the DM sector $X'\propto\epsilon_{\Delta\chi}\propto y_\chi^2$ (Eq.~\eqref{eq:cpld-beq2}), hence for a fixed $y_\chi$ the final asymmetry does not change. However, a larger $y_\varphi$ results in a larger decay width or equivalently, a shorter lifetime for the inflaton. As a consequence, we see the inflaton abundance falls earlier for a larger $y_\varphi$ (red dot-dashed curve) showing a smaller plateau. This also, in turn, affects the RHN yield. For a smaller $y_\varphi$ (that correspond to a smaller $T_\text{RH}$) we see a wider plateau in $\widetilde{\rho}_N$ (blue dashed curve). As in this case, the inflaton decay occurs over a longer period, the RHN production remains comparable to its decay over a longer epoch. For larger $y_\varphi$ the RHN yield also rises compared to other cases since $\widetilde{\rho}_N\propto\Gamma_\varphi\propto y_\varphi^2$. Since the RHN mass is fixed (so is the Dirac Yukawa coupling), hence the final $B-L$ asymmetries also converge for different choices of $y_\varphi$ as seen from the top right panel plot. Note that, both the yield $X$ and $\tilde{N}_{B-L}$ increase initially with the increase in $y_\varphi$, simply because the RHN yield rises. But as soon as the inflaton decay gets completed they converge to fixed asymptotic values. Similar to the behaviour seen in Fig.~\ref{fig:yld-T}, we again see the asymmetries saturating the moment inflaton decay is complete.  In the middle panel we show the dependence of relevant yields on $y_\chi$, by keeping $y_\varphi$ fixed to a constant value. Here we find, increasing $y_\chi$ results in an enhanced DM asymmetry but has a negligible effect on $B-L$ asymmetry. Since $y_\varphi$ is kept fixed (along with RHN mass), the inflaton decay width also remains fixed due to which there is no visible change in $\tilde{\rho}_\varphi$ (blue dot-dashed curve). The influence of $y_\chi$ on $B-L$ asymmetry is even minute as shown by the middle right panel plot, where different coloured solid curves are almost inseparable. Finally in the bottom panel we show the effect of having different RHN masses keeping the Yukawas fixed. First of all, here we see, different masses of the RHNs do not affect the evolution of inflaton energy density since the inflaton decay width remains approximately unchanged. However, a smaller $M_1$ results in a smaller plateau for the RHN (as shown by the red dashed curve) yield. This can be attributed to the fact that for heavier RHN, the corresponding Dirac Yukawa coupling is comparatively large, hence its decay starts competing with the production from a much earlier epoch leading to a wider plateau region. However, the RHN decay gets completed at the same time immaterial of their masses because of the fact that the reheating temperature remains fixed for a fixed $y_\varphi$. Also, for a low mass RHN, the yield in its density is larger since a smaller mass results in a smaller decay width which in turn implies that the decay rate starts competing with the production at a much later epoch. This results in a larger yield for RHN with smaller mass. Since a larger $M_1$ corresponds to a larger Dirac Yukawa coupling, hence the branching ratio of $N_1$ decay to SM final states become dominant when $M_1$ is comparatively large. This reduces the final DM yield as one can see from the solid curves in the bottom left panel. On the other hand, different $M_1$'s do not affect the asymptotic yield of the $B-L$ asymmetry as one can see from the bottom right panel plot, since for a fixed $y_\varphi$, the reheating temperature $T_\text{RH}$ does not change, and so is the inflaton decay width. Therefore, the $B-L$ production stops as soon the inflaton decay is complete irrespective of RHN mass. Before moving on we would like to mention that the final baryon asymmetry in the present scenario can be analytically determined, assuming the inflaton decays instantaneously after getting produced from inflaton decay as~\cite{Asaka:1999yd, Lazarides:1991wu,Giudice:1999fb,Asaka:1999jb,Hamaguchi:2001gw}

\bea
Y_B\simeq -\frac{28}{79}\,\frac{3}{2}\,\epsilon_1\,\frac{T_\text{RH}}{m_\varphi}\,.
\eea

\noindent We find, using $y_\varphi=10^{-7}\,, y_\chi=10^{-2}\,, M_1=10^{12}~\text{GeV}$ and $m_\varphi=10^{14}~\text{GeV}$ our full numerical calculation provides $Y_B^\text{numerical}=4.75\times 10^{-12}$, whereas from the approximately analytical expression we obtain $Y_B^\text{analytical}=4.25\times 10^{-13}$
for $T_\text{RH}\simeq 10^9~\rm GeV$. The difference of $\sim\mathcal{O}(1)$ in the magnitude of final baryon asymmetry is expected since in the full numerical analysis we have considered the effect of $T_\text{max}(> T_\text{RH})$ incorporating the non-instantaneous inflaton decay. 

\begin{figure}[htb!]
$$
\includegraphics[scale=0.37]{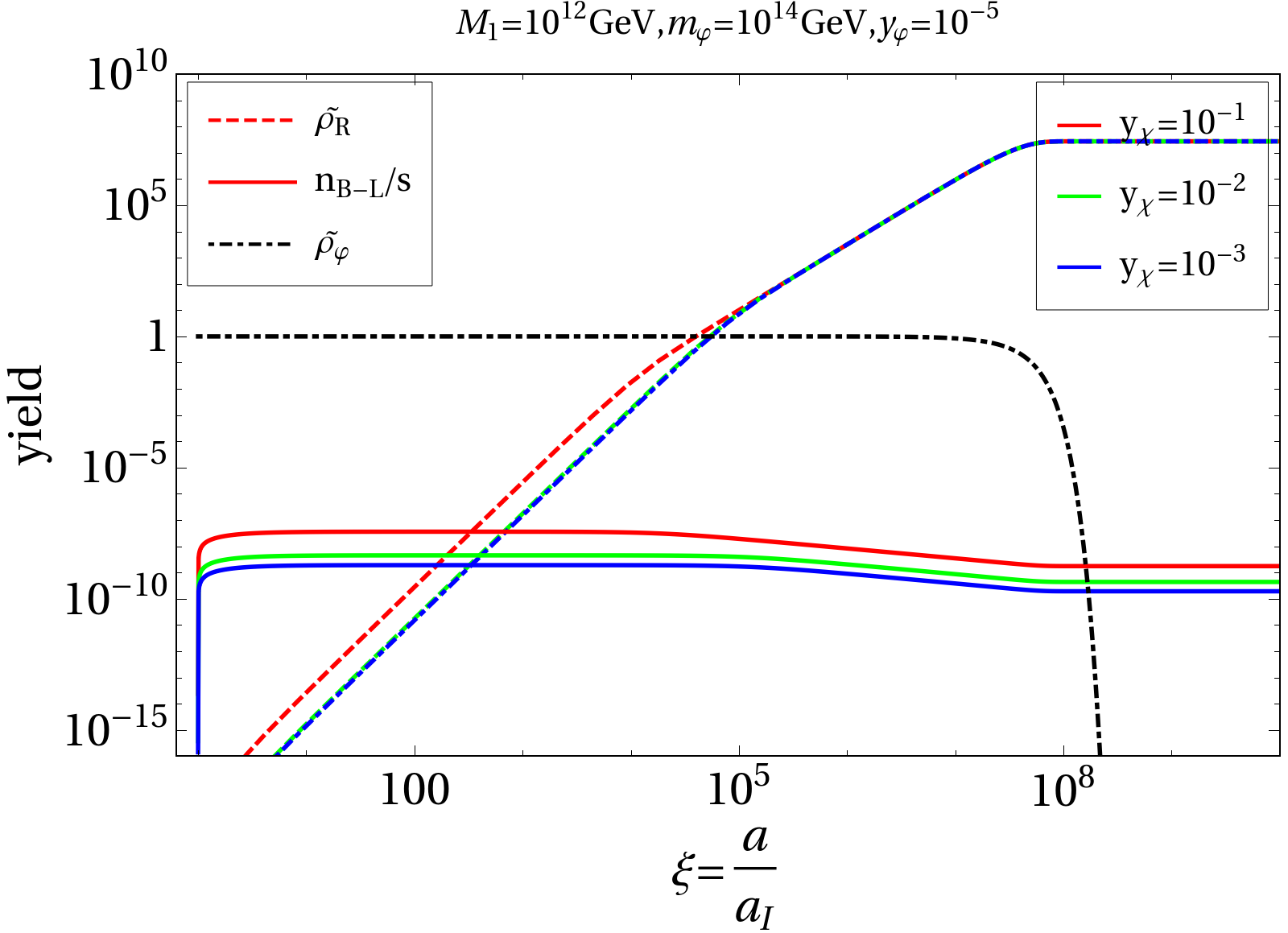}~~~~
\includegraphics[scale=0.37]{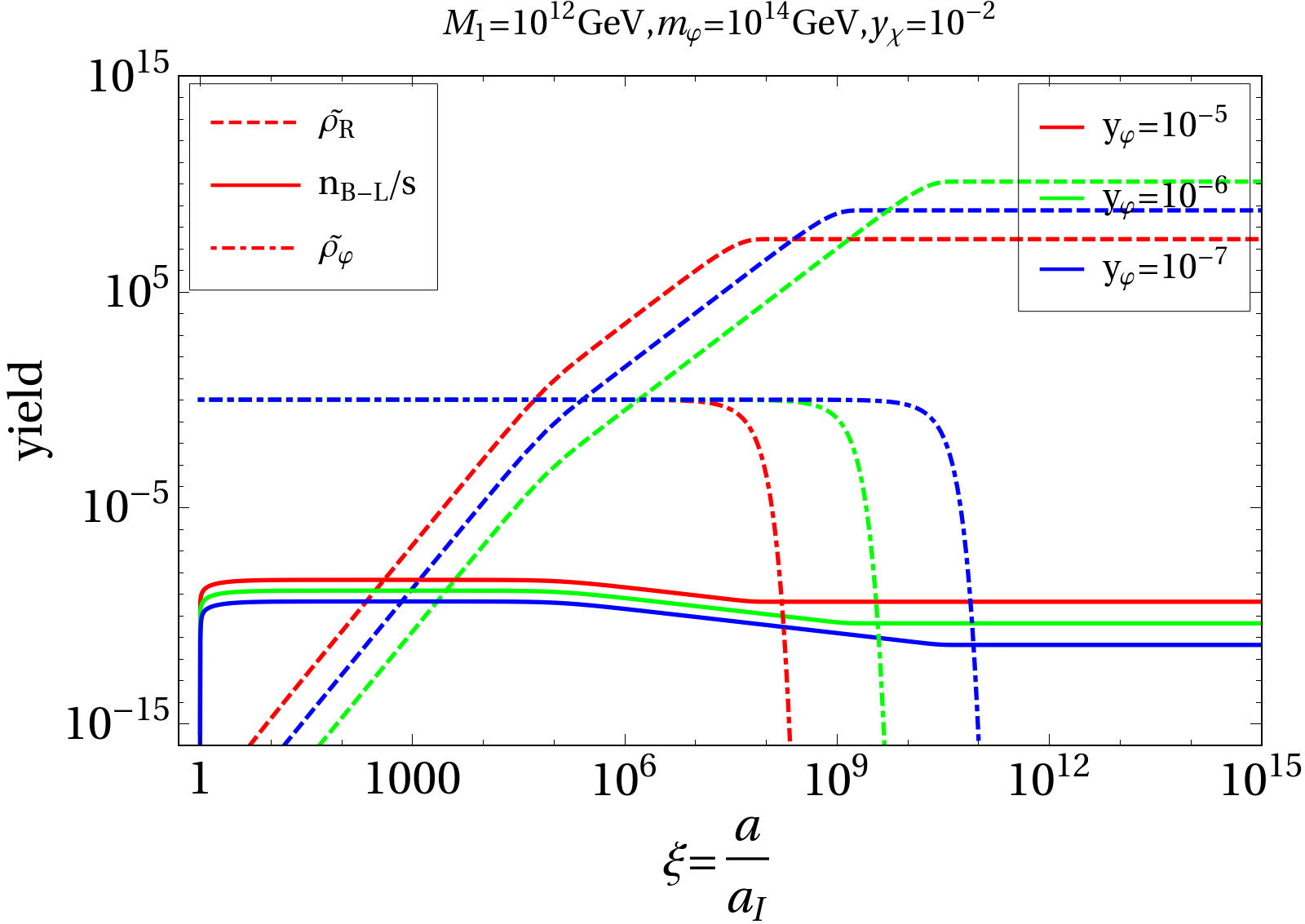}
$$
$$
\includegraphics[scale=0.37]{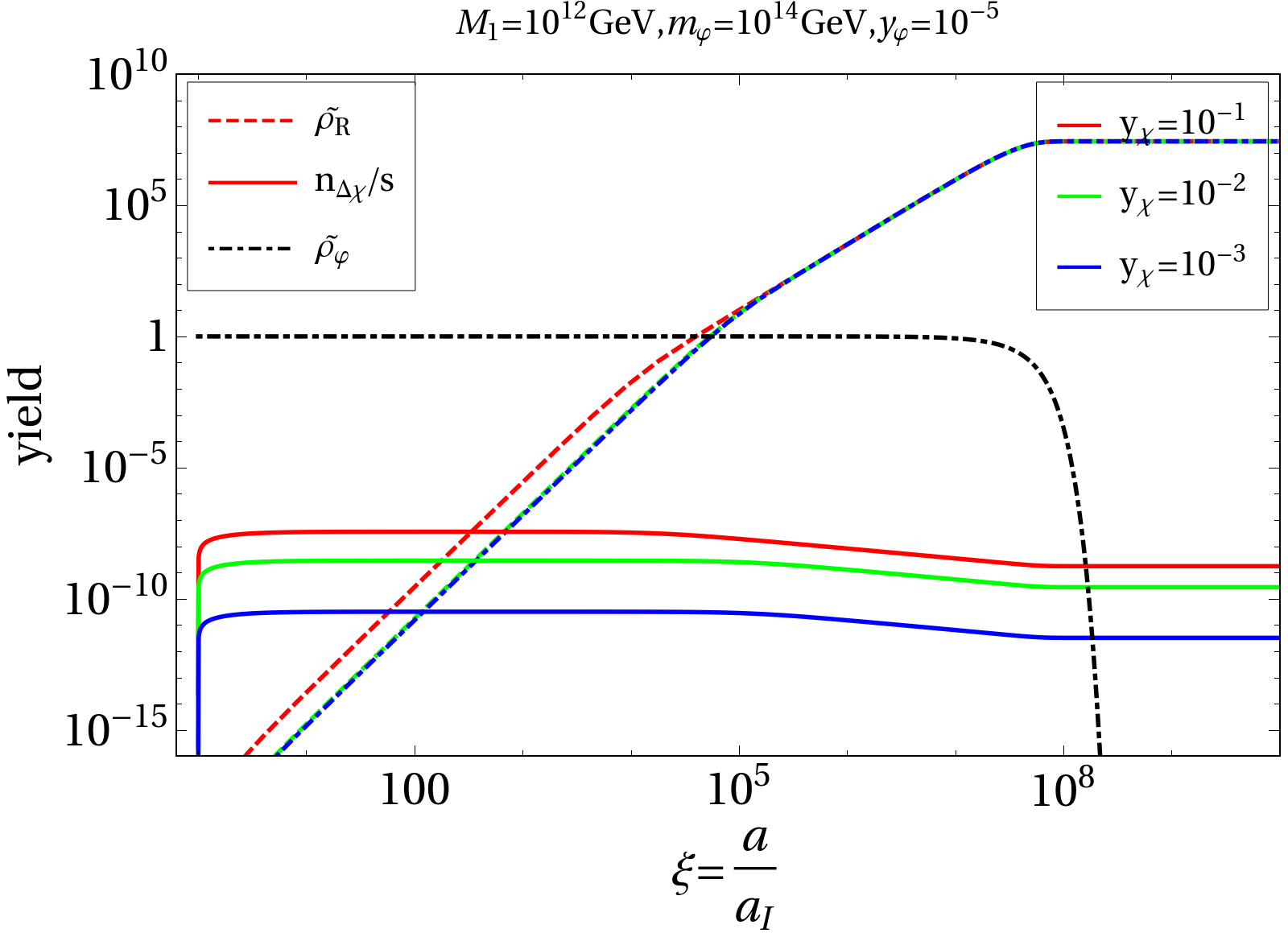}~~~~
\includegraphics[scale=0.37]{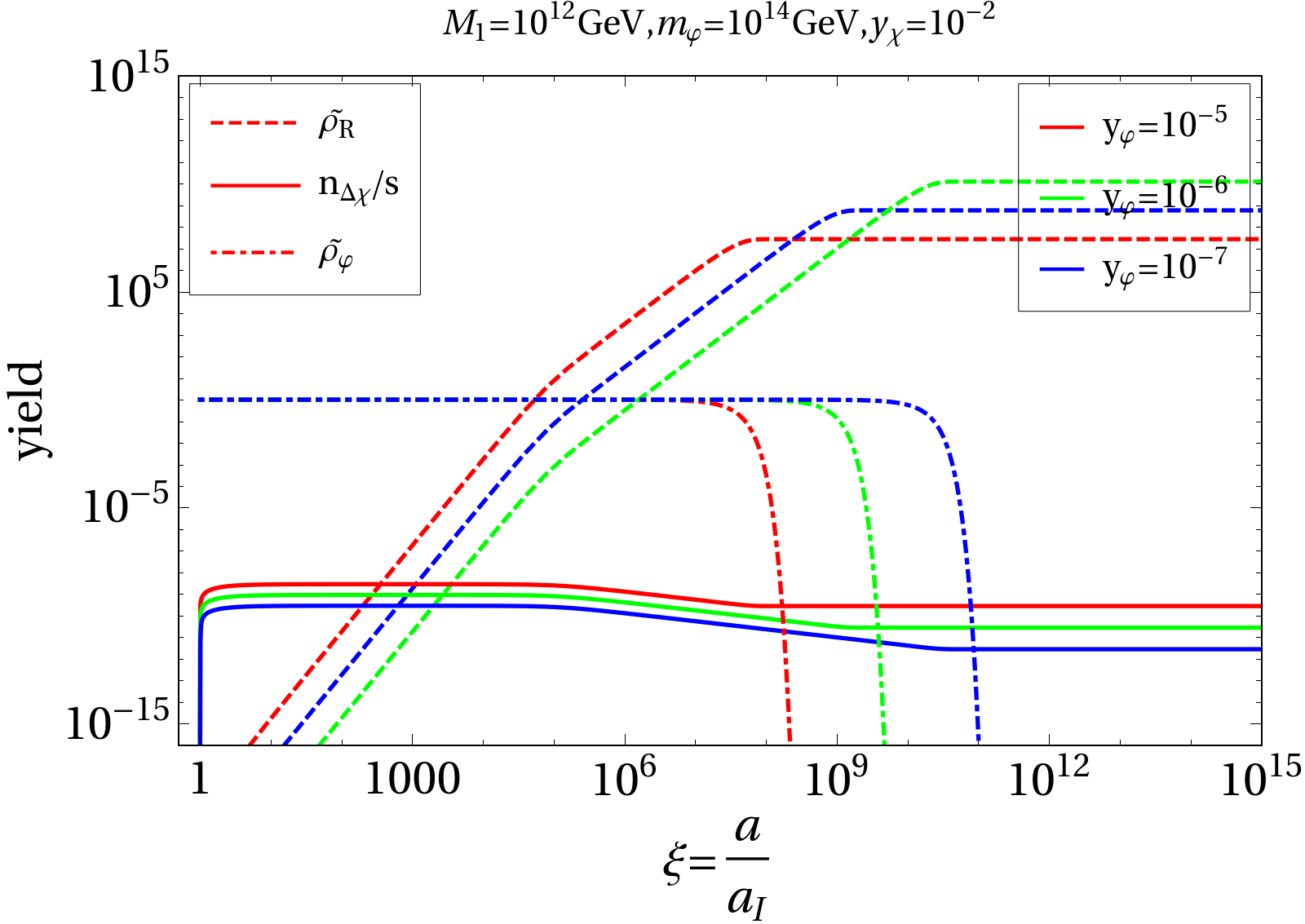}
$$
\caption{Top panel: Evolution of radiation, inflaton energy density and $B-L$ asymmetry with $\xi$ for different choices of $y_\chi$ (left) and $y_\varphi$ (right) shown in different colours. Bottom panel: Same as top but with asymmetry in the DM sector.}
\label{fig:asym-evol}
\end{figure}

In Fig.~\ref{fig:asym-evol} we have shown how the asymmetries in the visible (top panel) and in the DM sector (bottom panel) evolve with time. We have fixed the RHN and the inflaton mass to $10^{12}$ GeV and $10^{14}$ GeV respectively, while considering some benchmark values of the Yukawa couplings $y_\chi$ and $y_\varphi$ to understand their effects. For a fixed $y_\varphi$, radiation and inflaton energy densities are not affected by different choices of $y_\chi$ as one can notice from the top left figure. However, the $B-L$ asymmetry slightly rises for increase in $y_\chi$ since it altogether increases RHN decay width although the branching in the visible sector gets diminished. One should note the change in slope of all the solid curves at $\xi\sim 10^7$, where the inflaton decay is complete. This change in slope in $n_{B-L}/s$ is due to the decay of the RHN that results in entropy injection in the thermal plasma, which stops at $\xi\sim 10^8$ once the RHN decay is complete and thereafter remains constant till today. On the other hand, different choices of $y_\varphi$ potentially affect the decay lifetime of the inflaton causing it to decay faster for a larger $y_\varphi$. This is evident from the red dot-dashed curve in the top right panel. Here the change of slope in the solid curves are even more prominent and they occur at different epoch as the decay lifetime of the inflaton keeps changing with $y_\varphi$. Again because of the common single source of asymmetry we find $n_{\Delta\chi}/s$ to behave exactly in the same manner as $n_{B-L}/s$, as one can perceive from the bottom panel plots. Therefore we do not elaborate them further.

\begin{figure}[htb!]
$$
\includegraphics[scale=0.57]{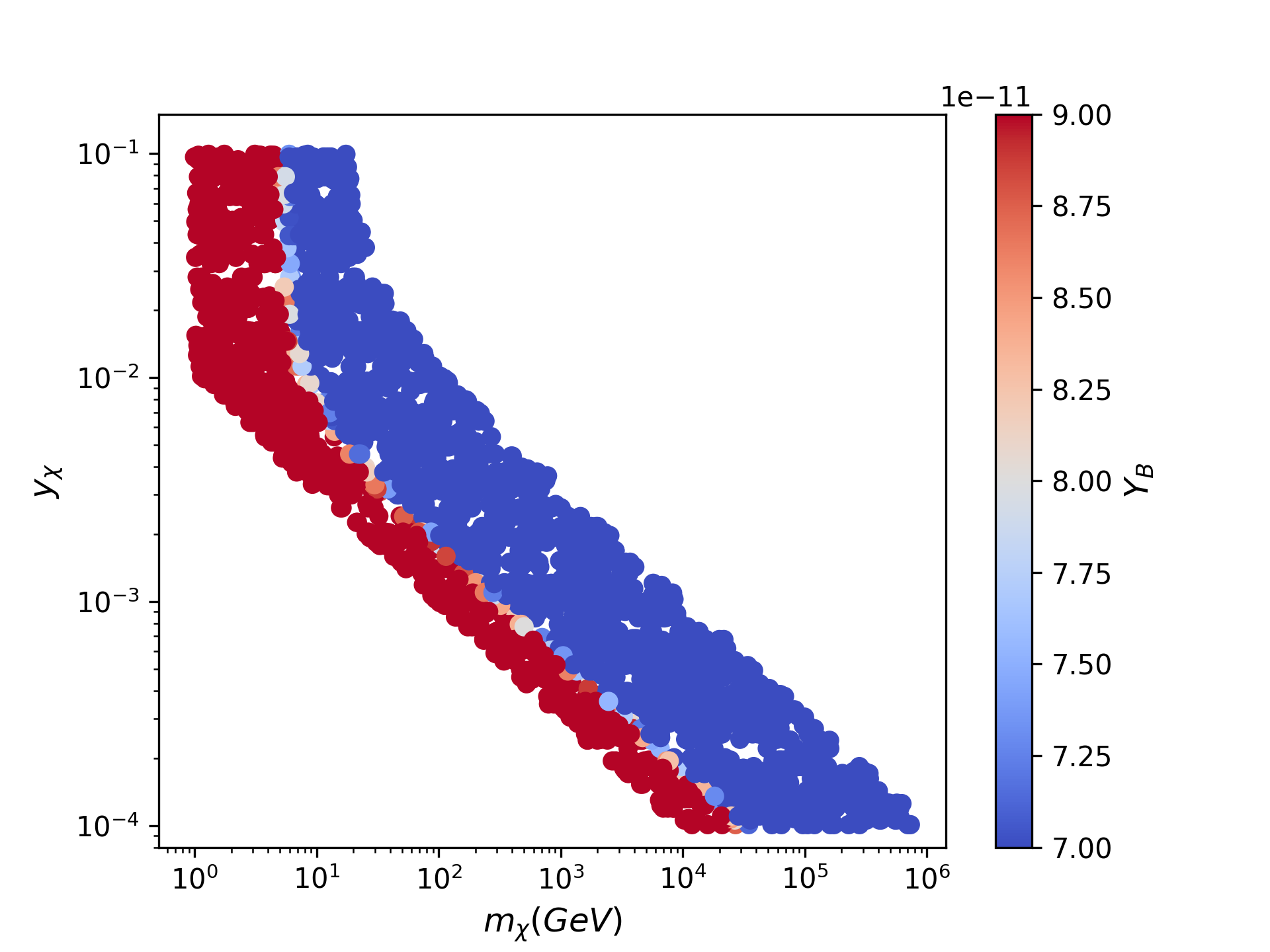}
\includegraphics[scale=0.57]{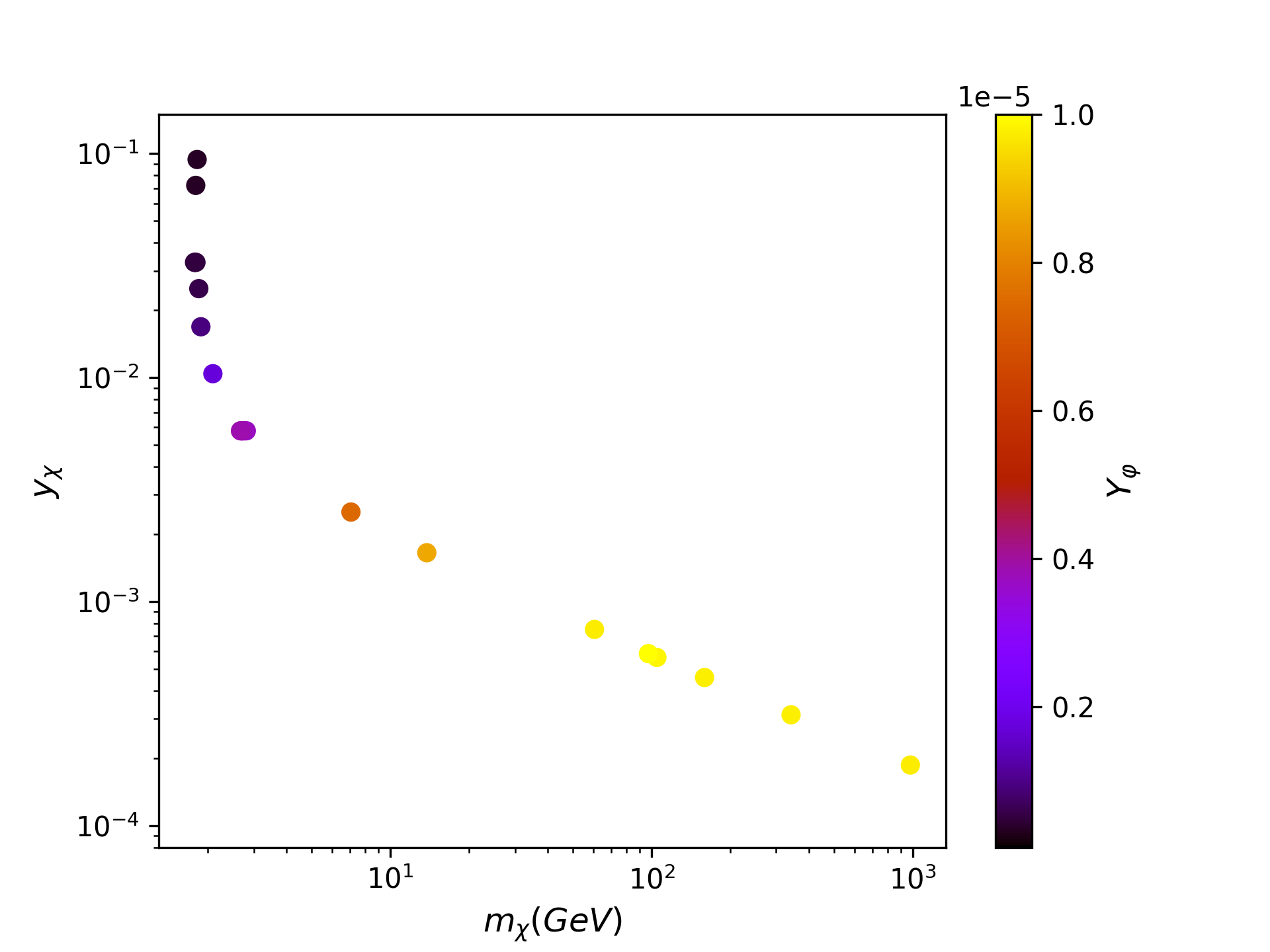}
$$
\caption{Left panel: Relic density allowed parameter space in $y_\chi-m_\chi$ plane where the colour bar shows the variation of $Y_B$, while all the points satisfy Planck observed relic abundance in the range $0.119\leq \Omega_\chi\,h^2\leq 0.122$. Right panel: Parameter space satisfying relic density and observed baryon asymmetry for $8.52\times 10^{-11}\leq Y_B\leq 8.98\times 10^{-11}$, where the colour code shows values of $y_\varphi$. Here we have fixed $M_1=10^{12}$ GeV and $m_\varphi=10^{14}$ GeV.}
\label{fig:param-space-inf}
\end{figure}

Finally, a scan of the viable parameter space can be done by varying the DM mass and the two Yukawa couplings over the ranges

\bea
m_\chi:\{1-10^8\}\,\text{GeV};\,y_\chi:\{10^{-4}-10^{-1}\};\,y_\varphi:\{10^{-7}-10^{-5}\}\,,
\eea

\noindent to satisfy both right relic abundance for the DM and also to produce the observed baryon asymmetry. The range of $y_\varphi$ is chosen in such a way that $\Gamma_\varphi\ll\Gamma_N$, and the universe has a very short RHN dominated epoch. One should also note, within this range of $y_\varphi$, the reheating temperature of the universe $T_\text{RH}\leq 10^{10}$ GeV that satisfies the condition for non-thermal leptogenesis since we are choosing $M_1=10^{12}$ GeV for all these scans. All coloured points in the left panel plot of Fig.~\ref{fig:param-space-inf} satisfy the Planck observed relic abundance of DM. Now, increasing $y_\chi$ results in a larger asymmetry in the DM sector because of larger branching ratio of corresponding $N_1$ decay, following Eq.~\eqref{eq:asyD}. This causes a larger final abundance for the DM number density since $X'\propto\epsilon_{\Delta\chi}$. Naturally, one requires a lighter DM to satisfy the observed abundance, since $\Omega_\chi\propto m_\chi\,X$. As a result, the parameter space shifts towards larger DM mass as we decrease $y_\chi$. It is possible to have DM of mass $\sim\mathcal{O}(\text{MeV})$ but at the expense of making the Yukawa couplings $y_{\chi,\varphi}$ larger. Thus, inflaton decay can give rise to DM of mass from a few MeVs up to several TeVs by tuning the Yukawa couplings within perturbative range. On the right panel we see only those points which satisfy both the relic abundance and the baryon asymmetry in the same two dimensional plane of $y_\chi-m_\chi$ as in the left, but now the colour coding is done with respect to $y_\varphi$. As we have already seen, a smaller $y_\chi$ requires a larger DM mass to satisfy the relic abundance, here we see for such points, in order to satisfy the observed $\eta_B$, a larger $y_\varphi$ is needed as well. Since a small $y_\chi$ implies a larger branching $\text{Br}\propto y_\chi/(y_\chi+y_N)$ of the RHN into the SM final states, resulting a larger asymmetry in the visible sector, hence one needs to have a smaller $y_\varphi$ to reduce the RHN yield. This will automatically result in a decrement of $\tilde{N}_{B-L}$ as $\tilde{N}_{B-L}'\propto \tilde{\rho}_N\propto\Gamma_\varphi$. Thus, a larger $y_\chi$ requires a smaller $y_\varphi$ and a larger $m_\chi$ to satisfy both the relic density and the right baryon asymmetry.

\begin{figure}[htb!]
$$
\includegraphics[scale=0.14]{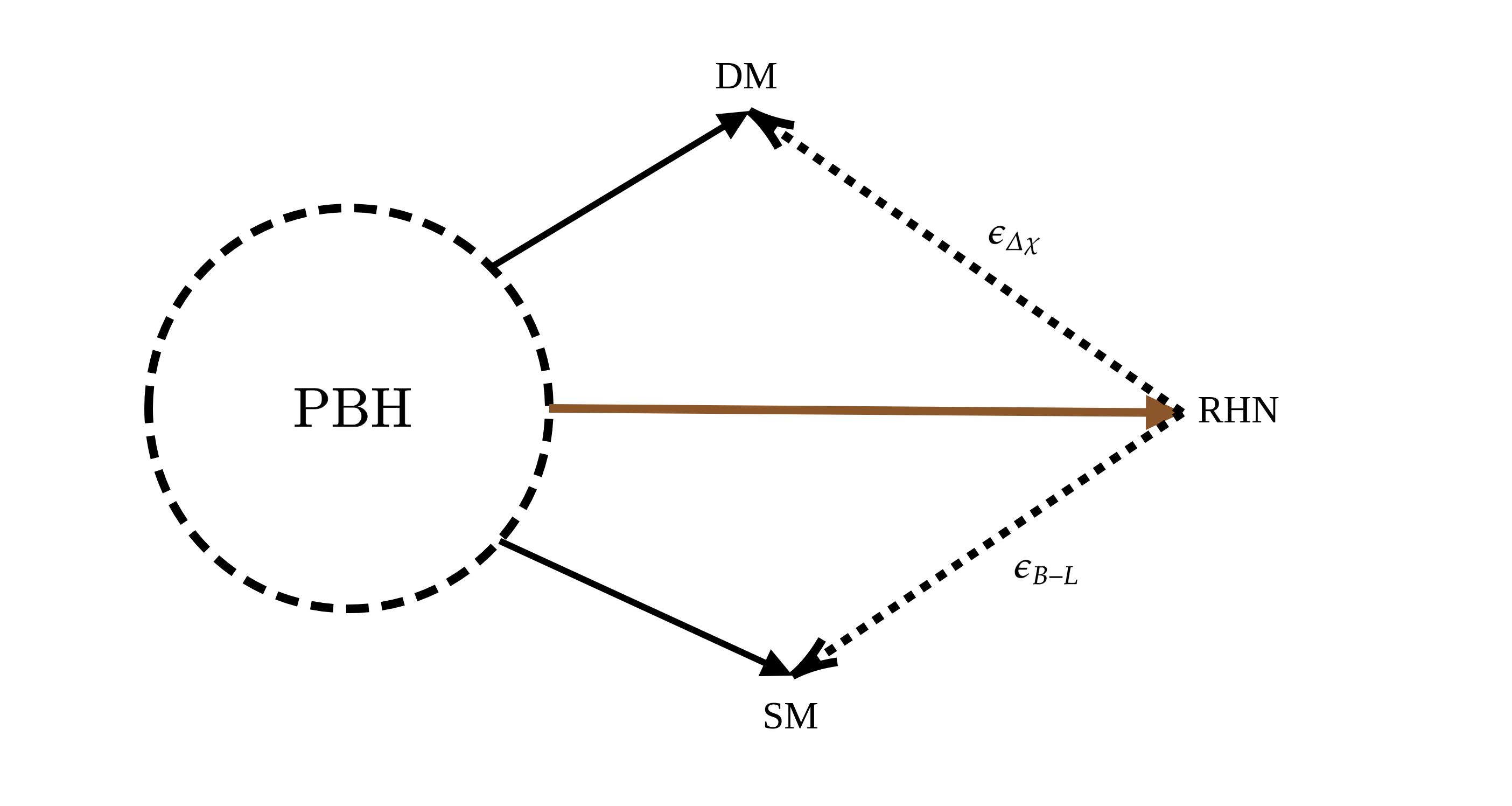}
$$
\caption{Schematic diagram of asymmetry production in the dark and in the visible sector in presence of primordial black holes.}\label{fig:scheme2}
\end{figure}

\section{Asymmetric Dark Matter from PBH evaporation}\label{sec:asdm2}

So far we have discussed the generation of asymmetries from the decay of inflaton in a model-agnostic way. In this section we will consider a scenario where asymmetries can emerge from the decay of the primordial black holes \footnote{A recent review of PBH may be found in \cite{Carr:2020gox}.}. To be more specific, the asymmetries result from out-of-equilibrium decay of RHN with the dominant contribution coming from non-thermal RHNs produced from the evaporation of PBH. However, the situation is a bit different here as unlike in the case of inflation where inflaton coupling could be tuned in a way such that it decays only into RHN, PBH, on the other hand, can emit {\it all} particles democratically. The scenario is schematically shown in Fig. \ref{fig:scheme2}. We discuss the interesting features of PBH briefly below followed by detailed discussion of its role in creation of asymmetries in dark as well as visible sectors.

\subsection{Primordial Black Hole: formation and constraints}

We assume PBHs are formed after inflation during the era of radiation domination. Assuming radiation domination, the mass of the black hole from gravitational collapse is typically close to the value enclosed by the post-inflation particle horizon and is given by~\cite{Fujita:2014hha,Masina:2020xhk}

\bea
m_\text{BH}=\frac{4}{3}\,\pi\,\gamma\,\Bigl(\frac{1}{\mathcal{H}\left(T_\text{in}\right)}\Bigr)^3\,\rho_\text{rad}\left(T_\text{in}\right)\,
\label{eq:pbh-mass}
\eea

\noindent with 

\bea
\rho_\text{rad}\left(T_\text{in}\right)=\frac{3}{8\,\pi}\,\mathcal{H}\left(T_\text{in}\right)^2\,M_\text{pl}^2\,
\eea

\noindent and $\gamma\simeq 0.2$ is a numerical factor which contains the uncertainty of the PBH formation.  As mentioned earlier,  PBHs are produced during the radiation dominated epoch, when the SM plasma has a temperature $T=T_\text{in}$ which is given by

\bea
T_\text{in}=\Biggl(\frac{45\,\gamma^2}{16\,\pi^3\,g_\star\left(T_\text{in}\right)}\Biggr)^{1/4}\,\sqrt{\frac{M_\text{pl}}{m_\text{BH}(T_\text{in})}}\,M_\text{pl}\,.
\label{eq:pbh-in}
\eea

Once formed, PBH can evaporate by emitting Hawking radiation \cite{Hawking:1974rv, Hawking:1975vcx}. A PBH can evaporate efficiently into particles lighter than its instantaneous temperature $T_\text{BH}$ defined as \cite{Hawking:1975vcx}
\begin{equation}
T_{\rm BH}=\frac{1}{8\pi\,G\, m_{\rm BH}}\approx 1.06~\left(\frac{10^{13}\; {\rm g}}{m_{\rm BH}}\right)~{\rm GeV}\,,
\end{equation}
\noindent where $G$ is the universal gravitational constant. The mass loss rate can be parametrised as  \cite{MacGibbon:1991tj}
\bea
\frac{dm_\text{BH}(t)}{dt}=-\frac{\mathcal{G}\,g_\star\left(T_\text{BH}\right)}{30720\,\pi}\,\frac{M_\text{pl}^4}{m_\text{in}(t)^2}\,,
\label{eq:pbh-dmdt}
\eea

\noindent where $\mathcal{G}\sim 4$ is the grey-body factor. Here we ignore the temperature dependence of $g_\star$ during PBH evolution, valid in the pre-sphaleron era. On integrating Eq.~\eqref{eq:pbh-dmdt} we end up with the PBH mass evolution equation as

\bea
m_\text{BH}(t)=m_\text{in}(T_\text{in})\Bigl(1-\frac{t-t_\text{in}}{\tau}\Bigr)^{1/3}\,,
\eea

\noindent with

\bea
\tau = \frac{10240\,\pi\,m_\text{in}^3}{\mathcal{G}\,g_\star(T_\text{BH})\,M_\text{pl}^4}\,,
\eea

\noindent as the PBH lifetime. Here onward we will use $m_\text{in}(T_\text{in})$ simply as $m_\text{in}$. The evaporation temperature can then be computed taking into account $H(T_\text{evap})\sim\frac{1}{\tau^2}\sim\rho_\text{rad}(T_\text{evap})$ as

\bea
T_\text{evap}\equiv\Bigl(\frac{45\,M_\text{pl}^2}{16\,\pi^3\,g_\star\left(T_\text{evap}\right)\,\tau^2}\Bigr)^{1/4}\,.
\label{eq:pbh-Tev}
\eea

\noindent However, if the PBH component dominates at some point the total energy density of the universe, the SM temperature just after the complete evaporation of PBHs is: $\overline{T}_\text{evap}=2/\sqrt{3}\,T_\text{evap}$~\cite{Bernal:2020bjf}. 



The initial PBH abundance is characterized by the dimensionless parameter $\beta$ that is defined as

\bea
\beta\equiv\frac{\rho_\text{BH}\left(T_\text{in}\right)}{\rho_\text{rad}\left(T_\text{in}\right)}\,,
\eea

\noindent that corresponds to the ratio of the initial PBH energy density to the SM energy density at the time of formation. Note that, $\beta$ steadily grows until PBH evaporation since the PBH energy density scales like non-relativistic matter $\sim a^{-3}$, while the radiation energy density scales as $\sim a^{-4}$. Therefore, an initially radiation-dominated universe will eventually become matter-dominated if the PBHs are still around. The condition of PBH evanescence during radiation domination can be expressed as~\cite{Masina:2020xhk}

\bea
\beta<\beta_\text{crit}\equiv \gamma^{-1/2}\,\sqrt{\frac{\mathcal{G}\,g_\star(T_\text{BH})}{10640\,\pi}}\,\frac{M_\text{pl}}{m_\text{in}}\,,
\label{eq:pbh-ev-rad}
\eea

\noindent  where $\beta_c$ is the critical PBH abundance that leads to early matter-dominated era. Note that for simplicity, we consider a monochromatic mass function of PBHs implying all PBHs to have identical masses. Additionally, the PBHs are assumed to be of Schwarzschild type without any spin and charge. The gravitational waves (GW) induced by large-scale density perturbations laid by PBHs could lead to a backreaction problem~\cite{Papanikolaou:2020qtd, Bernal:2020bjf}, that can be avoided if the energy contained in GWs never overtakes the one of the background universe or in other words if

\bea
\beta < 10^{-4}\,\Bigl(\frac{10^9\text{g}}{m_\text{in}}\Bigr)^{1/4}\,.
\eea

\begin{figure}[htb!]
$$
\includegraphics[scale=0.4]{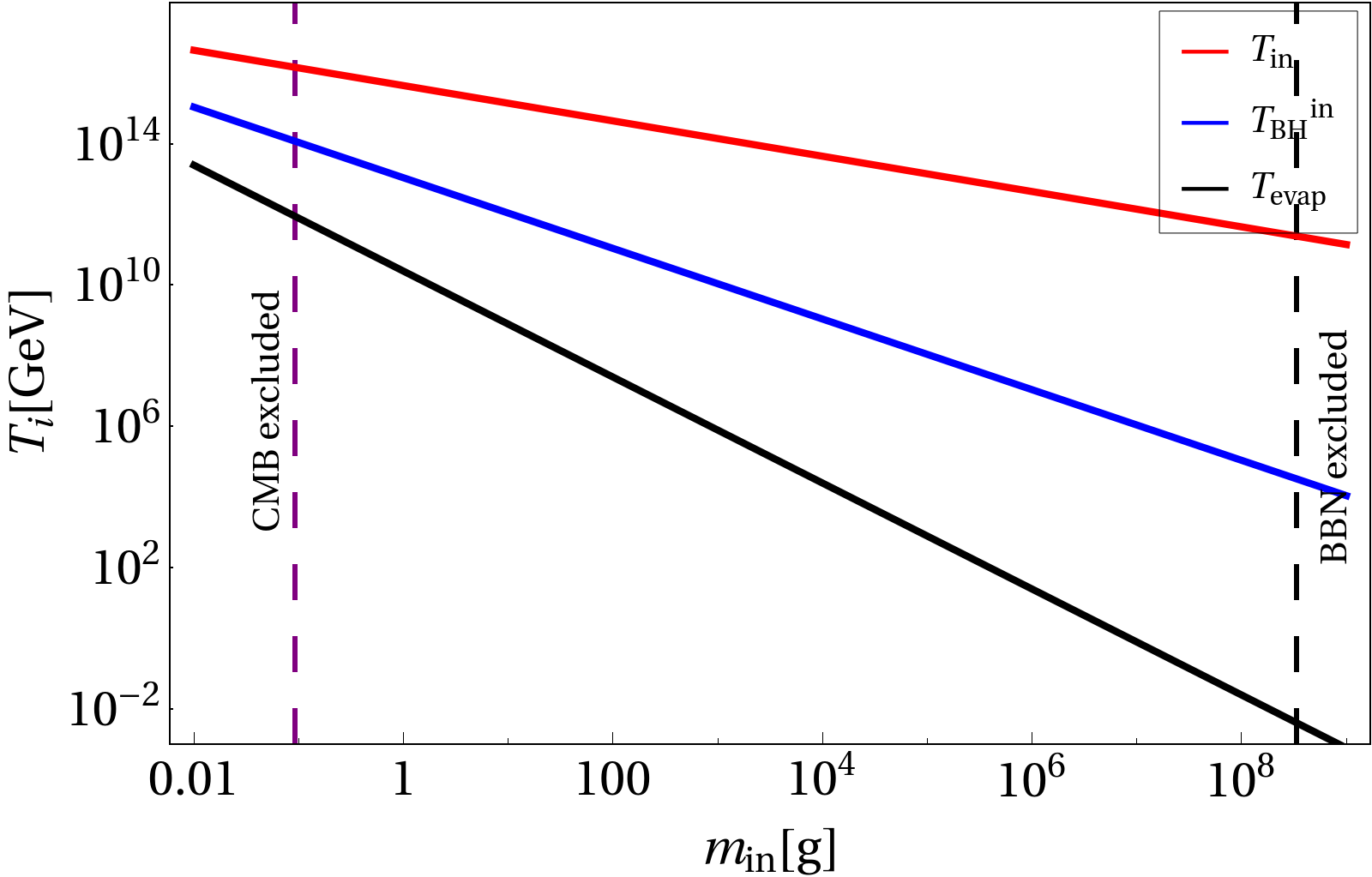}~~~~
\includegraphics[scale=0.4]{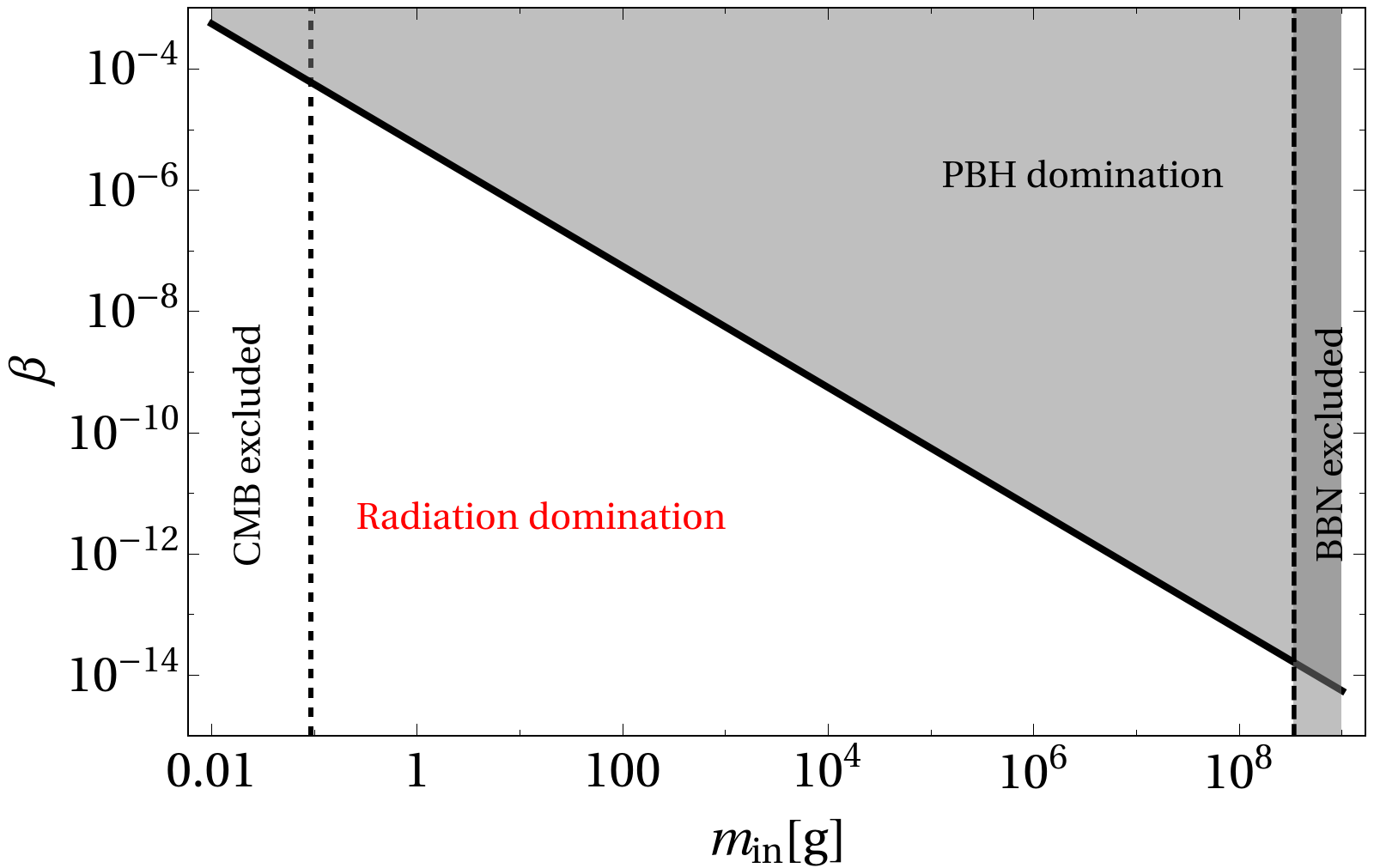}
$$
\caption{Left panel: PBH formation (red), evaporation (black) and the Hawking temperature (blue) as a function of the PBH mass. The purple and black dashed vertical lines correspond to the lower and upper bounds from CMB and BBN (see text). Right panel: $\beta_c$ as a function of PBH mass where the black thick diagonal line segregates radiation domination vs PBH domination.}
\label{fig:pbh-T}
\end{figure}

\noindent In left panel of Fig.~\ref{fig:pbh-T} we have shown the dependence of $T_\text{BH}$, $T_\text{in}$ and $T_\text{evap}$ on the PBH mass. As one can see, $T_\text{evap}$ falls fast with the rise in PBH mass (black solid line), while lighter PBHs have larger formation temperature. From the right panel, on the other hand, we find that in order to have PBH domination, $\beta$ has to be large enough for lighter PBH. As we will see, to produce observed baryon asymmetry together with right DM abundance, we need to rely upon ultralight PBHs and a comparatively large $\beta$ ensuring PBH domination. Since PBH evaporation produces all particles, including radiation that can disturb the successful predictions of BBN, hence we require $T_\text{evap}>T_\text{BBN}\simeq 4$ MeV. This can be translated into an upper bound on the PBH mass. On the other hand, a lower bound on PBH mass can be obtained from the CMB bound on the scale of inflation \cite{Planck:2018jri} : $\mathcal{H}_I\equiv \mathcal{H}(T_\text{in})\leq 2.5\times 10^{-5}\,M_\text{pl}$, where $\mathcal{H}(T_\text{in})=\frac{1}{2\,t_\text{in}}$ with $t(T_\text{in})=\frac{m_\text{in}}{M_\text{pl}^2\,\gamma}$ (as obtained from Eq.~\eqref{eq:pbh-mass}). Using these BBN and CMB bounds together, we have a window for allowed initial mass for PBH that reads

\bea
0.1\,\text{g}\lesssim m_\text{in}\lesssim 3.4\times 10^8\,\text{g}\,.
\eea

\noindent The range of PBH masses between these bounds is at present generically unconstrained~\cite{Carr:2020gox}. While PBH can evaporate by Hawking radiation, it can be stable on cosmological scales if sufficiently heavy, potentially giving rise to some or all of DM ~\cite{Carr:2020xqk}. The bounds and signatures of such heavy PBHs can be very different from the ones mentioned above and we do not discuss such cosmologically long-lived PBH any further.

\subsection{Right handed neutrino from PBH: Dark Matter and Baryogenesis}

Initially proposed by Hawking ~\cite{Hawking:1974rv, Hawking:1975vcx}, PBH can have several interesting consequences in cosmology \cite{Chapline:1975ojl, Carr:1976zz}. Even though the light PBHs of our interest are not long lived enough to be DM, they can still play non-trivial roles in genesis of DM as well as baryogenesis. Since PBH evaporate to all particles, irrespective of their SM gauge interactions, it can lead to production of DM, leptons, baryons etc as well as other heavy particles like RHN in our model. Although the evaporation of PBH in such a minimal scenario can not produce dark or visible sector asymmetries\footnote{One can also generate a chemical potential directly from PBH evaporation, as discussed within the framework of gravitational baryogenesis \cite{Smyth:2021lkn}.} on its own, it can produce heavy particles like RHNs whose subsequent decay can produce the required asymmetries. Such a role of PBH evaporation on baryogenesis was first pointed out in \cite{Hawking:1974rv, Carr:1976zz} followed by some detailed study in \cite{Baumann:2007yr} and recently it has been taken up by several authors in different contexts~\cite{Hook:2014mla, Fujita:2014hha, Hamada:2016jnq, Morrison:2018xla, Hooper:2020otu, Perez-Gonzalez:2020vnz, Datta:2020bht, JyotiDas:2021shi}\footnote{ In \cite{DeLuca:2021oer}, the authors found that the rate of baryon number violation via sphaleron transitions in the standard model can be enhanced in the presence of PBH.}. On the other hand, the role of PBH evaporation on DM genesis have been studied for different DM scenarios~\cite{Morrison:2018xla, Gondolo:2020uqv, Bernal:2020bjf, Bernal:2021bbv, Bernal:2021yyb}. 
Thus, PBH evaporation can lead to the generation of both RHNs and DM, depending on the PBH mass (see, for example,~\cite{Fujita:2014hha, Morrison:2018xla, Hooper:2019gtx, Lunardini:2019zob, Masina:2020xhk, Hooper:2020otu, Datta:2020bht, JyotiDas:2021shi, Schiavone:2021imu, Bernal:2021yyb, Bernal:2021bbv}). However, in the present framework, we are not interested in DM generation from direct PBH evaporation, rather we are interested in the scenario of DM production from the asymmetry generated in the dark sector via RHN decay. While we can not prevent DM generation from PBH evaporation, eventually DM abundance is dictated by its asymmetric component only, which is generated by the RHN decay only.


Before doing the complete numerical analysis, we first show the key features of such a scenario by using approximate analytical expressions. The total number of RHNs $\mathcal{N}$ with mass $M_N$ emitted during PBH evaporation can be estimated using \cite{Baumann:2007yr, Lunardini:2019zob}

\bea
d\mathcal{N} = -\frac{d\left(m_\text{BH}\right)}{3\,T_\text{BH}}\,,
\eea

\noindent which gives rise to

\begin{equation}
    \mathcal{N} = \frac{g_\mathcal{N}}{g_\star(T_\text{BH})}
    \begin{cases}
       \frac{4\,\pi}{3}\,\Bigl(\frac{m_\text{in}}{M_\text{pl}}\Bigr)^2 &\text{for } M_N < T_\text{BH}^\text{in}\,,\\[8pt]
        \frac{1}{48\,\pi}\,\Bigl(\frac{M_\text{pl}}{M_\mathcal{N}}\Bigr)^2 &\text{for } M_N > T_\text{BH}^\text{in}\,,
    \end{cases}\label{eq:pbh-num}
\end{equation}

\noindent where $g_\mathcal{N}$ is the number of degrees of freedom for the RHN and $T_\text{BH}^\text{in}=T_\text{BH}(t=t_\text{in})$ is the initial PBH temperature. Note that, for $M_N<T_\text{BH}^\text{in}$, PBH emits RHNs from the beginning, namely the formation of PBHs. In the opposite case, PBH emits RHNs only after its Hawking temperature reaches $M_N$.

The PBHs emit RHNs (along with all the SM particles and DM), and the CP-violating decays of such non-thermal RHNs produce the lepton asymmetry. This lepton asymmetry is then further converted into the observed baryon asymmetry via sphaleron transition like in standard leptogenesis scenario. If $\mathcal{N}$ is the number of RHNs emitted from a single PBH then the present baryon number yield can be written as~\cite{Baumann:2007yr,Fujita:2014hha,Datta:2020bht}

\bea
\frac{n_B}{s}\left(T_0\right)=\mathcal{N}\,\epsilon_1\,a_\text{sph}\,\frac{n_\text{PBH}}{s}\Big|_{T_\text{evap}}\,,
\label{eq:pbh-nb}
\eea

\noindent where we assume no further entropy production after PBH evaporation.  
\begin{figure}[htb!]
$$
\includegraphics[scale=0.14]{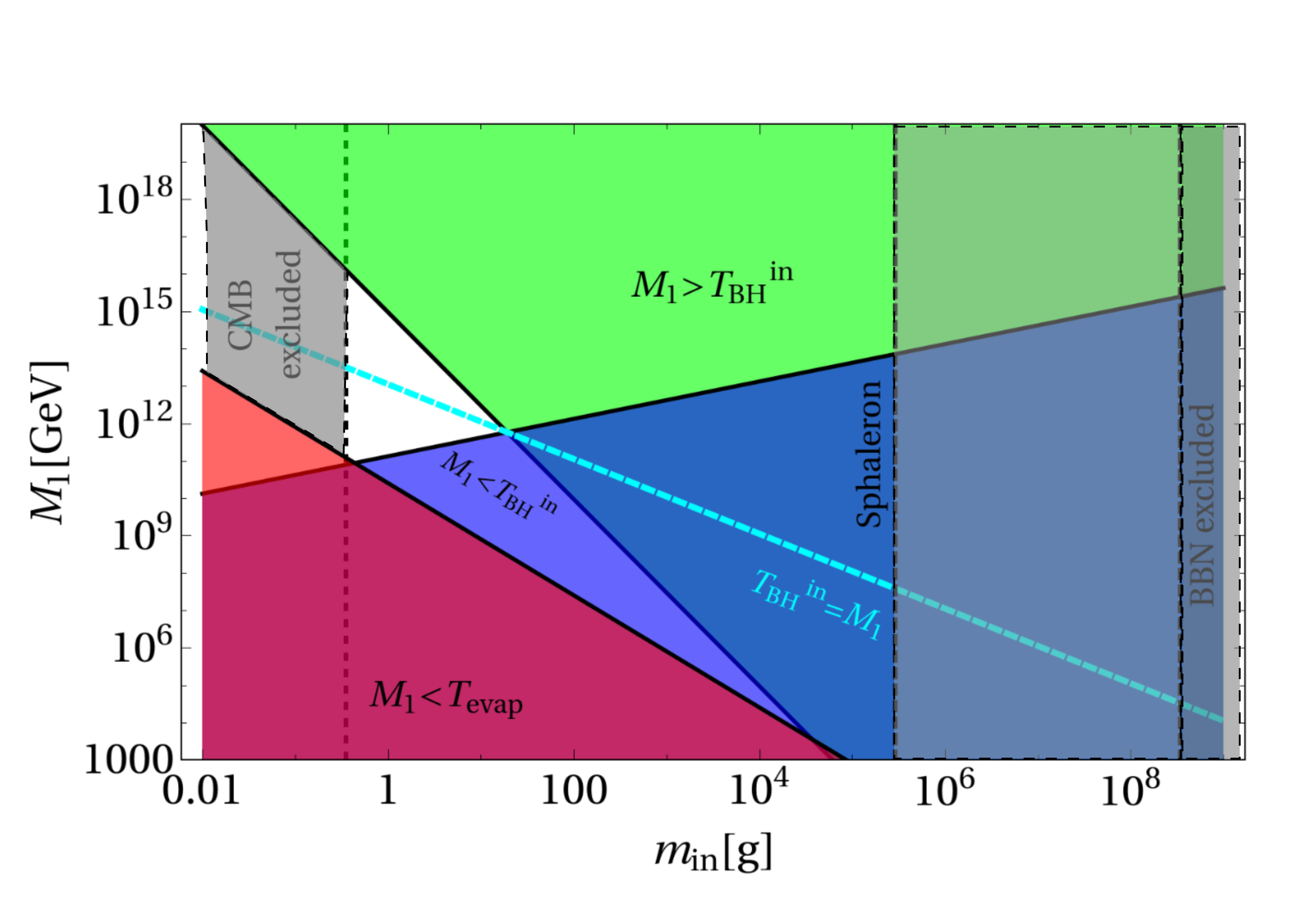}
$$
\caption{Bound on RHN mass from the requirement of obtaining observed baryon asymmetry from PBH evaporation. All the coloured regions are discarded from the bounds derived in Eq.~\eqref{eq:m1bound1} and Eq.~\eqref{eq:m1bound2}. The vertical black dashed line corresponds to (from left to right) the bound from the scale of inflation (CMB), sphaleron transition and BBN. The white triangular region in the middle is the region that is allowed.}
\label{fig:m1-bound}
\end{figure}

It is possible to analytically derive the mass range of RHNs emitted from PBH evaporation that can provide the required lepton asymmetry. In the Type-I seesaw mechanism, the quantity $\epsilon$ has an upper bound~\cite{Davidson:2002qv,Samanta:2020gdw}

\bea
\epsilon\lesssim \frac{3}{16\,\pi}\,\frac{M_1\,m_{\nu,\text{max}}}{v^2}\,,
\eea

\noindent where $v=246$ GeV is the SM Higgs VEV and $m_{\nu,\text{max}}$ is the mass of the heaviest light neutrino. On the other hand, the final asymmetry produced from PBH evaporation as computed from Eq.~\eqref{eq:pbh-nb},  $Y_B=n_B/s\Big|_{T_0}\simeq 8.7\times 10^{-4}$~\cite{Planck:2018jri}. These together constrain the mass of the RHN  produced from PBH evaporation both from above and from below 

\begin{equation}
    M_1 
    \begin{cases}
        > \frac{2\,g_\star(T_\text{BH})}{a_\text{sph}\,g_N}\,\frac{M_\text{pl}^2\,v^2}{m_{\nu,\text{max}\,m_\text{in}^2}}\,Y_B(T_0)\,\frac{n_\text{PBH}}{s}\Big|_{T_\text{evap}} &\text{for } M_1 < T_\text{BH}^\text{in}\,;\\[8pt]
        < \frac{a_\text{sph}\,g_N}{128\,\pi\,g_\star(T_\text{BH})}\,\frac{M_\text{pl}^2\,m_{\nu,\text{max}}}{v^2}\frac{1}{Y_B(T_0)}\,\frac{n_\text{PBH}}{s}\Big|_{T_\text{evap}} &\text{for } M_1 > T_\text{BH}^\text{in}\,,
    \end{cases}\label{eq:m1bound1}
\end{equation}

\noindent where we have used

\bea
n_\text{PBH}\left(T_\text{evap}\right)=\frac{1}{m_\text{in}}\,\frac{\pi^2}{30}\,g_\star\left(T_\text{evap}\right)\,T_\text{evap}^4\,.
\eea

Another bound comes from the fact that if $M_1<T_\text{evap}$, then the RHNs produced from PBH evaporation are in thermal bath and then washout processes are in effect. Hence, to ensure non-thermal production of baryon asymmetry one must follow~\cite{Fujita:2014hha}

\bea
M_1 > T_\text{evap}\implies M_1 \gtrsim 3\times 10^{-3}\,\Bigl(\frac{\mathcal{G}^2\,g_\star\left(T_\text{evap}\right)\,M_\text{pl}^{10}}{m_\text{in}^6}\Bigr)^{1/4}\,.
\label{eq:m1bound2}
\eea
\noindent For scenarios where both thermal and non-thermal RHNs were taken into account for generation of lepton asymmetry, one may refer to \cite{Perez-Gonzalez:2020vnz, JyotiDas:2021shi}.

Lastly, in order for lepton asymmetry to be sufficiently generated from RHNs produced from PBH evaporation, one requires evaporation to be over before sphaleron transition $T_\text{evap}\gtrsim T_\text{EW}$, which translates into the corresponding bound on initial PBH mass

\bea
m_\text{in}\lesssim 3\times 10^5\,\text{g}\,.
\eea

\noindent However, this bound similar to the BBN bound is naturally satisfied as one can see from Fig.~\ref{fig:m1-bound}. We thus find that the observed baryon asymmetry is produced over a very tiny region for $10^{11}\lesssim M_1\lesssim 10^{16}$ GeV and $0.5\lesssim m_\text{in}\lesssim 10$ g, depicted by the white triangular region in Fig.~\ref{fig:m1-bound}.


While the PBH evaporation can not create baryon asymmetry directly in our minimal scenario, it can create DM directly, as discussed in different contexts \cite{Morrison:2018xla, Gondolo:2020uqv, Bernal:2020bjf, Green:1999yh, Khlopov:2004tn, Dai:2009hx, Allahverdi:2017sks, Lennon:2017tqq, Hooper:2019gtx, Chaudhuri:2020wjo, Masina:2020xhk, Baldes:2020nuv, Bernal:2020ili,
Bernal:2020kse, Lacki:2010zf, Boucenna:2017ghj, Adamek:2019gns, Carr:2020mqm, Masina:2021zpu, Sandick:2021gew, Cheek:2021cfe, Cheek:2021odj}. However, in asymmetric DM scenario, the final DM abundance is dictated by the dark sector asymmetry which is created only by the out-of-equilibrium decay of RHN where the latter is produced dominantly from PBH evaporation. Thus, in the present scenario asymmetric DM yield can be expressed as

\begin{equation}
Y_\text{DM}\left(T_0\right) = \epsilon_{\Delta\chi}\,\mathcal{N}\,\frac{n_\text{PBH}}{s}\left(T_\text{evap}\right)\,,
\label{eq:pbh-dm-yld}
\end{equation}

\noindent that leads to DM abundance 

\begin{equation}
\Omega_\text{DM}\,h^2 = \frac{m_\text{DM}\,s_0}{\rho_c}\,Y_\text{DM}\left(T_0\right)\,,
\label{eq:pbh-dm-om}
\end{equation}

\noindent where $\mathcal{N}$ is defined via Eq.~\eqref{eq:pbh-num}, which results in

\begin{equation}
   \Omega_\text{DM}\,h^2 = \frac{g_\text{DM}}{g_\star\left(T_\text{evap}\right)}\,\frac{m_\text{DM}\,s_0}{\rho_c}\,\epsilon_{\Delta\chi}\,\frac{n_\text{PBH}}{s}\Bigr|_{T_\text{evap}}  
    \begin{cases}
       \frac{4\,\pi}{3}\,\left(m_\text{in}/M_\text{pl}\right)^2 &\text{for } T_\text{BH}^\text{in} > M_1\,;\\[8pt]
       \frac{1}{48\,\pi}\,\left(M_\text{pl}/M_1\right)^2 &\text{for } T_\text{BH}^\text{in} < M_1\,.
    \end{cases}\label{eq:pbh-dm-rel}
\end{equation}

\begin{figure}[htb!]
$$
\includegraphics[scale=0.38]{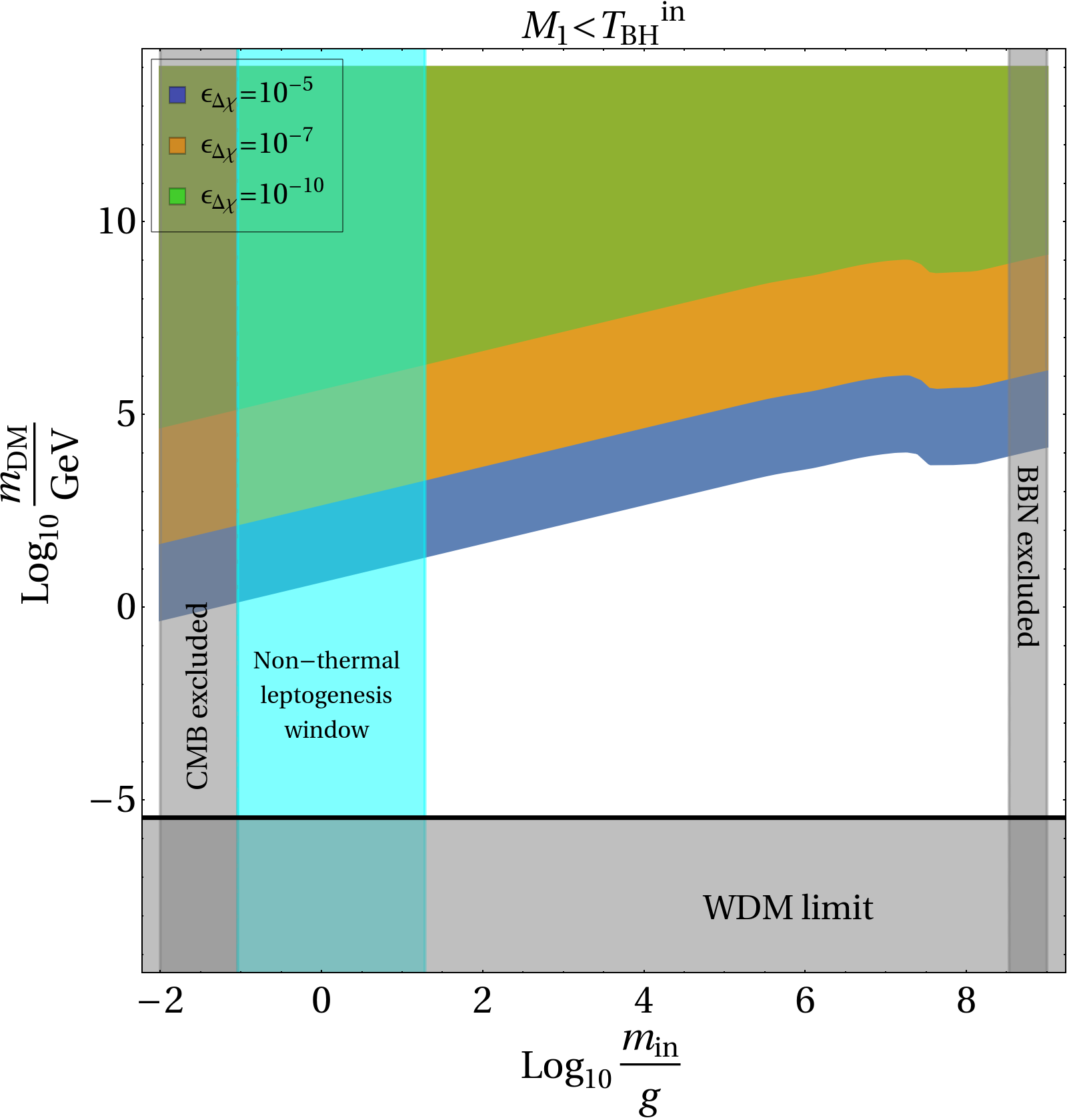}~~~~
\includegraphics[scale=0.38]{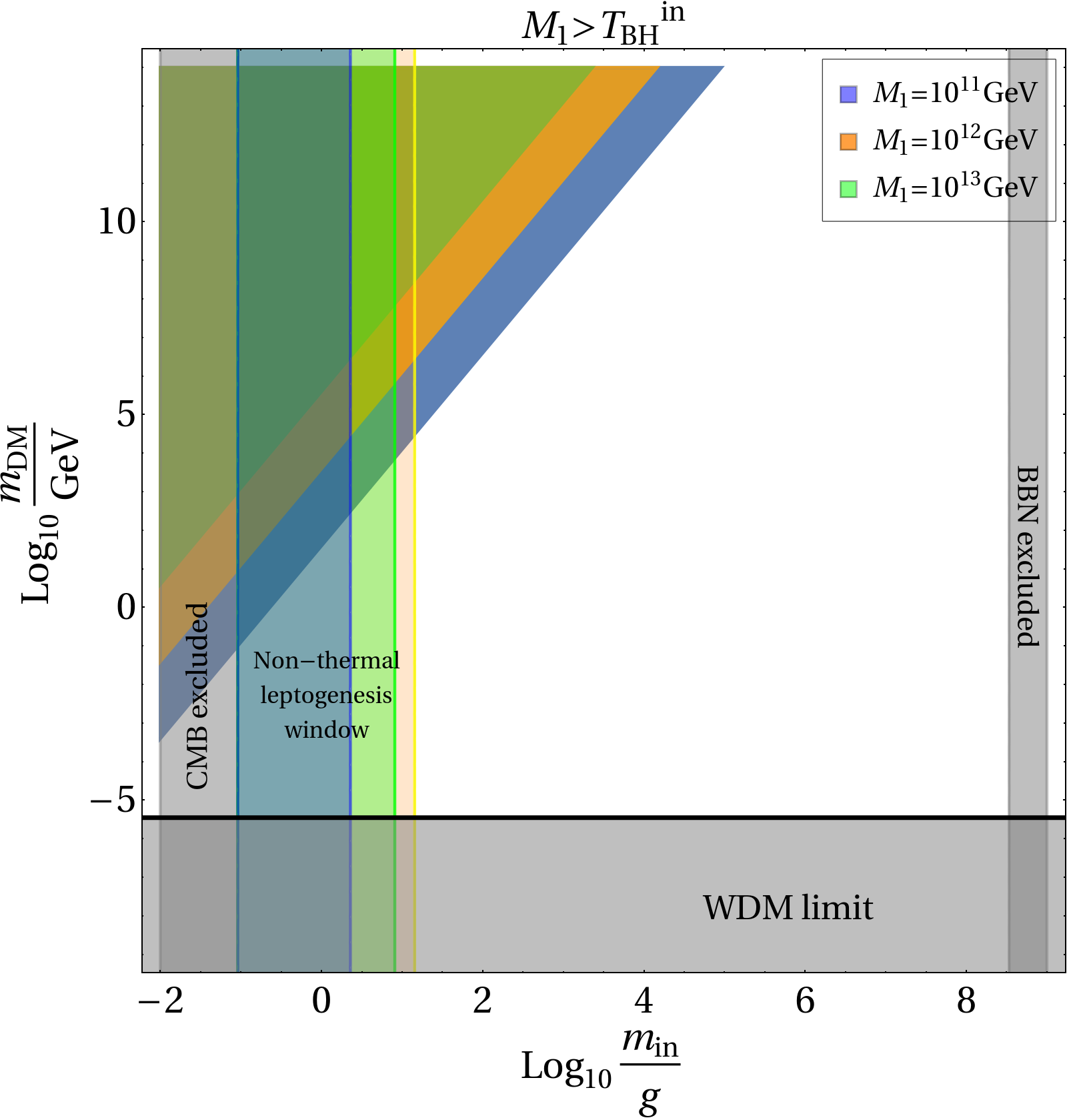}
$$
\caption{The dark green, orange and blue coloured regions are excluded from DM overproduction due to different choices of DM asymmetry (left panel) and RHN mass (right panel) shown by different colours. In the right panel we have chosen $\epsilon_{\Delta\chi}=10^{-10}$. The gray shaded regions are disallowed from CMB, BBN and warm DM limit, while the cyan band (left) is where non-thermal leptogenesis from PBH is allowed for $M_1\simeq 10^{12}$ GeV (see text).}
\label{fig:mdm-bound}
\end{figure}

\noindent Note that the final DM asymmetry depends on the mass of RHN as expected. To match the observed DM abundance $\Omega_\text{DM} h^2 \simeq 0.12$, the DM yield has to be fixed so that $m_\text{DM}\,Y_0 = \Omega_\text{DM} h^2 \frac{1}{s_0} \frac{\rho_c}{h^2} \simeq 4.3 \times 10^{-10}$~GeV, where $\rho_c \simeq 1.1 \times 10^{-5} h^2$~GeV/cm$^3$ is the critical energy density and $s_0 \simeq 2.9 \times 10^3$~cm$^{-3}$ is the entropy density at present~\cite{Aghanim:2018eyx}. Fig.~\ref{fig:mdm-bound} depicts the allowed mass range for the DM obtained analytically using Eq.~\eqref{eq:pbh-dm-rel}. In the left panel, different regions shown by the green, orange and blue colours correspond to different choices of DM asymmetry $\epsilon_{\Delta\chi}$, for the case where $T_\text{BH}^\text{in}>M_\text{1}$. All these regions correspond to DM overabundance and hence discarded. Here we see larger asymmetry imposes tighter constraint on the DM mass. This is expected since a larger asymmetry results in larger asymmetric DM abundance as per Eq.~\eqref{eq:pbh-dm-rel}. Hence slight increase in the DM mass results in overabundance. The slight distortion in the large DM mass region is due to the change in the number of light degrees of freedom around $T_\text{evap}\simeq 150$ MeV, i.e., around the time of QCD phase transition. In the opposite limit $T_\text{BH}^\text{in} < M_\text{1}$, shown in the right panel, we see a lighter RHN imposes tighter bound since $\Omega_\text{DM}\propto 1/M_1^2$ for a fixed CP asymmetry $\epsilon_{\Delta\chi}=10^{-10}$. 

In order not to spoil the structure formation, a fermion DM candidate which is part of the thermal bath or produced from the thermal bath should have mass above a few keV in order to give required free-streaming of DM as constrained from Lyman-$\alpha$ flux-power spectra~\cite{Irsic:2017ixq, Ballesteros:2020adh, DEramo:2020gpr}. Such light DM of keV scale leads to a warm dark matter (WDM) scenario having free-streaming length within that of cold and hot DM. If such light DM is also produced from PBH evaporation, it leads to a potential hot component in total DM abundance, tightly constrained by observations related to the CMB and baryon acoustic oscillation (BAO) leading to an upper bound on the fraction of this hot component with respect to the total DM, depending on the value of DM mass \cite{Diamanti:2017xfo}. A conservative $10~\%$ upper bound on such hot dark matter (HDM) component \cite{Bernal:2020bjf} can lead to similar constraints on DM mass along with PBH initial fraction. The requirement of producing right relic abundance, together with these lower limits on the DM mass put tight constraint on the DM mass emitted by the PBH. This is shown by the gray shaded regions in Fig.~\ref{fig:mdm-bound}. We also show the window of PBH mass in cyan where successful non-thermal leptogenesis from RHN emitted by PBH is possible (Fig.~\ref{fig:m1-bound}) for $M_1\simeq 10^{12}$ GeV. In the right panel we have chosen different masses for the RHN, corresponding to which the allowed mass window for non-thermal leptogenesis changes (maximum for $M_1=10^{12}$ GeV shown in orange) as denoted by different coloured vertical bands. The upshot of Fig.~\ref{fig:mdm-bound} is that, it is possible to generate observed asymmetry in visible sector, together with right relic abundance of asymmetric DM with ultralight PBH and for DM mass $\gtrsim 10^{-5}$ GeV. The upper bound on the DM mass depends on the size of the asymmetry generated within the dark sector, which depends also on the RHN mass scale. Finally, it is also importance to note that a large initial abundance of PBHs increases the DM capture rate and makes the DM under-abundant. However, this is found to be significant only for superheavy DM with mass $\sim 10^8$ GeV (for fermion) and $\beta\gtrsim 10^{-4}$~\cite{Bernal:2020bjf,Gondolo:2020uqv}.

\subsection{Results and Discussions}

In order to compute the asymmetries we will now perform a full numerical analysis considering a set of coupled BEQs accounting for the energy and number densities of different components. We focus on the production of the observed baryon asymmetry via non-thermal leptogenesis, together with the correct relic abundance for the asymmetric DM. Thus, we track the evolution of the comoving number densities of the RHN, PBH, lepton and DM asymmetries and the radiation energy density via their coupled BEQs. What is crucial here is the fact that since the PBHs are assumed to be produced during the radiation dominated era, hence thermal contribution to leptogenesis can not be overlooked. The evolution equation for PBH mass, energy densities and bath temperature\footnote{In principle, the BEQ for radiation should also contain the contribution from RHN decay into the thermal, but such contributions are negligible compared to the PBH contribution and can be ignored.} in presence of PBH reads~\cite{Perez-Gonzalez:2020vnz,JyotiDas:2021shi}


\begin{equation}\begin{split}
&  \frac{dm_\text{BH}}{d\xi} = -\frac{\kappa}{\xi\,\mathcal{H}}\,\epsilon(m_\text{BH})\,\left(\frac{1\text{g}}{m_{\rm BH}}\right)^2\,,
\\&
\frac{d\widetilde{\rho}_R}{d\xi}=-\frac{\epsilon_\text{SM}(m_\text{BH})}{\epsilon(m_\text{BH})}\,\frac{\xi}{m_\text{BH}}\,\frac{dm_\text{BH}}{d\xi}\,\widetilde{\rho}_\text{BH}\,,
\\& 
\frac{d\widetilde{\rho}_\text{BH}}{d\xi}=\frac{1}{m_\text{BH}}\,\frac{dm_\text{BH}}{d\xi}\,\widetilde{\rho}_\text{BH}\,,
\\&
\frac{dT}{d\xi}=-\frac{T}{\Delta}\Biggl[\frac{1}{\xi}+\frac{\epsilon_\text{SM}(m_\text{BH})}{\epsilon(m_\text{BH})}\,\frac{1}{m_\text{BH}}\,\frac{dm_\text{BH}}{d\xi}\,\frac{g_\star(T)}{g_{\star s}(T)}\,\xi\,\frac{\widetilde{\rho}_{BH}}{4\,\widetilde{\rho}_{R}}\Biggr]\,,
\end{split}
\end{equation}

\noindent where

\begin{equation}
\Delta = 1+\frac{T}{3\,g_{\star s}(T)}\,\frac{dg_{\star s}(T)}{dT}\,, 
\end{equation}

\noindent takes care of the variation of the total number of DOFs with temperature and $\xi=a/a_I$ (with $a_I=1$) as defined earlier. The evaporation function $\epsilon(m_\text{in})$ is taken from~\cite{Lunardini:2019zob, MacGibbon:1991tj}. The coupled BEQs for the evolution of the RHN number density, DM and lepton asymmetries, on the other hand, are given by~\cite{Perez-Gonzalez:2020vnz,JyotiDas:2021shi}

\begin{equation}\begin{split}
& a\mathcal{H} \frac{d\widetilde{n}_{N_1}^T}{d\xi}= -\left(\widetilde{n}_{N_1}^T-\widetilde{n}_{N_1}^\text{eq}\right)\,\Gamma^T_{N_1}\, ,
\\&
a\mathcal{H}\frac{d\widetilde{n}_{N_1}^\text{BH}}{d\xi}=-\widetilde{n}_{N_1}^\text{BH}\,\Gamma_{N_1}^\text{BH}+\Gamma_{\text{BH}\to N_1}\,\frac{\widetilde{\rho}_\text{BH}}{m_\text{BH}}\, ,
\\&
a\mathcal{H}\frac{d\widetilde{N}_{B-L}}{d\xi}=\epsilon_{\Delta L}\,\Biggl[\left(\widetilde{n}_{N_1}^T-\widetilde{n}_{N_1}^\text{eq}\right)\,\Gamma^T_{N_1}+\widetilde{n}_{N_1}^\text{BH}\,\Gamma_{N_1}^\text{BH}\Biggr]-\text{Br}_\text{SM}\,\mathcal{W}\,\widetilde{N}_{B-L}\, ,
\\&
a\mathcal{H}\frac{dX}{d\xi}=\epsilon_{\Delta \chi}\,\Biggl[\left(\widetilde{n}_{N_1}^T-\widetilde{n}_{N_1}^\text{eq}\right)\,\Gamma^T_{N_1}+\widetilde{n}_{N_1}^\text{BH}\,\Gamma_{N_1}^\text{BH}\Biggr]-\text{Br}_\text{DM}\,\mathcal{W}\,X\,,
\end{split}
\end{equation}

\noindent where all $\widetilde{n}_{N_1}$'s are comoving number densities of $N_1$ produced from the bath (denoted by superscript $T$) and PBH (denoted by superscript BH) and $\text{Br}$ stands for the branching ratio of RHN into leptons (denoted by subscript SM) and DM (denoted by subscript DM). Similarly $\widetilde{N}_{B-L}, X$ are comoving densities of $B-L$ and dark sector asymmetries respectively. Here we would like to mention that the RHNs produced from PBH evaporation never come into thermal equilibrium with the SM bath. In order to ensure that, we computed the thermally averaged cross-section for scattering of RHNs produced from the PBH evaporation against the bath particles, e.g., $N_1\ell\to N_1\ell$ following the prescription in~\cite{Cheek:2021cfe} (the detailed derivation is given in Appendix.~\ref{sec:app-sigmav}), and compared the corresponding rate with the Hubble rate. We found that for $T\simeq T_\text{evap}$, the RHN interaction rate is several orders of magnitude less than the Hubble expansion rate, typically $ n_\text{eq}\,\langle\sigma v\rangle/\mathcal{H}\lesssim\ 10^{-7}$. This shows that the RHNs produced from PBH are genuinely non-thermal. The thermally averaged decay rate of $N_1$ is denoted by $\Gamma^T_{N_1}$ and $\widetilde{n}_{N_1}^\text{eq}$ is the equilibrium number density. The Hubble parameter $\mathcal{H}$ entering in the Boltzmann equations is given by

\bea
\mathcal{\mathcal{H}} = \sqrt{\frac{8\,\pi}{3\,M_\text{pl}^2}\,\frac{\widetilde{\rho}_\text{BH}\,a_I\,\xi+\widetilde{\rho}_R}{a_I^4\,\xi^4}}\,,
\eea

\noindent and $\Gamma_{{\rm BH}\rightarrow{N_1}}$ is the non-thermal production term for $N_{1}$ (originating from PBH evaporation) and can be written as \cite{Lunardini:2019zob, Perez-Gonzalez:2020vnz}

\begin{equation}
\Gamma_{{\rm BH}\rightarrow{N_1}}=\int_0^\infty\,\frac{d^2\,\mathcal{N}}{dp\,dt}\,dp\simeq\dfrac{27\,T_{\rm BH}}{32\,\pi^{2}}\left(-z_{\rm BH}\,{\rm Li}_{2}(-e^{-z_{\rm BH}})-{\rm Li}_{3}(-e^{-z_{\rm BH}}) \right)\,,
\end{equation} 

\noindent where ${\rm Li}_{s}(z)$ are the poly-logarithm functions of order $s$ and $z_{\rm BH}=M_{1}/T_{\rm BH}$. $\Gamma_{1}^{\rm BH}$ is the decay width corrected by an average time dilation factor~\cite{Perez-Gonzalez:2020vnz} 

\begin{equation}
\Gamma_{N_1}^{\rm BH}=\Biggl\langle\frac{M_1}{E_1}\Biggr\rangle_\text{BH}\,\Gamma_1\approx\dfrac{K_{1}\left(M_{1}/T_{\rm BH}\right)}{K_{2}\left(M_{1}/T_{BH}\right)}\,\Gamma_{1}\,, 
\end{equation}

\noindent where $K_{1,2}[...]$ are the modified Bessel functions of second kind and the thermal average is obtained assuming that the Hawking spectrum has a  Maxwell-Boltzmann form, while $\Gamma_1$ is the RHN decay width given by Eq.~\eqref{eq:N1decay}. Finally, the washout factor $\mathcal{W}$ reads~\cite{Buchmuller:2004nz}

\bea
\mathcal{W} = \frac{1}{4}\,\Gamma_{N_1}^T\,K_2\left(z\right)\,z^2\,,
\eea

\noindent where $z\equiv M_1/T$ and we are ignoring the flavour effects as well as the scattering processes leading to washouts. The generation of lepton asymmetry has thermal and non-thermal sources stemming from the plasma and PBH evanescence respectively. On the other hand, DM can be present in the thermal bath while its asymmetric component arises from RHN decay and eventually only the asymmetric component survives. From Fig.~\ref{fig:m1-bound} we have already realized that non-thermal leptogenesis from PBH necessarily requires ultralight PBH with $M_1\gtrsim 10^{12}$ GeV. On the other hand, it is clear from the right panel of Fig.~\ref{fig:pbh-T}, for very light PBHs to dominate the energy density, the initial energy fraction of PBH density should be much higher. Hence, a long period of PBH domination is preferred for purely non-thermal leptogenesis from PBH~\cite{Datta:2020bht}. Otherwise, the asymmetry production will be dominated by thermally generated RHNs with PBH leading to subsequent entropy dilution only \cite{Perez-Gonzalez:2020vnz, JyotiDas:2021shi}. For $N_2$ leptogenesis, one can as well get an enhancement of asymmetry in the presence of PBH compared to the usual thermal case \cite{JyotiDas:2021shi}. However, we restrict ourselves to $N_1$ leptogenesis only and consider the production to be dominant from non-thermal RHNs produced from PBH evaporation. In the rest of the analysis we will thus restrict the RHN mass to be $M_1=10^{12}$ GeV unless otherwise specified. We will first look at the impact of having PBHs on the energy densities and the yield of the asymmetries in visible and dark sectors. For this we consider some benchmark masses of the PBH in 1-100 g range falling in the allowed region of Fig. \ref{fig:m1-bound}. We also fix the $N_1-$DM Yukawa coupling $y_\chi=0.1$ that determines the asymmetry in dark sector. 

\begin{figure}[htb!]
$$
\includegraphics[scale=0.25]{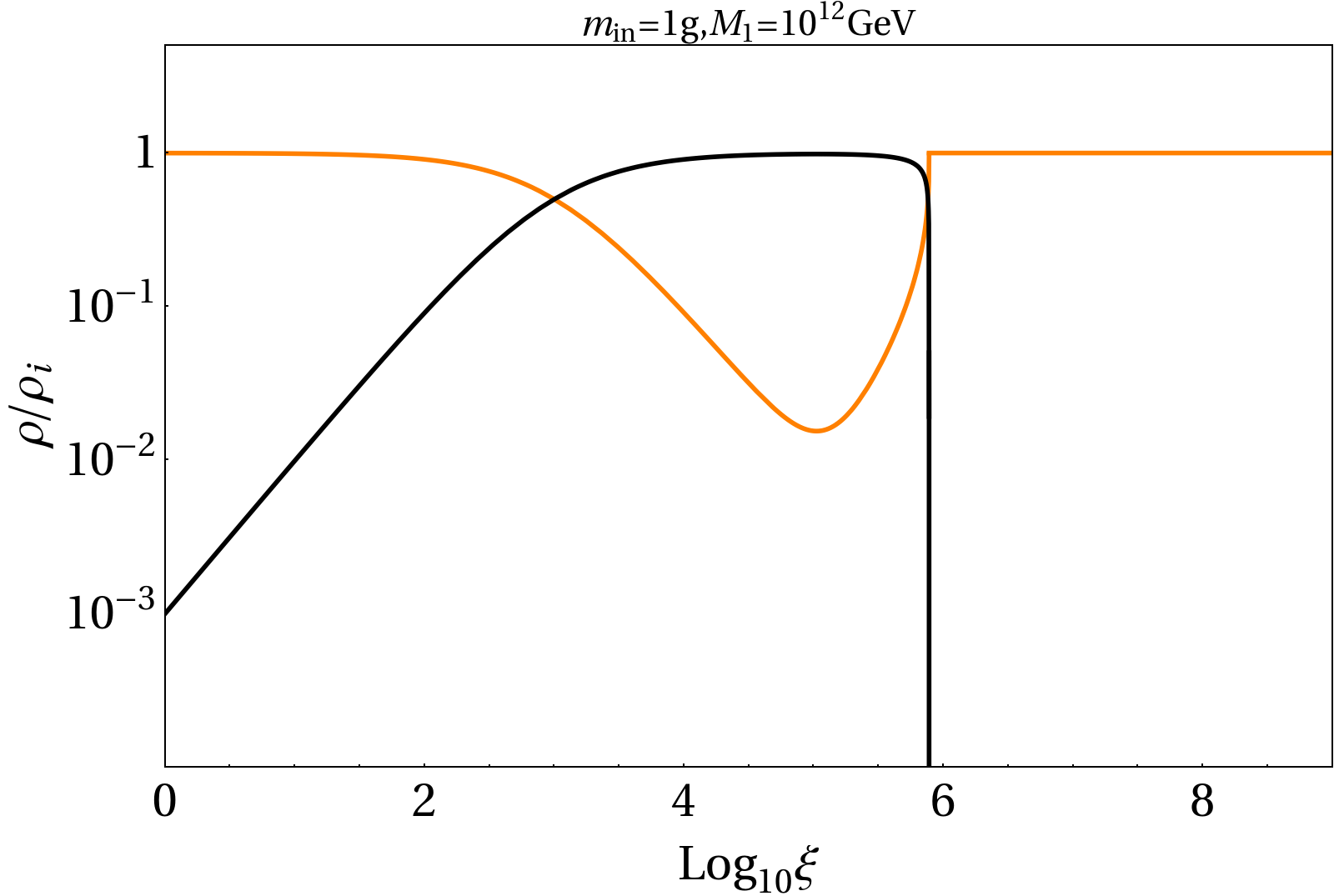}~~\includegraphics[scale=0.25]{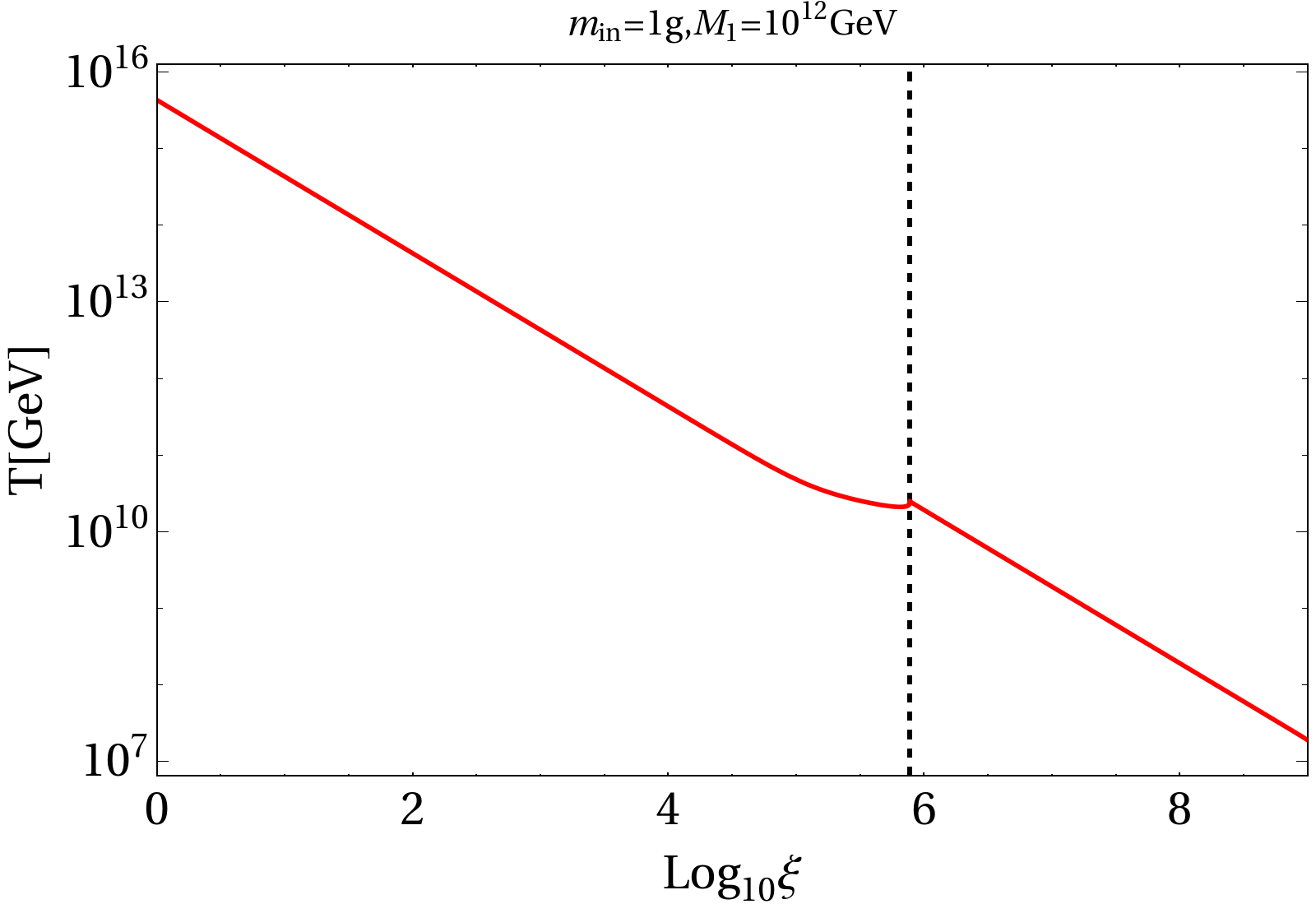}~~\includegraphics[scale=0.25]{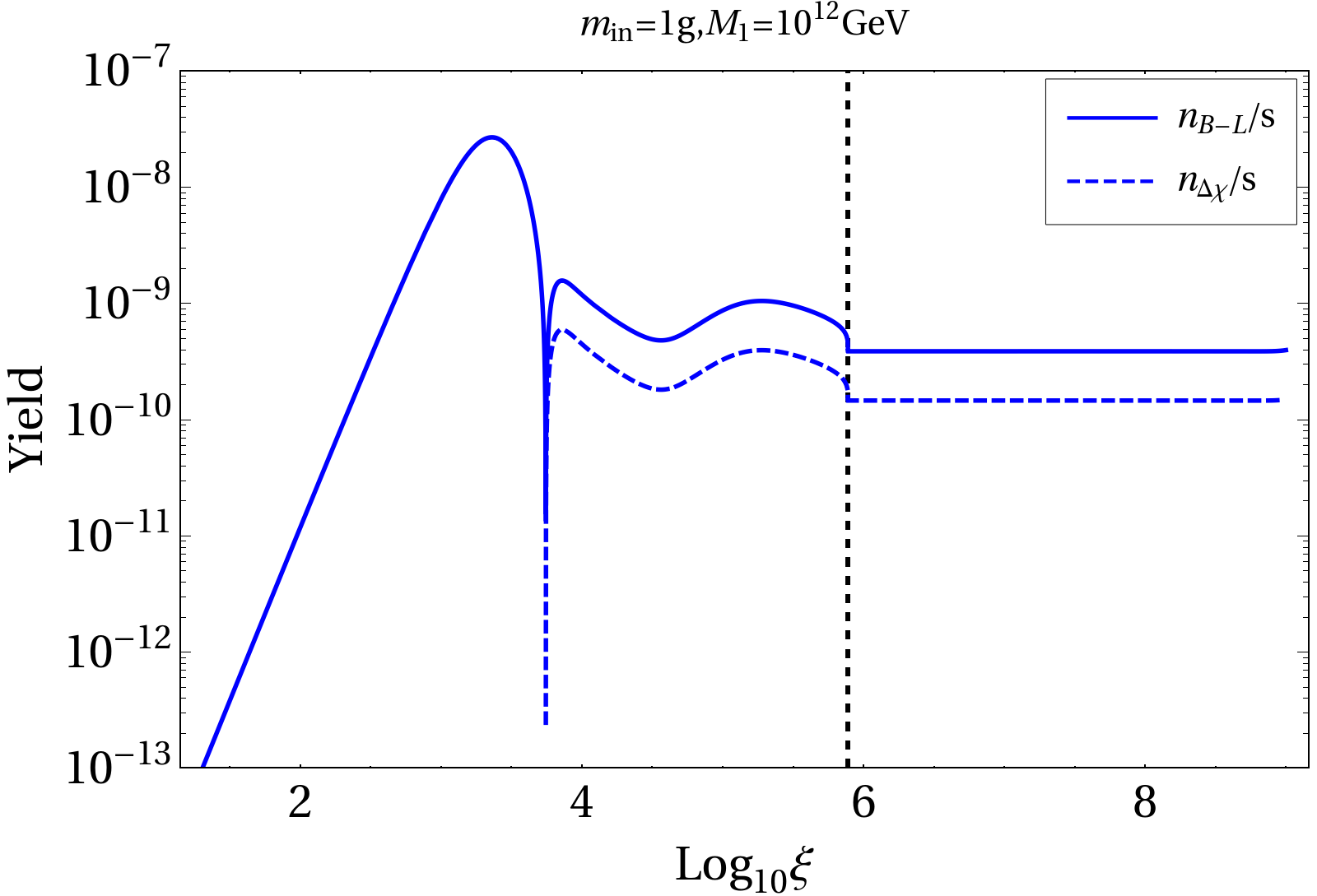}
$$
$$
\includegraphics[scale=0.25]{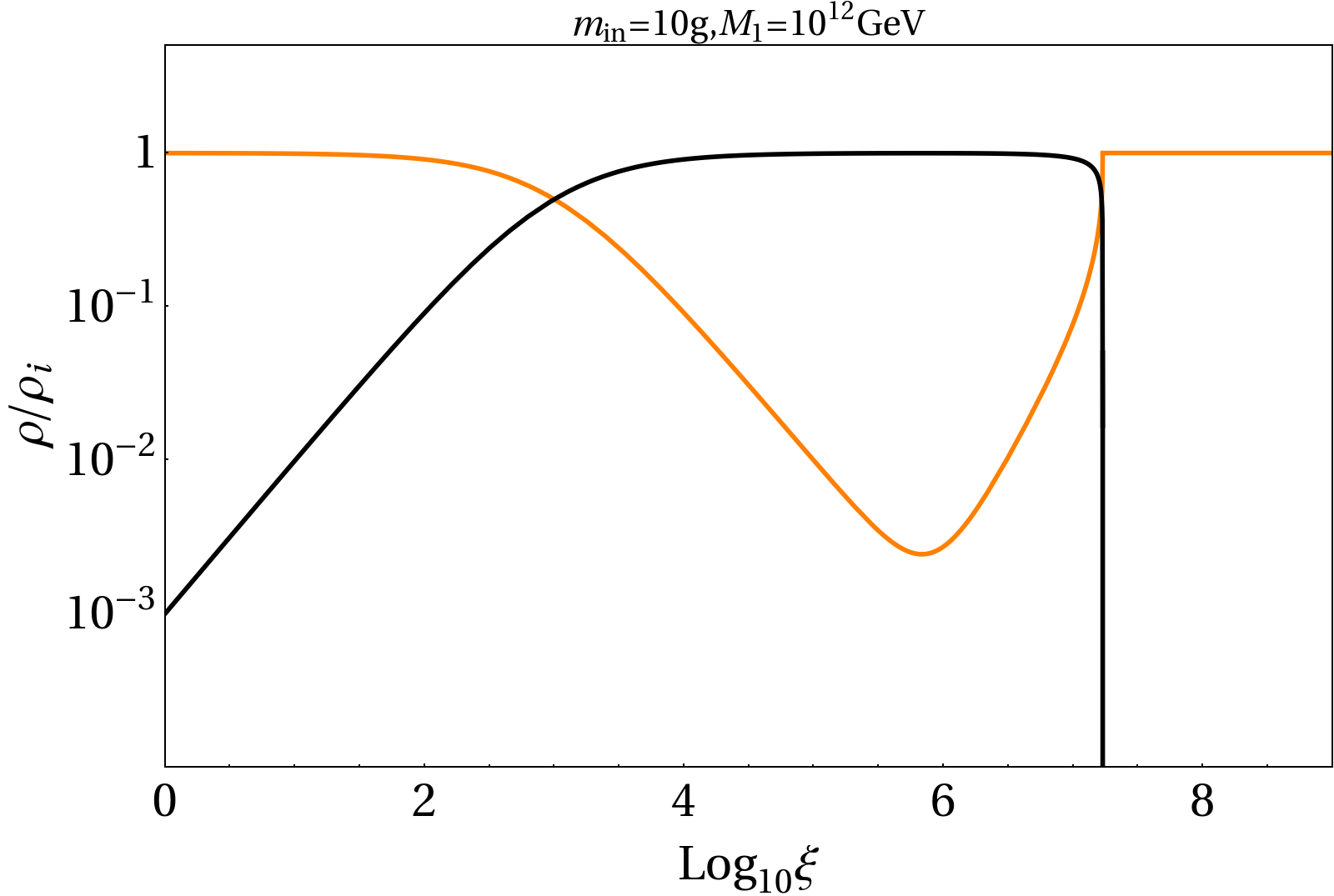}~~\includegraphics[scale=0.25]{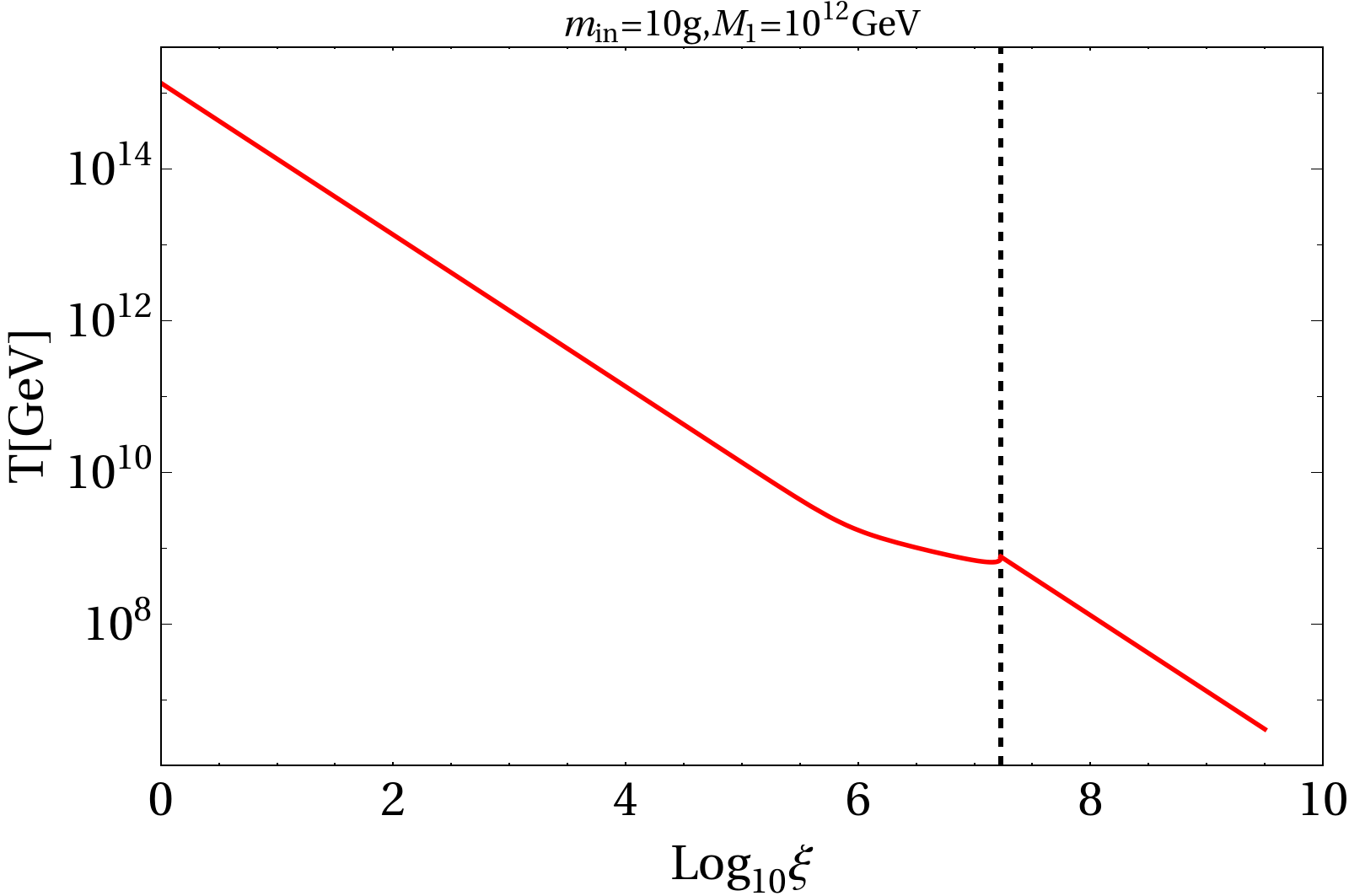}~~\includegraphics[scale=0.25]{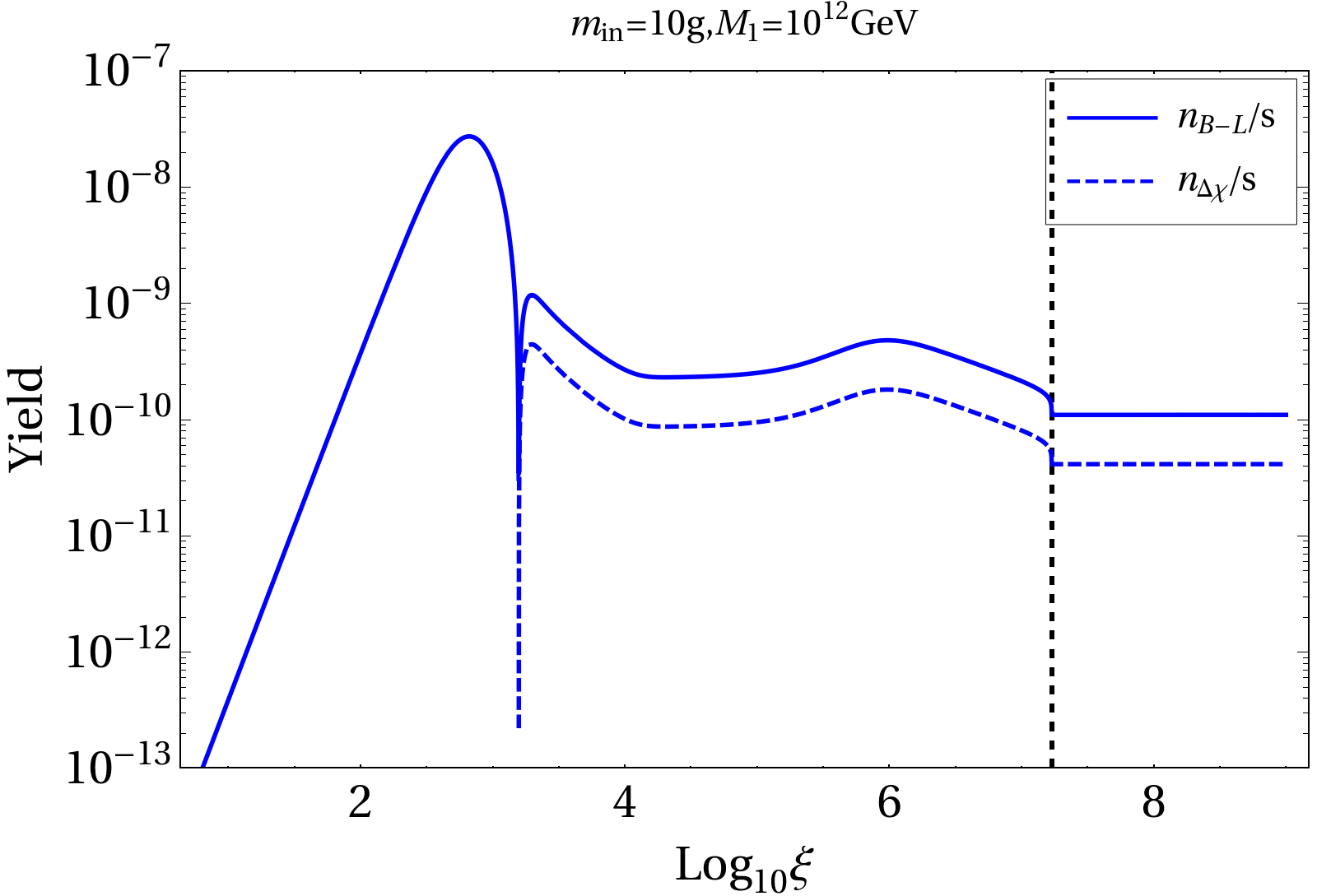}
$$
$$
\includegraphics[scale=0.25]{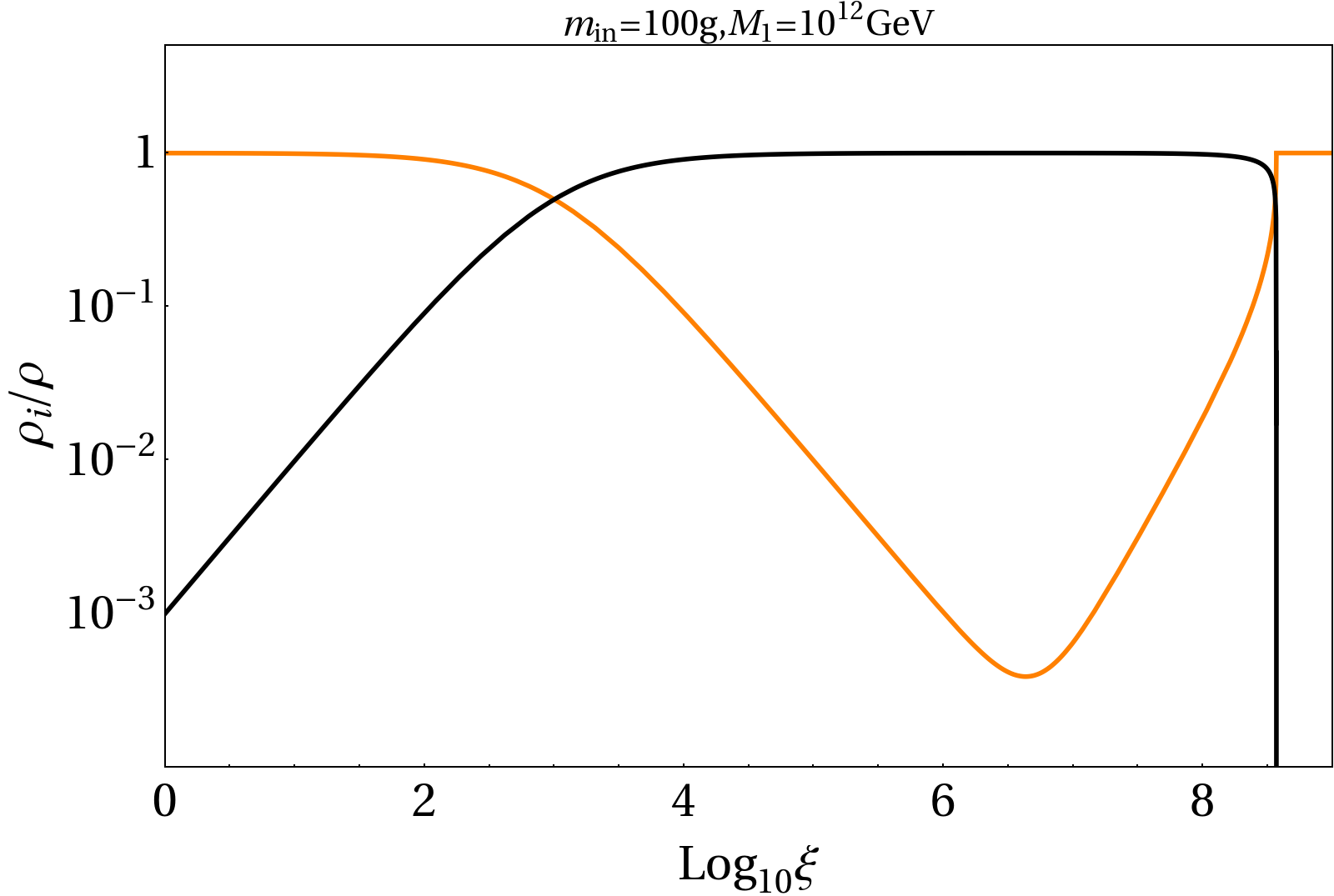}~~\includegraphics[scale=0.25]{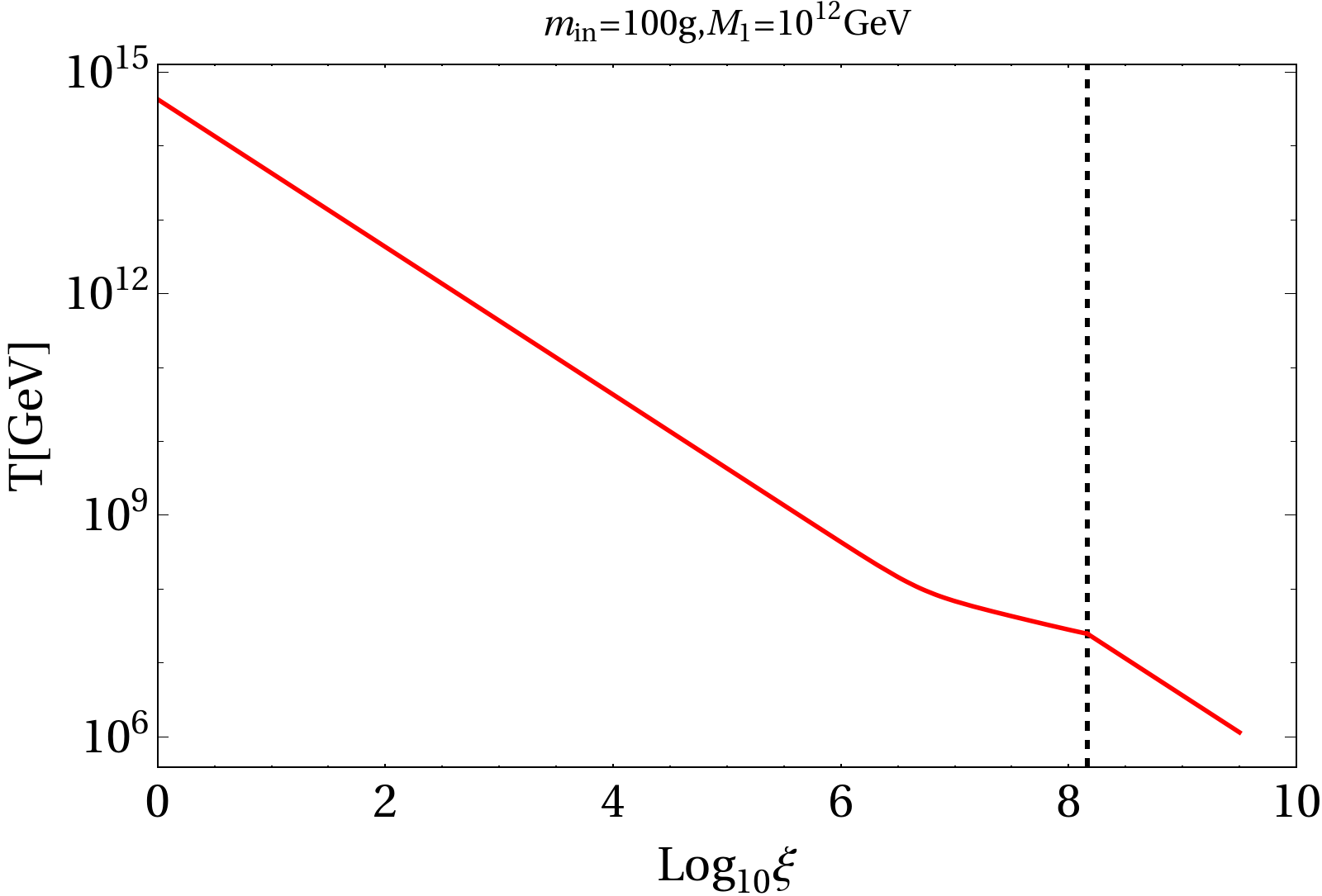}~~\includegraphics[scale=0.25]{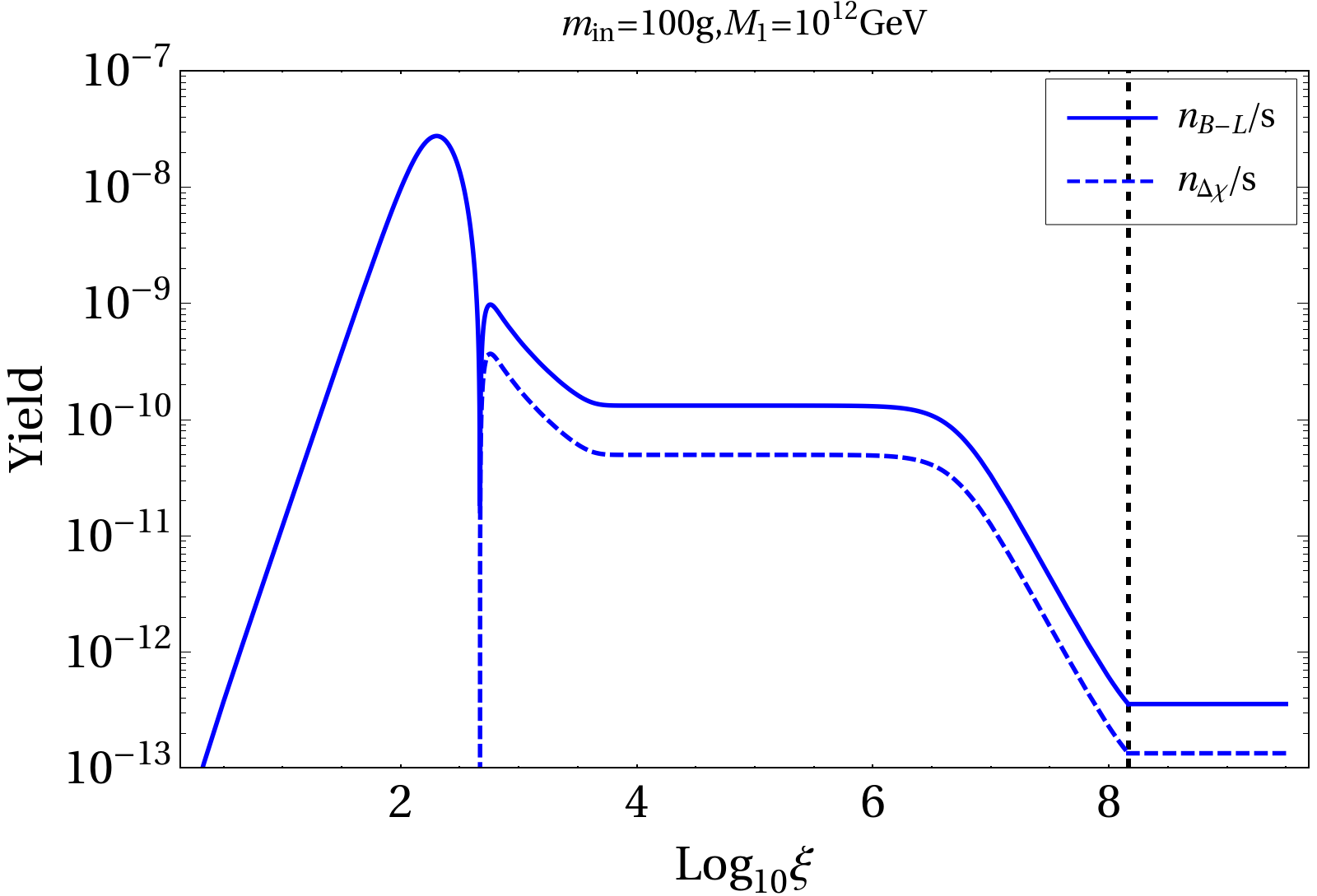}
$$
\caption{Energy density (left column), temperature of thermal bath (middle column) and yield of asymmetries (right column) as a function of scale factor for a three different choices of PBH mass and for a fixed RHN mass as mentioned in the plot label. In all cases we have considered PBH domination by considering $\beta=10^{-3}$, $M_1=10^{12}$ GeV and $y_\chi=10^{-1}$ . The black dashed vertical line in each case denotes the PBH evaporation time.}\label{fig:pbh-yield}
\end{figure}

\begin{figure}[htb!]
$$
\includegraphics[scale=0.38]{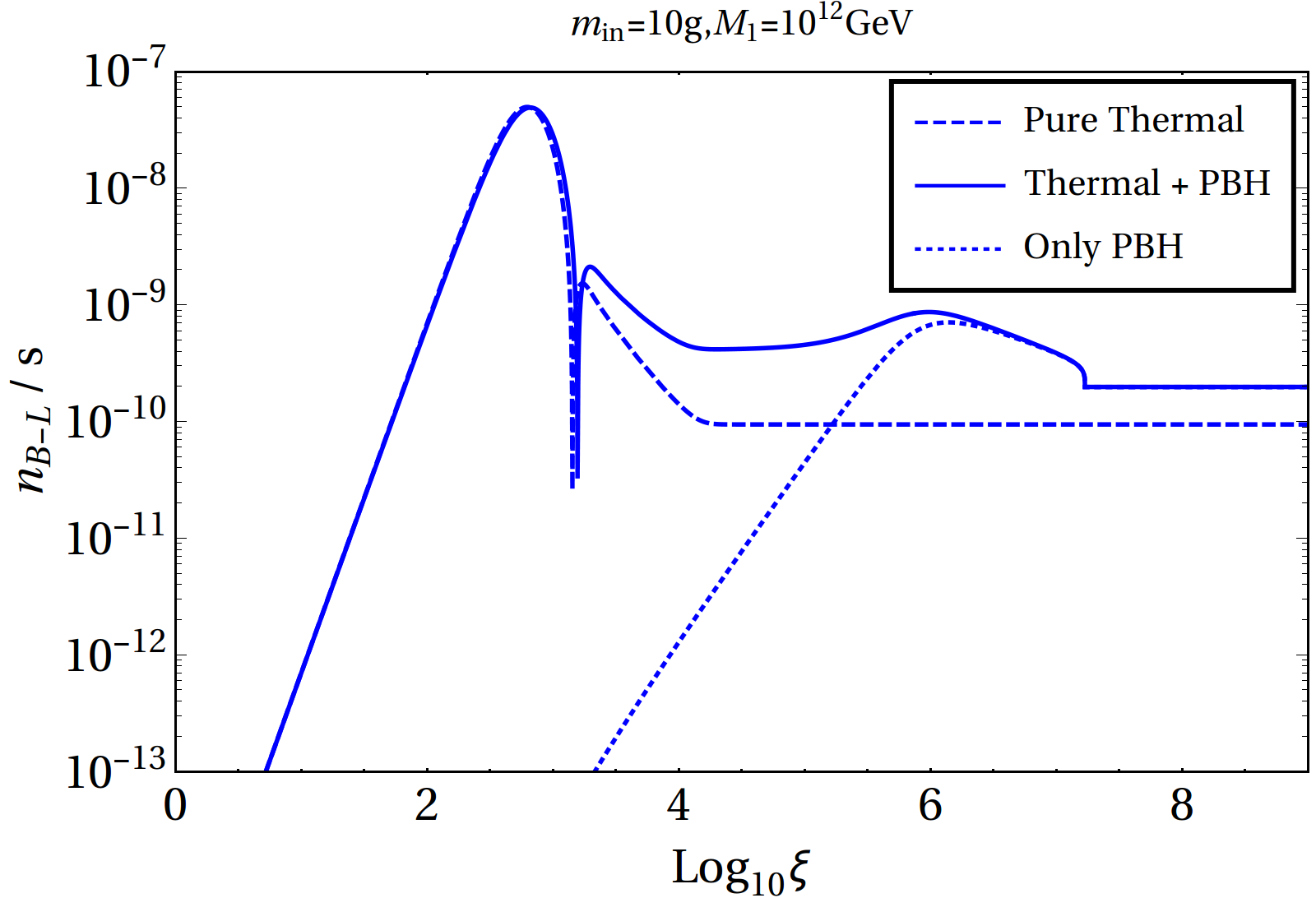}~~~~
\includegraphics[scale=0.38]{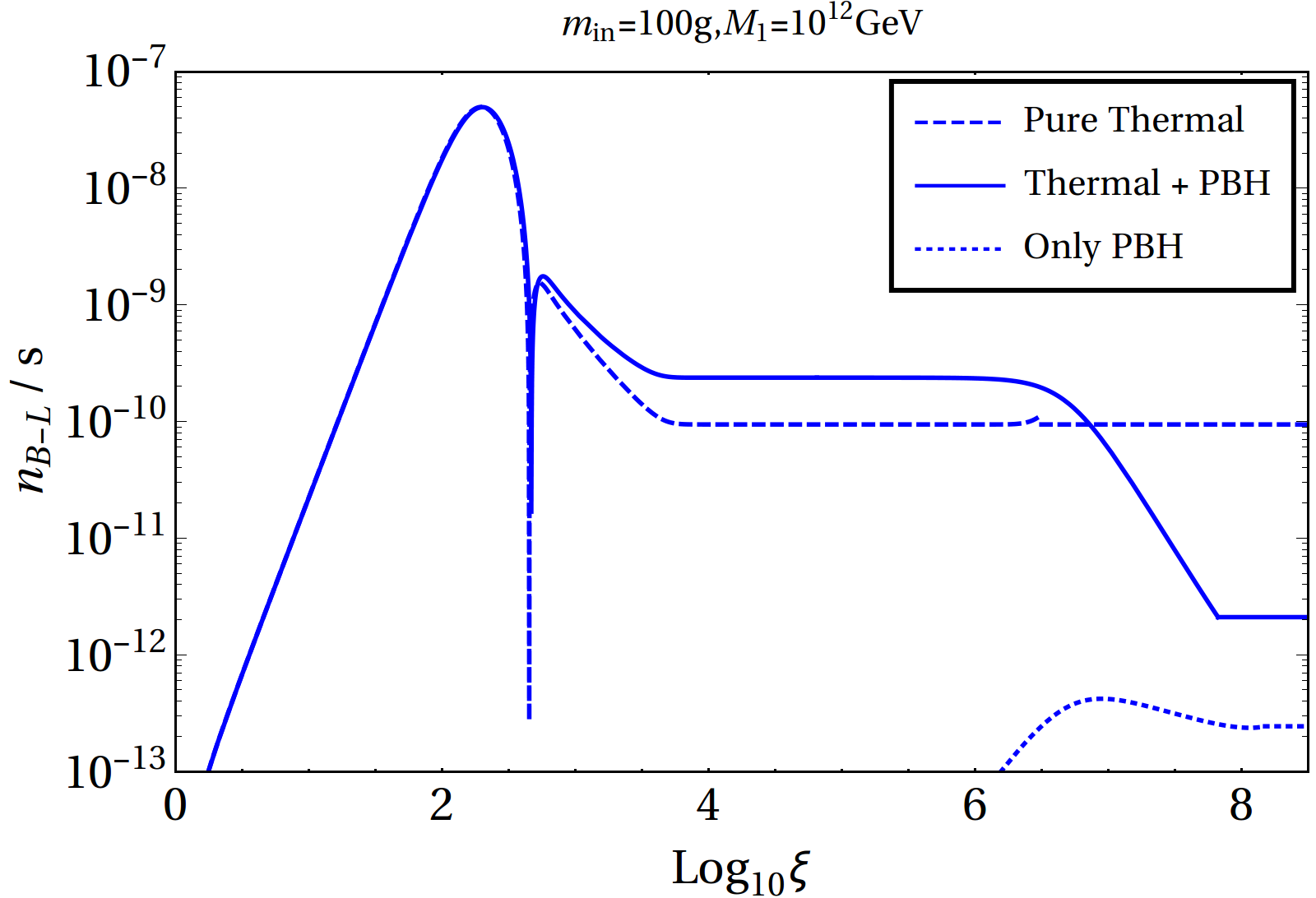}
$$
\caption{Evolution of the yield of asymmetries for three scenarios  : (i) purely thermal, (ii) only non-thermal contribution from PBH and (iii) both thermal as well as non-thermal contributions. Here, $\beta=10^{-3}$, $M_1=10^{12}$ GeV   and $y_\chi=10^{-1}$. The left panel represents a lighter PBH mass of $10$ g, whereas the right panel shows the evolution for a higher PBH mass of $100$ g.}\label{fig:pbh-comparision}
\end{figure}

The top left panel of Fig.~\ref{fig:pbh-yield} shows the evolution of radiation (orange) and PBH (black) energy densities with scale factor for PBH mass of 1 g. Here we see with time the PBH energy density rises compared to that of radiation (as $\rho_\text{BH}\sim a^{-3}$) and at around $\xi\sim 10^3$ the PBH energy density overtakes the radiation density. This corresponds to a bath temperature $T\sim 10^{12}$ GeV that can be read off from the adjacent panel on the right. Slightly beyond this point, the energy density in PBH shows a plateau. This plateau region gets broadened as the PBH mass increases (middle and bottom left panels) since a larger mass corresponds to a smaller evaporation temperature (Eq.~\eqref{eq:pbh-Tev}). As a result, for ultralight masses, PBH domination era gets over earlier. The PBH dominated era ends\footnote{PBH evaporation process, which is effectively instantaneous in sudden reheating approximation, transforms large density fluctuations into radiation and yields large pressure waves~\cite{Inomata:2019ivs,Domenech:2021wkk}.} at $\xi\sim 10^6$ for upper left panel plot where we see the black curve falls sharply. As the PBH evaporation dumps a huge amount of entropy into the thermal plasma, the plasma temperature shows a rise as one can notice from the kink in the red curve of upper middle panel plot. As the PBH mass increases, evaporation takes place at a later epoch, hence the kink in the red curve also shifts to a smaller temperatures as one can see from the lower middle panel plots. Plots shown in extreme right columns of Fig.~\ref{fig:pbh-yield} depict the evolution of yield of the asymmetries. The asymmetries in both the sectors evolve identically because of the same source. The asymmetries first increase because of the thermal contribution, and then diminish for the washout effect due to the inverse decay of thermal RHNs. Afterwards, they remain unchanged till the time the production of non-thermal RHN from PBH overtakes the thermal contribution. Then during the period of PBH evaporation dilution effect becomes significant (depending on the PBH mass) because of entropy injection in the thermal bath. Finally, the asymmetries saturate once the PBH is completely evaporated. The effect of entropy injection becomes more prominent for comparatively massive PBH as the period of evaporation becomes longer. For better understanding of the asymmetry evolution, in Fig.~\ref{fig:pbh-comparision}, we compare three scenarios where the baryon asymmetry results from the decay of (i) only thermal RHN (dashed line) in the absence of PBH, (ii) only non-thermal RHN produced from PBH (dotted line) and (iii) both thermal as well as non-thermal RHN (solid line) for two different PBH masses. These plots clearly show the difference in the yield of baryon asymmetry for these three cases. For the lighter PBH mass as shown in the left panel of Fig.~\ref{fig:pbh-comparision}, one finds that even after the entropy injection due to the PBH evaporation, the final baryon asymmetry remains larger than the one obtained in the scenario with no PBH\footnote{For comparatively lighter RHN, the effect of PBH on the final asymmetry is negligible~\cite{Datta:2020bht}, however to ensure non-thermal production we stick to $M_1\gtrsim 10^{11}$ GeV following Fig.~\ref{fig:m1-bound}.}. On the other hand, for a heavier PBH as shown in the right panel of Fig.~\ref{fig:pbh-comparision}, one expects a relatively larger entropy injection resulting in a larger dilution of the final baryon asymmetry, making it lesser in comparison to the one produced in thermal leptogenesis. Since asymmetries in dark and visible sectors evolve similarly, one can also expect that for lighter PBH masses, the asymmetric DM production will be more compared to a purely thermal scenario discussed in earlier works.

\begin{figure}[htb!]
$$
\includegraphics[scale=0.54]{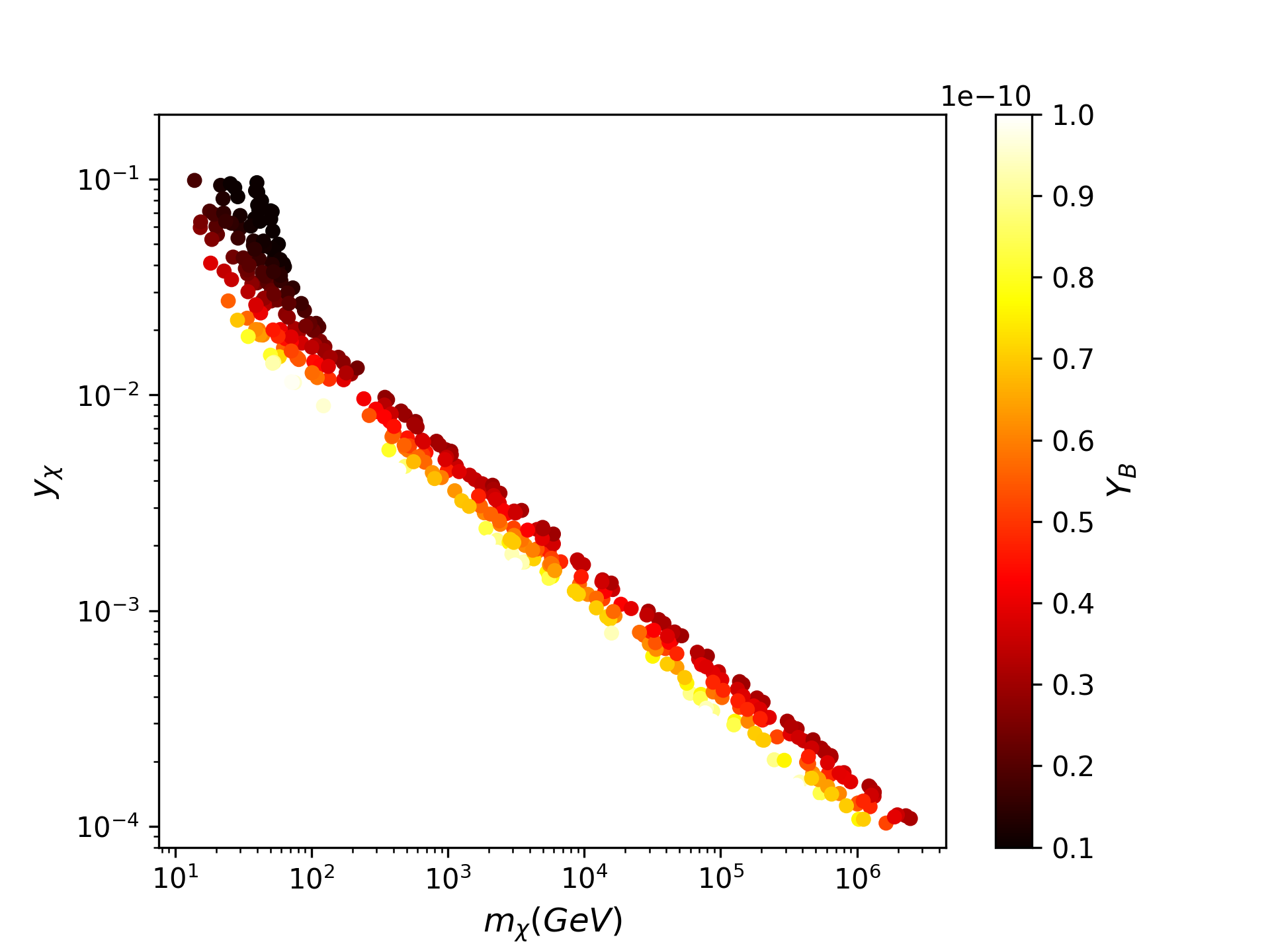}~
\includegraphics[scale=0.54]{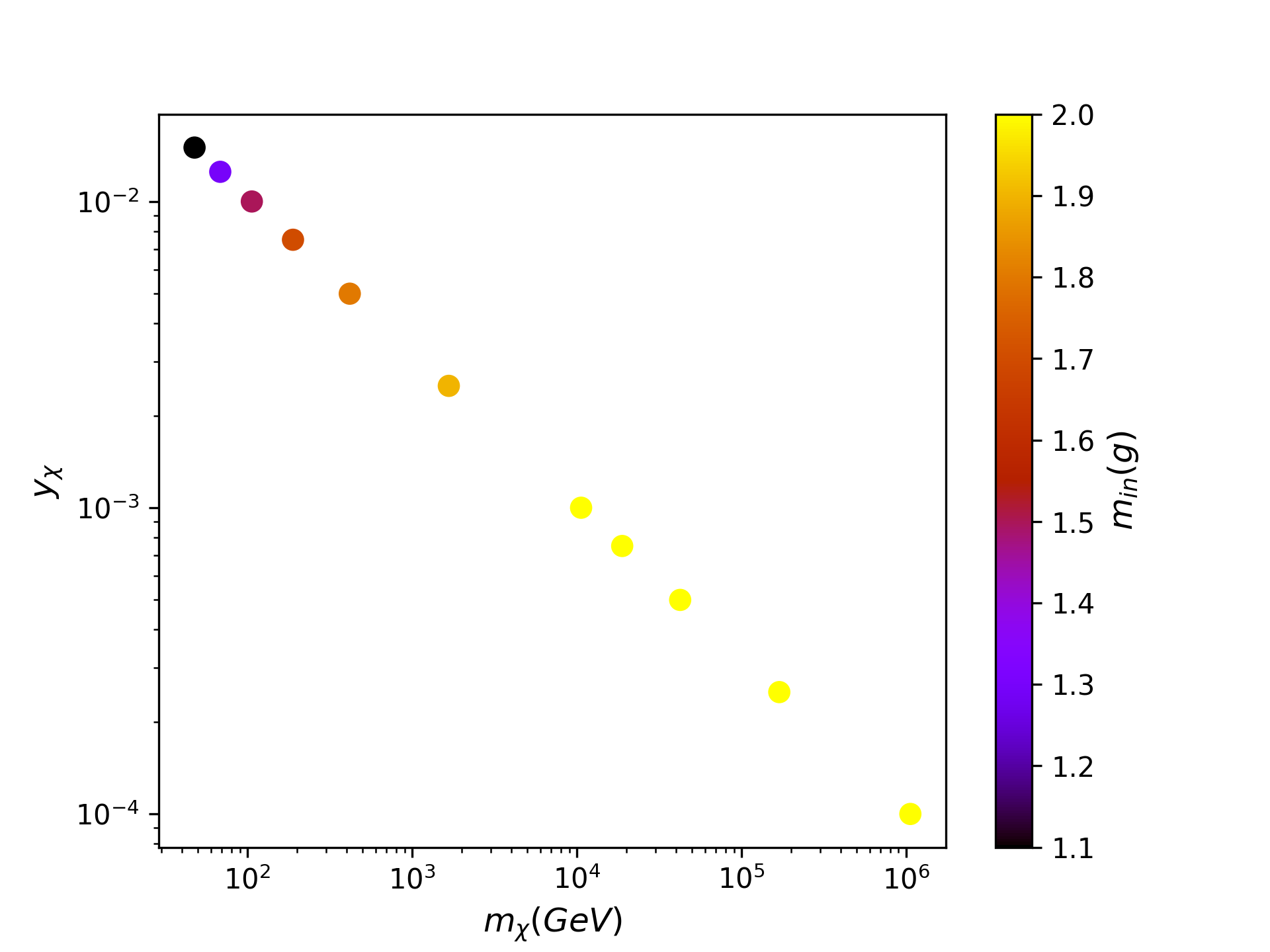}
$$
\caption{Left panel: Parameter space satisfying observed DM abundance in the bi-dimensional plane of $y_\chi-m_\chi$, where the colour coding is done with respect to $Y_B$. Right panel: Points satisfying both observed relic abundance and baryon asymmetry in $y_\chi-m_\chi$ plane, where the colour code shows variation of PBH mass. We have considered $\beta=8\times 10^{-3}$ to ensure PBH domination.}\label{fig:pbh-paramspace}
\end{figure}

In Fig.~\ref{fig:pbh-paramspace} we have illustrated the viable parameter space where asymmetric DM relic abundance and observed baryon asymmetry can be produced entirely from RHNs emitted due to PBH evaporation by solving the set of BEQs numerically. We have performed a scan over the following parameters

\begin{equation}
m_\chi: \{1-10^5\}~\text{GeV};\,m_\text{BH}:\{1-12\}~\text{g};\,y_\chi:\{10^{-4}-10^{-1}\}\,, 
\end{equation}

\noindent by keeping $M_1=10^{12}=M_2/50$ GeV and $\beta=8\times 10^{-3}$ to be fixed to ensure PBH domination as discussed before. The Dirac Yukawa coupling of neutrinos get fixed via Casas-Ibarra parametrisation mentioned earlier, after using the best-fit values of light neutrino parameters \cite{Zyla:2020zbs}. In the left panel of Fig.~\ref{fig:pbh-paramspace}, we show the allowed parameter space giving correct ADM abundance in $y_\chi-m_\chi$ plane while the colour code denotes the baryon asymmetry generated. As can be seen from this plot, for heavier DM mass, one requires smaller Yukawa coupling $y_\chi$. This is because heavier DM requires smaller dark asymmetry $\epsilon_{\Delta\chi}$ and hence smaller $y_\chi$ to generate correct relic abundance, following Eq.~\eqref{eq:pbh-dm-rel}. The parameter space gets broadened as the PBH mass keeps varying. The right panel shows, to satisfy the right ADM relic along with observed baryon asymmetry via non-thermal leptogenesis, one has to necessarily rely on ultralight PBH which we have already realized from Fig.~\ref{fig:m1-bound} and \ref{fig:mdm-bound}. The fact that heavier PBH mass requires heavier DM mass and hence smaller Yukawa $y_\chi$ to get smaller $\epsilon_{\Delta\chi}$, can also be understood from Fig.~\ref{fig:mdm-bound} based on approximate analysis. The results of complete numerical analysis also matches with this pattern as seen from the right panel plot of Fig.~\ref{fig:pbh-paramspace}.

\section{Possible UV completion of the Dark Sector}
\label{sec:uvmodel}
In the minimal setup discussed above, we have considered the dark sector to be composed of a Dirac fermion $\chi$ and a singlet scalar assuming them to be odd under a $\mathbb{Z}_2$ symmetry. The singlet scalar, being heavier than DM $\chi$, can decay into DM and SM particles. Thus, the only renormalisable interaction of DM $\chi$ is via RHN portal. The same RHN decays eventually produce the asymmetry in dark sector. Since the final relic abundance of DM is dictated by the asymmetric component only, it is important to make sure that the symmetric part annihilates away sufficiently in the early universe. This demands some new interactions of the DM as the RHN portal interaction is not strong enough to ensure that, specially when both RHN and the singlet scalar are both heavier compared to DM.

The simplest UV completion for dark sector is a dark Abelian gauge symmetry $U(1)_D$ under which DM $\chi$ and the singlet scalar $\mathcal{S}$ have equal and opposite charges, assumed to be $\pm 1$ for simplicity. The relevant dark sector Lagrangian can be written as
\begin{equation}
	\label{Lagrangian}
	\mathcal{L}_{\rm DM} \supset i ~\overline{\chi}~ \gamma^\mu~ D_\mu~ \chi + \frac{\epsilon}{2}B^{\alpha \beta}Y_{\alpha\beta}\,, 
\end{equation}
where $D_\mu = \partial_\mu + i g_D Z'_\mu$ is the covariant derivative and $B^{\alpha\beta}, Y_{\alpha \beta}$ are the field strength tensors of $U(1)_D, U(1)_Y$ respectively with $\epsilon$ being the kinetic mixing between them. Since the singlet scalar can have sizeable quartic interactions with SM Higgs, it can be produced in equilibrium, and can also bring DM into equilibrium with the SM bath for the choices of Yukawa couplings $y_\chi$ discussed before. Additionally, for sizeable kinetic mixing ($\epsilon \geq 10^{-5}$)\footnote{For a recent review on kinetic mixing bounds see~\cite{Caputo:2021eaa}.}, the DM can be in thermal equilibrium with the SM bath in the early universe via gauge portal interactions. 
Now, assuming a light $Z'$, DM and anti-DM can preferentially pair-annihilate into $Z'$ pairs with a cross-section
\begin{equation}
	\langle\sigma v\rangle \sim \frac{\pi \alpha^2_D}{m^2_{\chi}}
\end{equation}
where $\alpha_D=g^2_D/(4\pi)$. Since DM cross-section can not violate the unitarity bound \cite{Griest:1989wd}, one gets an upper bound on its mass around $10^5$ GeV so that the symmetric part can be annihilated away. One can also obtain a lower bound on ADM mass of around a few keV, from the requirement of perturbative unitarity of Yukawa couplings. We have already noticed that lighter DM requires sizeable Yukawa coupling $y_\chi$ in order to create a large asymmetry. It can be checked that for DM mass below a few keV, this Yukawa coupling will become non-perturbative. Coincidentally, this bound on DM mass is also similar to the lower bound we have in PBH scenario so that the hot DM component is restricted within $10\%$ for large PBH initial fraction $\beta$. One can put a more conservative lower bound on such DM from the requirement that the symmetric component annihilates away before the onset of BBN, in order not to inject late time entropies due to release of $Z'$ and their subsequent decays into light SM degrees of freedom. Therefore, DM mass of a GeV or heavier should be safe from such restrictions. While similar bounds exist in asymmetric DM scenarios in general (without inflation or PBH), here we have extended such scenarios to realise non-thermal origin of asymmetries aided by inflaton and PBH. 

The $U(1)_D$ sector not only ensures the annihilation of symmetric DM part, but also lead to a stable DM candidate without requiring additional discrete symmetries, depending on the charge assignments. It also forbids the Majorana mass term of $\chi$ keeping it Dirac with a conserved quantum number. While we did not discuss the origin of $Z'$ mass, it can be done either by spontaneous symmetry breaking or the Stueckelberg mechanism~\cite{Stueckelberg:1938hvi} without affecting rest of the discussions. Depending upon the gauge kinetic mixing, DM can also show up in direct detection experiments. In addition, for light mediator $Z'$, one can also realise the scenario of self-interacting dark matter (see, for example, \cite{Kaplinghat:2013yxa, Tulin:2017ara}) motivated from small-scale structure issues of ordinary cold DM\footnote{Some recent works in the direction of self-interacting DM in $U(1)_D$ model can be found in~\cite{Dutta:2021wbn, Borah:2021yek, Borah:2021pet}.}. To add to the complementarity, such dark $U(1)_D$ gauge symmetry can also lead to a first order phase transition in the early universe with observational consequences~\cite{Jinno:2016knw, Chiang:2017zbz, Hashino:2018zsi,Hasegawa:2019amx,  Mohamadnejad:2019vzg,Kim:2019ogz, Borah:2021ocu}.


\section{Conclusion}\label{sec:concl}
Asymmetric dark matter (DM) has been a well-studied framework motivated from explaining the baryon-DM coincidence problem dynamically. While the minimal frameworks to realise such possibility considers a heavy particle present in the thermal bath whose CP violating decays into visible and dark sectors generate the respective asymmetries, we consider the possibility of non-thermal origin of these asymmetries. In order to keep it minimal and also to connect to the origin of light neutrino masses, we consider the extension of Type-I seesaw model with dark sector particles~\cite{Falkowski:2011xh} so that the lightest right handed neutrino can play the role of creating the dark and visible sector asymmetries. The DM is assumed to be a singlet Dirac fermion $\chi$ which couples to the right handed neutrinos (RHNs) through another scalar singlet $\mathcal{S}$. While the interaction of the RHNs with the SM Higgs and leptons is responsible for generating the neutrino mass via Type-I seesaw mechanism, its simultaneous decay to the visible and the dark sector generates asymmetries in both the sectors. A fraction of lepton asymmetry is converted to baryon asymmetry via sphaleron transition, while the asymmetric component of $\chi$ survives and accounts for the observed DM relic. While thermal cogenesis has been discussed extensively in the literature, we consider the possibility of non-thermal RHNs by invoking the presence of additional sources. In the first attempt, we consider an inflaton field in post slow-roll stage to be the source of RHNs. The RHN subsequently decays not only to produce the dark and visible asymmetries, but leads to a brief reheating period of the universe as well. We consider the inflaton to couple only to the RHNs, while being agnostic about the details of inflaton potential and other interactions. We numerically solve a set of coupled Boltzmann equations to find the abundance of dark, visible sectors along with the reheat temperature of the universe. While a wide range of DM mass remains allowed, the RHN-DM coupling $y_\chi$ is practically a free parameter and tuned accordingly to obtain the right relic density. The RHN-SM couplings are determined from the requirement of fitting light
neutrino data. We choose the mass and couplings of the inflaton and the RHNs in such a way that the non-thermal leptogenesis scenario remains valid by requiring the RHN mass to be above the reheating temperature of the universe. Thus, this scenario connects the cogenesis of dark and visible sector asymmetries to the reheat temperature of the universe as the same non-thermal RHNs produced from inflaton decay plays non-trivial role in reheating and cogenesis. 

In the second scenario, we extend our prescription by considering a framework where the RHNs are sourced from evaporating primordial black holes that are produced in the radiation dominated era with a monochromatic mass spectrum. While RHNs can be produced from the thermal bath as well, we show that the asymmetries produced from non-thermal RHNs dominate over the thermal one, specially for lighter PBH masses. Keeping the parameter space within such ballpark where non-thermal leptogenesis from PBH evaporation dominates over the thermal contribution in generating the $B-L$ asymmetry, we find that the observed baryon asymmetry is obtainable only for ultralight PBH of mass $\lesssim 15$ g and RHN mass $M_1\gtrsim 10^{11}$ GeV. Coincidentally, the same bound on the RHN mass scale is also obtained for the inflaton case discussed before. PBH mass in such a ballpark necessarily requires a prolonged period of PBH domination, typically requiring a large initial fraction $\beta\gtrsim 10^{-3}$. Considering bounds from CMB, BBN and astrophysical constraints, we show that PBH evaporation is also capable of producing required asymmetry in the dark sector leading to correct relic abundance for asymmetric DM as massive as $\sim 10^5$ GeV, depending on the choice of the Yukawa coupling $y_\chi$.

Several complementary prospects of detection for asymmetric DM can be realised depending on the UV completion of the dark sector, which we have not investigated in this minimal setup. In addition to the discovery potential for the particular particle physics framework, the ultra-light PBH leading to early matter domination can itself have observational consequences like emission of gravitational waves via Hawking radiation \cite{Anantua:2008am} or other ways \cite{Saito:2008jc, Hooper:2020evu, Papanikolaou:2020qtd} which can have interesting detection prospects at both high and low frequency GW experiments \cite{Kozaczuk:2021wcl}. Another interesting future prospects could be to study a complete framework for baryon DM cogenesis which incorporate the details of inflationary potential or the origin of ultralight PBHs. We leave such interesting possibilities to future works.

\section*{Acknowledgements}
BB received funding from the Patrimonio Autónomo - Fondo Nacional de Financiamiento para la Ciencia, la Tecnología y la Innovación Francisco José de Caldas (MinCiencias - Colombia) grant 80740-465-2020. This project has received funding /support from the European Union's Horizon 2020 research and innovation programme under the Marie Sklodowska-Curie grant agreement No 860881-HIDDeN. 
\appendix
\section{Light neutrino mass \& Casas-Ibarra Parametrisation}\label{appen1}

The extension of the SM particle spectrum with singlet RHN allows us to write its Yukawa interaction with the SM lepton doublet and Higgs (second term in Eq.~\eqref{eq:lgrng}). As the neutral component of the SM Higgs doublet acquires a VEV leading to the spontaneous breaking of the SM gauge symmetry, neutrinos in the SM obtain a Dirac mass that can be written as

\bea
m_D= \frac{y_{N}}{\sqrt{2}}v.
\eea

\noindent The Dirac mass $m_D$ together with the RHN bare mass $M_N$, can explain the nonzero light neutrino masses with the help of Type-I seesaw~\cite{GellMann:1980vs, Mohapatra:1979ia,MINKOWSKI1977421}. Here, the light-neutrino masses can be expressed as,

\bea
m_{\nu}\simeq m_{D}^T~M^{-1}~m_{D}.
\eea

The mass eigenvalues and mixing are then obtained by diagonalising the light-neutrino mass matrix as

\bea
m_{\nu}=\mathcal{U}^* m_{\nu}^d \mathcal{U}^{\dagger}\,,
\eea

\noindent with $m_{\nu}^d=dia(m_1,m_2,m_3)$ consisting of the mass eigenvalues and $\mathcal{U}$ being the Pontecorvo-Maki-Nakagawa-Sakata matrix~\cite{Zyla:2020zbs}\footnote{The charged lepton mass matrix is considered to be diagonal.}. The interesting aspect of leptogenesis lies in the fact that the same Yukawa couplings involved in neutrino mass generation also play a non-trivial role in determining the lepton asymmetry of the universe as they dictate the decay width of RHNs into the SM leptons. In order to obtain a complex structure of the Yukawa coupling which is essential from the perspective of leptogenesis, we use the well-known Casas-Ibarra (CI) parametrisation~\cite{Casas:2001sr}. Using this one can write the Yukawa coupling $y_N$ as,

\bea
y_N = \frac{\sqrt{2}}{v}\sqrt{M}~\mathbb{R}~\sqrt{m_{\nu}^d}~\mathcal{U}^{\dagger}\,,
\label{CI}
\eea

\noindent where $\mathbb{R}$ is a complex orthogonal matrix $\mathbb{R}^T \mathbb{R} = I$, which we choose as

\begin{align}
\mathbb{R} =
\begin{pmatrix}
0 & \cos{z} & \sin{z}\\
0 & -\sin{z} & \cos{z}
\end{pmatrix}\,,
\label{eq:rot-mat}
\end{align} 

\noindent where $z=a+ib$ is a complex angle. The above structure of $\mathbb{R}$ can be justified by considering two RHNs or considering the third RHN $N_3$ to be very heavy and effectively decoupled from the bath. In such a scenario our neutrino Yukawa matrix becomes of dimension $2\times3$. Such a scenario also predicts the lightest active neutrino to be exactly massless. The digonal light neutrino mass matrix $m_{\nu}^d$ is calculable using the best fit values of solar and atmospheric mass  obtained from the latest neutrino oscillation data~\cite{Zyla:2020zbs}. Now, the elements of Yukawa coupling matrix $y_N$ for a specific value of $z$, can be obtained for different choices of the heavy neutrino masses. For example, with $M_1=10^{12}$ GeV and $\{a,b\}=\{10^{-3},10^{-2}\}$ we obtain the following structure

\bea
y_N=\left(
\begin{array}{ccc}
 0.00913\, -0.00003 i & 0.01102\, -0.00052 i &  -0.00884-0.00059 i \\
 -0.00607-0.01209 i & 0.05962\, -0.00024 i & 0.06772\, +0.00019 i \\
\end{array}
\right)\,,
\eea

\noindent which satisfies the light neutrino mass, as well as produces desired CP asymmetry in the visible sector, as we discuss below. The complex angle $z$ can be chosen in a way that the CP asymmetry is enhanced, while keeping the Yukawa couplings within perturbative limits.

\section{Thermally-Averaged cross-section with different temperatures}
\label{sec:app-sigmav}

We define the thermally averaged cross-section as

\begin{align}
&\langle\sigma v_\text{mol}\rangle=\frac{\int\sigma\,v_\text{mol}\,e^{-E_1/T_1}\,e^{-E_2/T_2}\,d^3p_1\,d^3p_2}{\int\,e^{-E_1/T_1}\,e^{-E_2/T_2}\,d^3p_1\,d^3p_2}\,.  
\label{eq:therm-av}
\end{align}

The momentum-space volume element is given by~\cite{Gondolo:1990dk}

\begin{equation}
d^3p_1\,d^3p_2=4\,\pi\,p_1\,dE_1\,4\,\pi\,p_2\,dE_2\,\frac{1}{2}\,d\cos\theta\,.
\end{equation}

We then perform the following variable transformation:

\begin{align}
& \zeta_+\equiv\frac{E_1}{T_1}+\frac{E_2}{T_2}\nonumber\\&
\zeta_-\equiv \frac{E_1}{T_1}-\frac{E_2}{T_2}\nonumber\\&
s \simeq M_1^2+2(E_1\,E_2-p_1\,p_2\,\cos\theta)\,.
\end{align}

Using the Jacobian transformation

\begin{equation}
\mathcal{J} = \frac{1}{2}\,\text{det} 
\begin{pmatrix}
T_1 & T_1 & 0 \\
T_2 & -T_2 & 0\\
0 & 0 & -\frac{1}{p_1\,p_2}
\end{pmatrix}=\frac{T_1\,T_2}{4\,p_1\,p_2}\,,
\end{equation}

\noindent the volume element turns out to be

\begin{equation}
d^3p_1\,d^3p_2 =  \frac{1}{2}\,16\pi^2\,p_1\,p_2\,E_1\,E_2\,\mathcal{J}\,d\zeta_+\,d\zeta_-\,ds = 2\pi^2\,E_1\,E_2\,T_1\,T_2\,d\zeta_+\,d\zeta_-\,ds\,.    
\end{equation}

Now, the viable integration region: $E_1\geq M_1\,,E_2\geq M_1\,,|\cos\theta|\leq 1$ can be translated to the integration limits on the new variables as

\begin{align}
&\left|\cos\theta\right| =\frac{M_1^2+2\,E_1\,E_2-s}{2\,p_1\,p_2} = \frac{2s-2M_1^2+T_1\,T_2\,(\zeta_-^2-\zeta_+^2)}{T_2\,(\zeta_--\zeta_+)\,\sqrt{T_1^2\,(\zeta_++\zeta_-)^2-4\,M_1^2}}
\nonumber\\&
\implies \frac{M_1^2\,T_2^2\,\zeta_+-\sqrt{T_2 \left(s-M_1^2\right)^2\, \left[M_1^2\,(T_1-T_2)+T_1 \left(T_1\,T_2\, \zeta_+^2-s\right)\right]}}{T_2 \left[M_1^2\,(T_2-T_1)+s T_1\right]}\leq\zeta_-
\nonumber\\&
\leq \frac{M_1^2\,T_2^2\,\zeta_++\sqrt{T_2 \left(s-M_1^2\right)^2\, \left[M_1^2\,(T_1-T_2)+T_1 \left(T_1\,T_2\, \zeta_+^2-s\right)\right]}}{T_2 \left[M_1^2\,(T_2-T_1)+s T_1\right]}\,,
\label{eq:zetmn}
\end{align}

\noindent where by demanding the expression inside the square root in the above equation to be real, we obtain

\begin{equation}
|\zeta_+| \geq \sqrt{\frac{s\,T_1+M_1^2\,(T_2-T_1)}{T_1^2\,T_2}}\,. 
\label{eq:zpl}
\end{equation}

Now, the numerator of Eq.~\eqref{eq:therm-av} reads

\begin{align}
& \int\sigma v_\text{mol}\,e^{-\zeta_+}\,2\pi^2\,E_1\,E_2\,T_1\,T_2\,d\zeta_+\,d\zeta_-\,ds
%
\nonumber\\&
= 2\,\pi^2\,T_1\,T_2\int_{M_1^2}^\infty ds\,\int_{\zeta_+^\text{min}}^\infty\,d\zeta_+\,\sigma\,\Biggl[\frac{s}{2}\,\sqrt{1-\frac{2\,M_1^2}{s}+\frac{M_1^4}{s^2}}\Biggr]\,\mathcal{G}(s,\zeta_+\,...)\,e^{-\zeta_+}\,,
\end{align}

\noindent where $\mathcal{G}(s,\zeta_+\,...)\equiv\,(\zeta_-^\text{max}-\zeta_-^\text{min})$ is obtained from Eq.~\eqref{eq:zetmn} and $\zeta_+^\text{min}$ comes from Eq.~\eqref{eq:zpl}.

The denominator can similarly be written as

\begin{align}
&\int\,d^3p_1\,d^3p_2\,e^{-(E_1/T_1+E_2/T_2)} 
= 32\,\pi ^2 M_1^2\,T_1\,T_2^3 K_2\left(\frac{M_1}{T_1}\right)\,.
\end{align}


Therefore, the final expression turns out

\begin{align}
& \langle\sigma v\rangle =\frac{1}{16\,M_1^2\,T_2^2}\,\frac{1}{K_2(M_1/T_1)}\,\int_{M_1^2}^\infty ds\,\int_{\zeta_+^\text{min}}^\infty\,d\zeta_+\,\sigma\,\Biggl[\frac{s}{2}\,\sqrt{1-\frac{2\,M_1^2}{s}+\frac{M_1^4}{s^2}}\Biggr]\,(\zeta_-^\text{max}-\zeta_-^\text{min})\,e^{-\zeta_+}\,,
\end{align}

\noindent where $\zeta_+^\text{min}$ can be obtained from Eq.~\eqref{eq:zpl}. We also verify that the above expression reproduces Eq.(B13) of Ref.~\cite{Cheek:2021cfe} upon considering both the initial state particles to be massive with degenerate masses. 




    
\bibliography{Bibliography}

\end{document}